\documentclass[%
 prd,
 twocolumn,
 superscriptaddress,
 numerical,
 showpacs,
 amsmath,amssymb,
 aps,
nofootinbib,
longbibliography,
 floatfix
]{revtex4-1}

\usepackage{bm}
\usepackage{bbm}
\usepackage{soul}
\usepackage{cancel}
\usepackage{braket}
\usepackage{slashed}
\usepackage{verbatim}
\usepackage{graphicx}
\usepackage{hyperref}
\usepackage{combelow} 
\usepackage{mathtools}
\usepackage[usenames,dvipsnames,svgnames,table]{xcolor}
\hypersetup{
  colorlinks,
  citecolor=Violet,
  linkcolor=Blue,
  urlcolor=teal}

\let\oldciteauthor=\citeauthor
\def\citeauthor#1{\hypersetup{citecolor=black}\oldciteauthor{#1}}

\let\oldciten=\onlinecite
\def\onlinecite#1{\hypersetup{citecolor=blue}\oldciten{#1}}

\let\oldcite=\cite
\def\cite#1{\hypersetup{citecolor=blue}\oldcite{#1}}

\newcommand{\beqn}{\begin{eqnarray}}
\newcommand{\eeqn}{\end{eqnarray}}
\newcommand{\beqs}{\begin{subequations}}
\newcommand{\eeqs}{\end{subequations}\\[-2mm]\noindent}

\definecolor{purple}{rgb}{0.8,0,0.6}
\definecolor{PURPLE}{rgb}{0.8,0,0.6}
\definecolor{orange}{rgb}{1,0.55,0}
\definecolor{limegreen}{rgb}{0.2,0.8,0.2}
\definecolor{yeswecolor}{rgb}{0.1,0.8,0.3}

\definecolor{battleshipgrey}{rgb}{0.2, 0.52, 0.51}

\allowdisplaybreaks
\begin{document}

\title{Firewall boundaries and mixed phases of rotating quark matter in linear sigma model}

\author{Sergio Morales Tejera}
\affiliation{Department of Physics, West University of Timi\cb{s}oara, Bd.~Vasile P\^arvan 4, Timi\cb{s}oara 300223, Romania}

\author{Victor E. Ambru\cb{s}}
\thanks{Corresponding author: victor.ambrus@e-uvt.ro.}
\affiliation{Department of Physics, West University of Timi\cb{s}oara, Bd.~Vasile P\^arvan 4, Timi\cb{s}oara 300223, Romania}

\author{Maxim N. Chernodub}
\thanks{Corresponding author: maxim.chernodub@univ-tours.fr}
\affiliation{Institut Denis Poisson, CNRS - UMR 7013, Universit\'e de Tours, 37200 France}
\affiliation{Department of Physics, West University of Timi\cb{s}oara, Bd.~Vasile P\^arvan 4, Timi\cb{s}oara 300223, Romania}

\begin{abstract}
A rigidly-rotating body in unbounded space is usually considered a pathological system since it leads to faster-than-light velocities and associated breaches of causality. However, numerical results on chiral symmetry breaking in rotating plasmas of interacting fermions reveal surprisingly close correspondence in predictions between the rigorous bounded and formal unbounded approaches. 
To provide insight into this correlation, we consider the linear sigma model coupled to quarks, undergoing rigid rotation in unbounded Minkowski space-time. 
Within the mean-field approach, we adopt three consecutive levels of approximation to the ground state of the system that feature uniform (model 1), weakly inhomogeneous (model 2) and fully inhomogeneous (model 3) condensates. Models 1 and 2 that do not take into account spatial gradients of the condensate show agreement with the Tolman-Ehrenfest law. Model 3 exhibits a deviation from the Tolman-Ehrenfest prediction due to the appearance of a new energy scale set by the inhomogeneity of the ground state.
Its boundary conditions are fixed by imposing regularity at the rotation axis and by demanding the global minimization of the grand potential. We dub the latter as ``firewall boundary conditions,'' translating into the requirement of vanishing condensate on the light cylinder, which follows from the fact that the system state formally diverges at the light cylinder.
In all models, we present the phase diagram of the system and point out that in models 2 and 3, the system resides either in a chirally-restored phase, or in a mixed phase that possesses spatially-separated chirally-restored and chirally-broken phases. Finally, we discuss the properties of the system under inhomogeneous rotation using the relativistic version of the Rankine vortex model.
\end{abstract}

\maketitle

\section{Introduction}\label{sec:intro}

The last decade has seen a renewed interest in rotating systems due to the experimental evidence of persistent polarization of the hadrons born out of the rotating quark-gluon plasma (QGP) formed in non-central ultrarelativistic heavy-ion collisions, measured by the STAR collaboration via the decay of hyperons at the relativistic heavy-ion collider (RHIC) at the Brookhaven National Laboratory (BNL) \cite{STAR:2017ckg}. One important question is related to the influence of rotation on the thermodynamics of strongly-interacting matter, namely whether rotation promotes or inhibits the chiral symmetry restoration and/or the deconfinement transition. Rigid rotation has emerged as the preferred toy model to investigate the properties of quantum matter under rotation.

The rigid rotation of a physical body requires that the system be bounded at a certain distance from the rotation axis in order to avoid the violation of causality. The broken causality becomes evident when a point on the rotating body surpasses the speed of light at a finite distance from the rotation axis. For uniform rigid rotation, this phenomenon manifests itself at the light cylinder, where the temporal component of the co-rotating metric becomes null, establishing the location of the singularity within the rotating space. By the Tolman-Ehrenfest law~\cite{Tolman:1930ona, Tolman:1930zza}, the local temperature diverges on the light cylinder. Disregarding the necessity of imposing boundary conditions is recognized as yielding artificial outcomes~\cite{Davies:1996ks,Nicolaevici:2001yy}.

In the absence of boundaries, a rotating system develops several pathologies. For a classical system (e.g., described in kinetic theory \cite{Cercignani:2002,Ambrus:2016ocv}), the temperature and related observables (e.g., the energy-momentum tensor) blow up at the light cylinder. A quantum system is more subtle. Without the transverse momentum quantization enforced by boundary conditions, the system supports infrared, ``superhorizon'' modes that extend beyond the light cylinder. For fermions, this leads to a discrepancy between the rotating and static vacua \cite{Iyer:1982ah,Ambrus:2014uqa}. For bosons, the consequence is more dramatic: since these finite-momentum superhorizon modes can have vanishing corotating energy, their number density can increase without bounds, leading to a catastrophic divergence of the state at all points in the system (including the spatial domain inside the light cylinder)~\cite{Vilenkin:1980zv,Frolov:1987dz,Ottewill:2000yr,Duffy:2002ss,Ambrus:2014uqa}.

Rotating states are characterized by a radial inhomogeneity of their thermal properties: the local temperature is proportional to the Lorentz factor of a corotating observer, increasing away from the rotation axis towards infinity on the light cylinder. This property has been confirmed also for the quantum scalar \cite{Duffy:2002ss} and Dirac~\cite{Ambrus:2014uqa, Ambrus:2015lfr} fields, and it has been further generalized to phenomenologically interesting models of interacting quarks~\cite{Wang:2019nhd, Zhang:2020hha, Chen:2022mhf}. For the parameter range that puts the system near a phase transition, these spatial inhomogeneities can promote the appearance of new mixed phases in which spatially separated regions with unequivalent symmetries and different condensates may coexist~\cite{Chernodub:2020qah, Braguta:2023iyx, Jiang:2024zsw, Braguta:2024zpi}. For a uniformly rotating medium, these regions are separated along the radial direction.

Restricting ourselves to effective models that mimic the behavior of the interacting quarks in QCD, such as the linear sigma model coupled to quarks~\cite{Gell-Mann:1960mvl} or Nambu--Jona-Lasinio (NJL) \cite{Nambu:1961tp,Nambu:1961fr} models -- one can find that several works disregarding the boundary conditions~\cite{Chen:2015hfc, Jiang:2016wvv, Sun:2023kuu, Wang:2018sur, Sun:2023yux, TabatabaeeMehr:2023tpt, Gaspar:2023nqk, Hernandez:2024nev} still obtain results consistent with the ones in which the boundary conditions are strictly imposed~\cite{Chernodub:2017ref, Chernodub:2016kxh, Zhang:2020hha, Sadooghi:2021upd, Mehr:2022tfq, Chen:2023cjt, Singha:2024tpo}. However, in the latter approaches, some ambiguity appears, as the type of the boundary affects the properties of the plasma in its vicinity. For example, the so-called MIT boundary conditions~\cite{Chodos1974} break the chiral symmetry explicitly and, therefore, lead to enhanced chiral symmetry breaking near the boundary of the system~\cite{Chernodub:2016kxh}. The spectral boundary conditions~\cite{Atiyah:1975jf} do not break the chiral symmetry explicitly but may still induce a mass gap in the bulk of the system due to the finite-size effects~\cite{Singha:2024tpo}. Notice that even without rotation, the presence of boundary conditions expectedly affects the finite-volume thermodynamics associated with the chiral symmetry~\cite{Chernodub:2016kxh, Chen:2022mhf, Kovacs:2023kbv}. It is important to remark that, besides avoiding singularities, boundary conditions also make the system closed in a thermodynamic sense.

The aforementioned studies were done in the mean-field approximation, under which the meson fields (in the LSM model) or the chiral condensate (in the NJL model) are approximated as classical fields. Most commonly, these condensates are treated as constant~\cite{Chernodub:2016kxh,Chernodub:2017ref,Singha:2024tpo} or slowly varying (i.e., their gradients are negligible)~\cite{Sun:2024anu,Wang:2018sur,Sun:2023yux,Sun:2023kuu,Chen:2023cjt} functions of spatial coordinates. Remarkably, studies performed with rigorous boundary conditions \cite{Chernodub:2016kxh,Chernodub:2017ref,Singha:2024tpo} give results that are in qualitative agreement with the approaches that do not impose boundary conditions
\cite{Sun:2024anu,Wang:2018sur,Sun:2023yux,Sun:2023kuu,Chen:2023cjt}. The latter studies are possible since, in the mean field approximation, the bosonic (meson and/or gluon) fields are classical fields; hence, the resulting quantum state is regular within the light cylinder.\footnote{The state is regular in the sense that the expectation values of quantum operators (e.g., fermion condensate or energy-momentum tensor) is finite up to the light cylinder \cite{Ambrus:2014uqa}.}

In this paper, we work within the linear sigma model coupled to quarks (LSM$_q$) with the aim of elucidating the impact of condensate inhomogeneities on the ground state of the system. We work within four major simplifying assumptions: first, the quantum fluctuations of the meson fields are neglected; second, we do not impose boundary conditions on the fermion fields, thereby ignoring boundary effects; third, we approximate the density operator $\hat{\rho}$ defining the rigidly-rotating quantum state by its local thermal equivalent, by which the state of rigid rotation is retained at a purely kinematic level, ignoring quantum corrections that are typically of quadratic or higher order in the rotation angular velocity $\Omega$. 
This approach is convenient from a computational point of view, as all expectation values can be computed in momentum-space, via a one-dimensional integral with respect to the particle energy. A fully-rigorous treatment of the quantum expectation values via the Bessel-Fourier series \cite{Ambrus:2014uqa,Ambrus:2019khr,Ambrus:2019ayb} would greatly increase the computational complexity of the project. We plan to further address the role of quantum corrections in a forthcoming publication.
The fourth approximation employed in our work refers to the evaluation of expectation values in the presence of an inhomogeneous condensate. This involves evaluating the path integral with point-dependent effective mass, which can be achieved, e.g., by constructing a basis that diagonalizes the corresponding Hamiltonian, as discussed in Refs.~\cite{Wang:2018zrn,Wang:2019nhd}. As in the case of incorporating quantum corrections, this approach would greatly increase the computational complexity of our project. We therefore work in the simplified approximation that the scales on which the condensate varies are much larger than the strong-interaction scale, and thus we evaluate expectation values as if the fermion field would have a constant mass given by the local value of the sigma condensate.

Our goal is to extend the mean-field approach by incorporating the spatial gradients of the condensate, which become important especially close to the light cylinder. Throughout this work, we focus on the mean-field value of the sigma meson and compare three models of increasing degree of accuracy. 

In model 1, we consider that the $\sigma$ condensate is constant, $\sigma(x)\to\bar{\sigma}$, its value being given by the requirement of minimizing the grand potential $\Phi$. The system itself is regarded as a cylinder of radius $R$. Within model 1, we study the chiral symmetry restoration for various system sizes, $0 < \Omega R \le 1$. We show that the thermodynamic phase of the system can be well understood based on a spatial average of the local temperature and chemical potential predicted by the Tolman-Ehrenfest law.

In model 2, we allow for $\sigma$ to be a function of the distance $\rho$ to the rotation axis. Its value is then obtained by demanding the minimization of the grand potential $\Phi$ locally, at each point inside the system. At the same time, within model 2, we assume that $\sigma(x)$ is slowly varying, thereby neglecting its spatial gradients. The local slowly-varying condensate $\sigma(x)$ can describe the possible inhomogeneous phases of the system, corresponding to the phase of the equivalent static system residing at the local temperature and chemical potential, as given by the Tolman-Ehrenfest law.

We now move on to the main focus of our work: in model 3, we take into account the spatial gradients of the $\sigma$ condensate. Imposing the local minimization of the grand potential $\Phi$ with respect to $\sigma(\rho)$ leads to the Klein-Gordon (KG) equation, which we solve numerically in order to find $\sigma(\rho)$. As a second-order differential equation, the KG equation has two integration constants. One of them is fixed by requiring regularity on the rotation axis. The second is fixed by demanding the global minimization of $\Phi$, computed over the system of size $R$. Focusing on the case when the system extends to the light cylinder, we find that the minimization of $\Phi$ imposes naturally that $\sigma$ vanishes on the light cylinder. We coin this emergent boundary condition as ``the firewall boundary condition,'' as it emerges naturally as a response to the kinematical state of the system.  As with model 2, the system resides either fully in the chirally-restored phase, or in a mixed phase, being chirally-broken in the vicinity of the rotation axis and chirally-restored towards the light cylinder. 

As opposed to models 1 and 2, where the rotation parameter $\Omega$ serves only to define a length scale, entering exclusively under combinations of the form $\rho \Omega$, in model 3, the spatial gradients in the Klein-Gordon equation allow $\Omega$ to enter on its own, as an energy scale. One important consequence is that the value $\sigma_0$ of the condensate on the rotation axis becomes a function of $\Omega$. This is contrary to model 2, where $\sigma_0$ is fixed only by the temperature and chemical potential on the rotation axis. We thus discuss the phase diagram of the system as a whole, with respect to the transition from a mixed phase to a chirally-restored phase, at the level of $\sigma_0$. We find that increasing $\Omega$ promotes the chiral restoration.  Moreover, when $\Omega$ exceeds a critical value $\Omega_c = 175$ MeV, the system is in the chirally-restored phase, regardless of its temperature or chemical potential.

Finally, we apply the above methodology to the more realistic case of an inhomogeneous rotation profile. Since states which are not rigidly rotating are not in thermal equilibrium, such profiles suffer viscous damping, as illustrated in Refs.~\cite{Florkowski:2018csl,Gabbana:2019twb}. We take for definiteness the simplest model known as the Rankine vortex, characterized by a localized vortex structure of constant angular velocity $\Omega_0$ (rigid rotation) up to a finite radius $R_c$, followed by a vorticity-free decay of the angular velocity, extending to infinity. We illustrate the discrepancies between Model 2 and Model 3 as the vortical region ($\Omega_0 R_c$) is increased. We also show how the Rankine configuration converges to the rigidly-rotating (firewall) configuration as $\Omega_0 R_c \to 1$.

The structure of the paper is as follows. In Section~\ref{sec:model}, we describe the linear sigma model coupled to quarks (LSM${}_q$) and formulate the approximations between models 1--3 discussed above. 
Then, in Sections~\ref{sec:model1}-\ref{sec:model3}, we explore the properties of these models under rotation. 
Section~\ref{sec:phasediagram} is devoted to discussing the phase diagram of model 3, concentrating on its mixed-phase structure generated by uniform rotation.  
The case of nonuniform rotation is addressed in Sec.~\ref{sec:inh}.
Finally, Sec.~\ref{sec:conc} concludes this paper.

\section{Model description}
\label{sec:model}

In this section, we present the models discussed in this paper. In Subsec.~\ref{sec:model:LSMq}, we introduce the linear sigma model coupled to quarks and discuss the mean-field approximation. Subsections~\ref{sec:model:model1}--\ref{sec:model:model3} introduce models 1--3 presented in the introduction. The physical content of these models is further explored in Sections~\ref{sec:model1}--\ref{sec:model3}. Subsection~\ref{sec:model:therm} discusses the computation of the fermionic path integral under rotation.

\subsection{\texorpdfstring{LSM$_q$}{LSMq} and the mean-field approximation}\label{sec:model:LSMq}

The Lagrangian of the LSM${}_q$ model,
\begin{equation}
 \mathcal{L} = \mathcal{L}_M + \mathcal{L}_q\,,
\end{equation}
is a sum of the interacting mesonic and quark contributions, respectively. The mesonic part,
\begin{equation}\label{eq:mesonLagrangian}
 \mathcal{L}_M = \frac{1}{2} (\partial_\mu \sigma \partial^\mu \sigma + \partial_\mu \vec{\pi} \cdot \partial^\mu \vec{\pi}) - U(\sigma,\vec{\pi}),
\end{equation}
possesses the kinetic terms for the meson fields, the sigma meson $\sigma$ and the pion $\vec \pi$, supplemented by the mesonic potential:
\begin{equation}
 U(\sigma,\vec{\pi}) = \frac{\lambda}{4} (\sigma^2 + \vec{\pi}^2 - v^2)^2 - h \sigma\,.
 \label{eq_U_mesonic}
\end{equation}

The quark Lagrangian reads
\begin{equation}\label{eq:quarkLagrangian}
 \mathcal{L}_q = \bar{\psi} \left[ \frac{i}{2} \overleftrightarrow{\slashed{\partial}} -g (\sigma + i \gamma^5 \vec{\tau} \cdot \vec{\pi}) \right] \psi,
\end{equation}
where $\slashed{\partial} = \gamma^\mu \partial_\mu$, $\gamma^5 = i \gamma^0 \gamma^1 \gamma^2 \gamma^3$, $g$ is the coupling strength, and $\vec{\tau}$ are the Pauli matrices acting on the flavor content of $\psi = (\psi_u, \psi_d)$. The parameters of the model $(\lambda, v, h, g)$ are obtained from matching the predictions of the model at vanishing temperature with the known results for the pion decay constant $f_\pi=93$ MeV and for the masses of the constituent quark $m_q = 307$ MeV, the $\sigma$ meson $m_\sigma = 600$ MeV, and the pion $m_\pi = 138$ MeV \cite{Scavenius:2000qd}. Explicitly, the values of the model parameters are as follows:
\begin{align}
    \lambda & =\dfrac{m_\sigma^2-m_\pi^2}{2f_\pi^2} \simeq 19.71, \qquad g = \dfrac{m_q}{f_\pi} \simeq 3.30,\nonumber\\
     \quad  h &= f_\pi m_\pi^2 \simeq 1.77\times 10^{6} \ \textrm{MeV}^3,\nonumber \\v&=f_\pi\sqrt{\dfrac{m_\sigma^2-3m_\pi^2}{m_\sigma^2-m_\pi^2}}\simeq 87.65\ \textrm{MeV}.
     \label{eq_mdl_prmts}
\end{align}

In thermal equilibrium, the model can be described by the partition function:
\begin{equation}
    \mathcal{Z} = \int \mathcal{D}\psi\mathcal{D}\overline{\psi}\mathcal{D}\sigma \mathcal{D}\vec{\pi}\ e^{i \int d^4 x \mathcal{L}}\, ,
\end{equation}
where the functional integral goes over the quark and mesonic fields.
The thermodynamic properties of the theory are naturally encoded in the analytical continuation to imaginary time ($t \to -i \tau$) of the path integral. In particular, the grand canonical potential is given by
\begin{align}
    \Phi &= -\dfrac{1}{\beta}\ln \mathcal{Z}_E \,,
    & 
    \mathcal{Z}_E &= \int \mathcal{D}\psi\mathcal{D}\overline{\psi}\mathcal{D}\sigma \mathcal{D}\vec{\pi} e^{-S_E}\,,
\end{align}
where the subscript ``$E$'' stands for the Euclidean version of the corresponding quantities, 
\begin{equation}
 S_E = \int_0^\beta d\tau \int_{\mathcal{M}_3} d^3 x \mathcal{L}_E, \quad \mathcal{L}_E(\tau,\vec{x}) = -\mathcal{L}(it,\vec{x}),
 \label{eq:E_from_M}
\end{equation}
while $\mathcal{M}_3$ denotes the spatial domain occupied by the system.

In the saddle point approximation, which corresponds to a classical minimum of the grand potential with respect to the fields, the dynamics of the fields $\sigma$, $\pi$ and $\psi$ are governed by the Euler-Lagrange equations,
\begin{align}
 [\Box + \lambda(\sigma^2 + \vec{\pi}^2 - v^2)] \sigma &= h - g \bar{\psi} \psi,\nonumber\\
 [\Box + \lambda(\sigma^2 + \vec{\pi}^2 - v^2)] \vec{\pi} &= -i g \bar{\psi} \gamma^5 \vec{\tau} \psi, \nonumber\\
 [i \slashed{\partial} - g(\sigma + i \gamma^5 \vec{\tau} \cdot \vec{\pi})] \psi &= 0.
\end{align}

In the following, we will work in the mean-field approximation, by which the quantum field $\hat{\sigma} = \langle \hat{\sigma} \rangle + \delta \hat{\sigma}$ is represented as a sum of its expectation value $\langle \hat{\sigma} \rangle$ and quantum fluctuations $\delta \hat{\sigma}$, with $\delta \hat{\sigma}$ being subsequently ignored. Similarly, we decompose the pion field in the same way: $\vec{\pi} \rightarrow \hat{\vec{\pi}} = \langle \vec{\pi} \rangle + \delta \hat{\vec{\pi}}$. For compactness in the notation, we shall write $\langle \hat{\sigma} \rangle \equiv \sigma$ and similarly for the field $\vec{\pi}$. Under the mean-field approximation, the vacuum expectation values $\sigma$ and $\vec{\pi}$ are taken as fixed quantities that define the thermodynamic ground state of the system:
\begin{equation}\label{eq:partition_mean_field}
    \mathcal{Z}_E^{\rm m.f.} = \int \mathcal{D}\psi\mathcal{D}\overline{\psi}e^{- S_E}\, ,
\end{equation}
where the superscript ``m.f.'' stands for the \textit{mean-field} result. A thermodynamically favorable solution is the one that minimizes the grand canonical potential, which is a functional of the fields:
\begin{equation}
    \Phi_{\rm m. f.}(\sigma,\vec{\pi}) = \int_{\mathcal{M}_3}d^3x \mathcal{L}_{E;M} - \frac{1}{\beta}\ln \mathcal{Z}_{E;q}\,.
    \label{eq_Phi_mf}
\end{equation}
Here, $\mathcal{Z}_{E;q}$ is the partition function of the quark fields that represent the only dynamical degree of freedom in the mean-field approximation.

We find the equations of motion by extremizing the grand canonical potential ~\eqref{eq_Phi_mf}. We shall consider three different approaches to determine the thermodynamic ground state of the system. Later, we will show that these approaches describe the ground state with increasing levels of accuracy.

\subsection{Model 1: Uniform \texorpdfstring{$\sigma$}{sigma} condensate}\label{sec:model:model1}

In the first approach, we assume that the effective mass takes a constant value globally: $\sigma\to{\bar \sigma} = {\rm const}$. Similarly, we assume that $\vec{\pi}\to\overline{\vec{\pi}} = {\rm const}$. Clearly, a uniform ground state in a rotating medium is not a true ground state of the system: even in a classical rotating fluid, the ground state is not homogeneous \cite{Cercignani:2002}. However, it is still worth starting our discussion with the uniform approximation to the ground state with a two-fold aim: to compare our results with other, mostly uniform-state approaches and to evaluate the effect of the inhomogeneous ground states considered later.

In order to find the condensates ${\bar\sigma}$ and $\overline{\vec{\pi}}$, we minimize the grand canonical potential. Taking a variation with respect to the field ${\bar \sigma}$, we get:
\begin{align}
    \dfrac{\delta \Phi_{\textrm{m.f.}}}{\delta \overline \sigma} &= -\dfrac{1}{\beta \mathcal{Z}^{\rm m.f.}_E}\dfrac{\delta \mathcal{Z}^{\rm m.f.}_E}{\delta {\bar\sigma}} \nonumber\\ 
    &=\dfrac{1}{\beta\mathcal{Z}^{\rm m.f.}_E} \int \mathcal{D}\psi\mathcal{D}\overline{\psi}e^{-S_E} \int d^4X \dfrac{\delta \mathcal{L}_E(X)}{\delta {\bar\sigma}}\,,
    \label{eq:extremization_constant}
\end{align}
where we used $X_\mu = X^\mu = (\tau, \mathbf{x})$ to denote Euclidean-space coordinates.
The variation \eqref{eq:extremization_constant} can be rewritten as
\begin{equation}
 \dfrac{\delta \Phi_{\rm m.f.}}{\delta {\bar \sigma}} =  V \frac{\partial U(\bar{\sigma},\overline{\vec{\pi}})}{\partial \bar{\sigma}}    
 + g \int d^3x  \langle\overline\psi \psi\rangle \,,
\end{equation}
where the mesonic potential is given in Eq.~\eqref{eq_U_mesonic} and we took into account the time-independence of the fermionic condensate in the ground state, whose expectation value is obtained as 
\begin{equation}
 \langle\overline\psi \psi\rangle(X)=\dfrac{1}{\mathcal{Z}^{\rm m.f.}_E}\int \mathcal{D}\psi\mathcal{D}\overline{\psi} \ \overline{\psi}\psi e^{- S_E}\,.
\end{equation}
The path integral over the fermionic fields gives us the thermal expectation value of the fermionic condensate in a given state, which we specify later. Imposing $\delta \Phi_{\rm m.f.} / \delta \bar{\sigma} = 0$, we arrive at the mass gap equation, 
\begin{equation}
 \lambda ({\bar \sigma}^2 + \overline{\vec{\pi}}^2-v^2) {\bar \sigma} = h - \dfrac{g}{V}\int_{\mathcal{M}_3} d^3x \langle \overline\psi \psi \rangle(\mathbf{x})\,.
\end{equation}

Taking now the variations of the grand canonical potential with respect to $\overline{\vec{\pi}}$, with similar manipulations we arrive at:
\begin{equation}
   \lambda ({\bar \sigma}^2 + \overline{\vec{\pi}}^2-v^2) \overline{\vec{\pi}}=  \dfrac{g}{V}\int_{\mathcal{M}_3} d^3x \langle -i \overline\psi \gamma^5 \vec\tau\psi \rangle(\mathbf{x})\,.
\end{equation}
The expectation value of the pseudoscalar condensate vanishes identically in the considered setup. As a consequence, the previous equation is solved by taking $\overline{\vec{\pi}}=0$. 

Finally, we consider a cylindrically-symmetric state, such that the fermion condensate $\langle\overline\psi \psi\rangle$ depends only on the transverse radial coordinate $\rho = \sqrt{x^2 + y^2}$. This choice of the ground state is justified by the geometry of the problem. Therefore, the ``averaged'' gap equation in the approximation of the uniform condensate is given by 
\begin{align}\label{eq:gap_averaged}
 \lambda ({\bar \sigma}^2 -v^2) {\bar \sigma} = h - g\dfrac{2}{R^2}\int_0^R d\rho \rho \langle \overline\psi \psi \rangle(\rho)\,,
\end{align}
where the integral is taken over the cylinder of radius $R$.

\subsection{Model 2: Slowly-varying \texorpdfstring{$\sigma$}{sigma} condensate}\label{sec:model:model2}

Taking the variation of the thermodynamic potential with respect to the local value of the condensate $\sigma$, we find
\begin{equation}
    \dfrac{\delta \Phi_{\rm m. f.}}{\delta \sigma(X')}    
    =\dfrac{1}{\beta \mathcal{Z}^{\rm m.f.}_E} \int \mathcal{D}\psi\mathcal{D}\overline{\psi}e^{-S_E} \int d^4X \dfrac{\delta \mathcal{L}_E(X)}{\delta \sigma(X')}\,,
\end{equation}
with $S_E = \int d^4 X \mathcal{L}_E$.
The variation of the quark Lagrangian $\mathcal{L}_{E;q} = -\mathcal{L}_q$ [see Eqs.~\eqref{eq:quarkLagrangian} and \eqref{eq:E_from_M}] with respect to $\sigma(X')$ can be obtained using
\begin{equation}
    \int d^4X \frac{\delta \mathcal{L}_{E;q}}{\delta \sigma(X')} = g\overline{\psi}(X')\psi(X')\,,
\end{equation}
where we have used $\delta\sigma(X)/\delta\sigma(X') = \delta^4(X-X')\,$. For the variations of the mesonic Lagrangian \eqref{eq:mesonLagrangian}, we obtain
\begin{align}\label{eq:variation_bdry}
 \int & d^4X \frac{\delta \mathcal{L}_{E;M}(X)}{\delta \sigma(X')} = 
 \frac{\partial U}{\partial \sigma} - \Box_E \sigma + \mathcal{B}, \nonumber\\
 &\mathcal{B} = \int d^4X\, \partial_\mu \left[\frac{\delta\sigma(X)}{\delta \sigma(X')} \partial_\mu \sigma(X)\right],
\end{align}
where the last boundary term $\mathcal{B}$ vanishes if we fix the value of $\sigma$ at the boundary $\partial\mathcal{M}_3$, i.e. $\delta\sigma_{\partial \mathcal{M}_3}=0$, provided that $\sigma$ is finite. We used the notation $\Box_E = -\Box = \partial^2_\tau + \boldsymbol{\nabla}^2$ for the Euclidean d'Alembert operator.

Imposing $\delta \Phi_{\rm m. f.} / \delta \sigma(X') = 0$ gives
\begin{equation}
    \left[\Box +\lambda (\sigma^2 + \vec{\pi}^2-v^2)\right] \sigma = h - g \langle \overline\psi \psi \rangle\,.
\end{equation}
Taking now the variations of the grand canonical potential for the pion fields $\vec\pi$ and performing similar steps as described above, we arrive at the equation of motion for $\vec{\pi}$:
\begin{equation}
    \left[\Box +\lambda (\sigma^2 + \vec{\pi}^2-v^2)\right] \vec\pi =  g \langle -i \overline\psi \gamma^5 \vec\tau\psi \rangle\,.
\end{equation}

As we already mentioned, we consider a cylindrically-symmetric state, for which $\sigma$ and $\vec{\pi}$ depend only on the transverse coordinate $\rho = \sqrt{x^2 + y^2}$, such that 
\begin{align}\label{eq:eoms_sigma_pi}
 \left[-\frac{1}{\rho} \frac{d}{d\rho} \rho \frac{d}{d\rho} + \lambda(\sigma^2 + \vec{\pi}^2 - v^2)\right] \sigma &= h - g \langle \bar{\psi} \psi \rangle,\nonumber\\
 \left[-\frac{1}{\rho} \frac{d}{d\rho} \rho \frac{d}{d\rho} + \lambda(\sigma^2 + \vec{\pi}^2 - v^2)\right] \vec{\pi} &= g \langle -i \bar{\psi} \gamma^5 \vec{\tau} \psi \rangle.
\end{align}
The expectation value of the pseudoscalar condensate, $\langle -i\bar{\psi} \gamma^5 \vec{\tau} \psi \rangle$, vanishes for the given state, such that $\vec{\pi} = 0$ is an acceptable solution. The equation for $\sigma$ is highly nonlinear since the scalar condensate $\langle \hat{\bar{\psi}} \hat{\psi} \rangle$ depends on the effective fermion mass, $M=g\sigma$. 

The slowly varying $\sigma$ approach consists of neglecting the effect of radial gradients in Eq. \eqref{eq:eoms_sigma_pi}. Therefore, the value of $\sigma$ at each point is given by the `local' gap equation:

\begin{equation}\label{eq:gap_local}
    \lambda(\sigma^2  - v^2) \sigma = h - g \langle \bar{\psi} \psi \rangle\,,
\end{equation}
where we have used $\vec{\pi}=0$. Both the $\sigma$ field and the expectation value of the fermion condensate depend on the transverse plane radial distance $\rho$. The main feature of the local mass gap equation \eqref{eq:gap_local} is that the chiral condensate $\langle\overline{\psi}\psi\rangle=\langle\overline{\psi}\psi\rangle(\rho)$ has to be evaluated at the radially-dependent condensate $\sigma=\sigma(\rho)$.

\subsection{Model 3: Strongly-inhomogeneous \texorpdfstring{$\sigma$}{sigma} condensate}\label{sec:model:model3}

The natural next step to consider in this model is to take into account the effect of the radial gradients in the condensate $\sigma$. The derivation of the equations of motion from the path integral presented in Sec. \ref{sec:model:model2} applies here as well, giving Eq. \eqref{eq:eoms_sigma_pi}. In particular, for vanishing $\vec{\pi}$, the differential equation for the $\sigma$ meson is given by
\begin{equation}\label{eq:model_diff}
    \left[-\frac{1}{\rho} \frac{d}{d\rho} \rho \frac{d}{d\rho} + \lambda(\sigma^2  - v^2)\right] \sigma = h - g \langle \bar{\psi} \psi \rangle\,.
\end{equation}

\subsection{Thermal rotating state}\label{sec:model:therm}

In thermal field theory, the expectation value of an operator $\widehat{A}$ can be computed using the density operator $\hat{\rho}$ via~\cite{Kapusta:2006pm}
\begin{equation}
 \langle \widehat{A} \rangle = \mathcal{Z}^{-1} {\rm Tr}(\hat{\rho} \widehat{A}), \quad 
 \mathcal{Z} = {\rm Tr}(\hat{\rho}),
\end{equation}
where $\mathcal{Z}$ is the partition function. Since we are interested in studying the LSM$_q$ model in the presence of rotation at finite temperature and chemical potential, we employ the density operator \cite{LL5,Vilenkin:1980zv,Becattini:2012tc}
\begin{equation}
 \hat{\rho} = e^{-\beta_0 (\widehat{H} - \boldsymbol{\Omega} \cdot \widehat{\mathbf{J}} - \mu_0 \widehat{Q})},
 \label{eq:rho_GTE}
\end{equation}
where $\widehat{H}$, $\widehat{\mathbf{J}}$ and $\widehat{Q}$ represent the Hamiltonian, angular momentum and charge operators, while $\beta_0$ and $\mu_0$ represent the inverse temperature and chemical potential, respectively.

More generically, $\hat{\rho}$ in Eq.~\eqref{eq:rho_GTE} can be obtained using the Zubarev method \cite{Zubarev:1979afm,vanWeert:1982,Becattini:2012tc},
\begin{equation}
 \hat{\rho} = \exp\left[- \int d\Sigma_\mu (\widehat{T}^{\mu\nu} \beta_\nu - \alpha \widehat{J}^\mu) \right].
\end{equation}
In global thermal equilibrium, the local four-temperature satisfies the Killing equation \cite{Cercignani:2002}, $\partial_\nu \beta_\mu + \partial_\mu \beta_\nu = 0$. The solution of the Killing equation corresponding to a rigidly-rotating state is $\beta^\mu \partial_\mu = \beta(\rho) u^\mu \partial_\mu = \beta_0(\partial_t + \Omega \partial_\varphi)$, with local inverse temperature
\begin{equation}
    \beta(\rho)\equiv \beta_\rho = \dfrac{1}{T_\rho} = \frac{\beta_0}{\Gamma_\rho}\,, 
    \label{eq_beta_rho}
\end{equation}
where 
\begin{equation}
    \Gamma_\rho = 1/\sqrt{1 - \rho^2 \Omega^2}\,,
    \label{eq_Gamma_rho}
\end{equation}
is the Lorentz factor. We denote quantities evaluated on the rotation axis with the subscript ``0''. In thermal equilibrium, $\alpha = \mu / T = \beta_0 \mu_0 = {\rm const}$. 

The inhomogeneous temperature profile given by Eqs.~\eqref{eq_beta_rho} and \eqref{eq_Gamma_rho} corresponds to the Tolman-Ehrenfest law~\cite{Tolman:1930ona, Tolman:1930zza}, which describes the thermodynamic equilibrium of matter in the background of a static gravitational field. In our case, the gravitational field corresponds to the centrifugal force generated by the uniform rotation, which favors lower (higher) local temperatures of the system near (far from) the axis of rotation.

When evaluating observables at a given point $x$, it is convenient to approximate the density operator in Eq.~\eqref{eq:rho_GTE} as $\hat{\rho} \to \hat{\rho}_{\rm LTE}(x)$, by writing the four-temperature $\beta^\mu(y) \simeq \beta^\mu(x) + O(\partial \beta)$, with $y$ being the integration variable over the hypersurface $d\Sigma_\mu$. In this case, the hypersurface integral can be performed using $\int d\Sigma_\mu \widehat{T}^{\mu\nu} = \widehat{P}^\nu$ and $\int d\Sigma_\mu \widehat{J}^\mu = \widehat{Q}$, with $\widehat{P}^\mu$ and $\widehat{Q}$ being the momentum and charge operators, such that \cite{Becattini:2014yxa,Becattini:2015nva,Buzzegoli:2018wpy}
\begin{align}
 \hat{\rho}_{\rm LTE}(x) &= \exp\left(- \beta_\nu(x) \int d\Sigma_\mu \widehat{T}^{\mu\nu} + \alpha \int d\Sigma_\mu \widehat{J}^\mu \right) \nonumber\\
 &= e^{-\beta_\mu \widehat{P}^\mu + \alpha \widehat{Q}}.
 \label{eq:rho_LTE}
\end{align}
The state described by $\hat{\rho}_{\rm LTE}$ corresponds to a static state, seen by a moving observer. Therefore, its properties can be determined in the rest frame defined by $u^\mu = T(x) \beta^\mu$. 
The local thermal equilibrium operator $\hat{\rho}_{\rm LTE}(x)$ is clearly missing quantum effects, such as spin-orbit coupling. As discussed in Refs.~\cite{Ambrus:2014uqa,Becattini:2015nva,Ambrus:2019cvr}, the quantum corrections are generally suppressed as $O[(\Omega / T)^2]$, however they become dominant close to the light cylinder \cite{Ambrus:2017opa}. We leave further investigation of their role in the thermodynamics of strongly-interacting matter under rotation for future work.

For a globally-static system of Dirac fermions, the grand potential reads 
\begin{align}
 \Phi^q_{\rm static} &= \int d^3x \phi^q_{\rm static}(x), \nonumber\\
 \phi^q_{\rm static} &= -\frac{N_f N_c}{\beta_0} \sum_{\varsigma,\lambda} \int \frac{d^3p}{(2\pi)^3} \ln(1 + e^{-\beta_0 E + \varsigma \alpha}),
\end{align}
where $\varsigma = \pm 1$ corresponds to particle/anti-particle states and $\lambda = \pm 1/2$ represents the polarization degree of freedom.\footnote{Note that $\lambda$ in fermionic expectation values represents the fermionic polarization degree of freedom, which is different from the parameter $\lambda$ appearing in the mesonic potential, Eq.~\eqref{eq_U_mesonic}.}
With the approximations described above, we write the local fermionic grand potential for a rotating system as
\begin{align}
 \phi^q_{\rm rot}(x) &= -\frac{N_f N_c}{\beta_0} \sum_{\varsigma,\lambda} \int \frac{d^3p}{(2\pi)^3} \, \ln(1 + e^{-\beta_\mu p^\mu + \varsigma \alpha}) \nonumber\\
 &= -\frac{N_f N_c}{\beta_\rho} \sum_{\varsigma,\lambda} \int dP  \, E\, \ln(1 + e^{-\beta_\rho E + \varsigma \alpha}).
 \label{eq:phiD_rot}
\end{align}
In deriving the above result, we first introduced the Lorentz-invariant integration measure in momentum space,
\begin{equation}
    dP = \dfrac{d^3p}{(2\pi)^3E}\,.
    \label{eq:dP}
\end{equation}
On the second line of Eq.~\eqref{eq:phiD_rot}, we changed the integration variable to $p'^\mu = \Lambda^\mu{}_\nu p^\nu$, with $\Lambda^\mu{}_\nu u^\nu = \delta^\mu_0$ defining the rest frame. Then, $\beta_\mu p^\mu = \beta_\rho u_\mu p^\mu = \beta_\rho E'$, while $E = p'^\nu \Lambda_\nu{}^0 = \Gamma_\rho E' + \mathbf{u} \cdot \mathbf{p}'$. Therefore, the grand canonical potential takes the following form:
\begin{multline}\label{eq_grand_potential}
   \Phi_{\rm m. f.} = \int_\mathcal{M} d^3x\left[\dfrac{1}{2}\partial_\mu\sigma\partial^\mu\sigma-\dfrac{\lambda}{4}(\sigma^2-v^2)^2 +h\sigma\right.\\
   \left.  -\frac{N_fN_c}{\beta_\rho} \sum_{\varsigma, \lambda} \int\dfrac{d^3p}{(2\pi)^3} \ln\left(1+e^{-\beta_\rho E+\varsigma \alpha}\right) \right]\,.
\end{multline}
The fermionic condensate at a distance $\rho$ can be evaluated as described above, 
\begin{align}
 \braket{\hat{\bar{\psi}} \hat{\psi}}  
 &= g \sigma(\rho) N_f N_c \sum_{\varsigma,\lambda} 
 \int \frac{dP}{e^{\beta_\rho (E - \varsigma \mu_\rho)} + 1}\,.
 \label{eq:tevs}
\end{align}

Note that, in the evaluation of the grand canonical potential \eqref{eq_grand_potential} and of the fermionic condensate \eqref{eq:tevs}, we considered $\sigma$ to have a constant value given by its actual local value, $\sigma(\rho)$. This should be understood as a proxy for the response of the system when the path integral is computed self-consistently with a $\rho$-dependent condensate $\sigma$ [see, e.g., Ref.~\cite{Wang:2018zrn}]. The fully self-consistent evaluation of the fermionic path integral, with a spacetime-dependent condensate, is challenging and will not be considered further in this work.
A second observation is related to the approximation made when going from Eq.~\eqref{eq:rho_GTE} to \eqref{eq:rho_LTE}: the latter operator gives access only to the ``classical'' contributions. The full operator would bring about ``quantum corrections'' \cite{Ambrus:2014uqa,Becattini:2015nva}, which become dominant close to the light cylinder \cite{Ambrus:2017opa}. While such terms would certainly be relevant, especially when considering Model 3, we leave an extension of our analysis down this path for future work.

\section{Model 1: Uniform ground state}\label{sec:model1}
 
\begin{figure}
\centering 
\begin{tabular}{c}
    \includegraphics[width=0.95\linewidth]{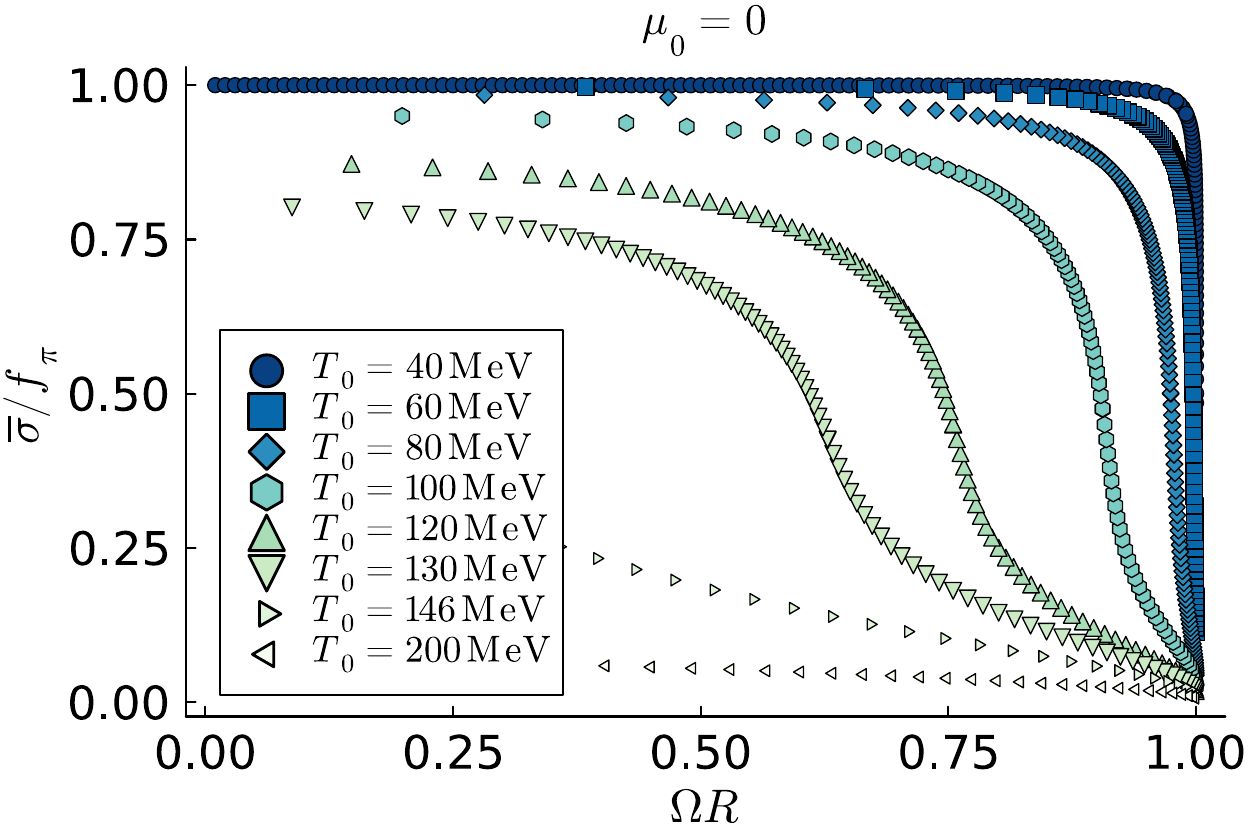} \\
    \includegraphics[width=0.95\linewidth]{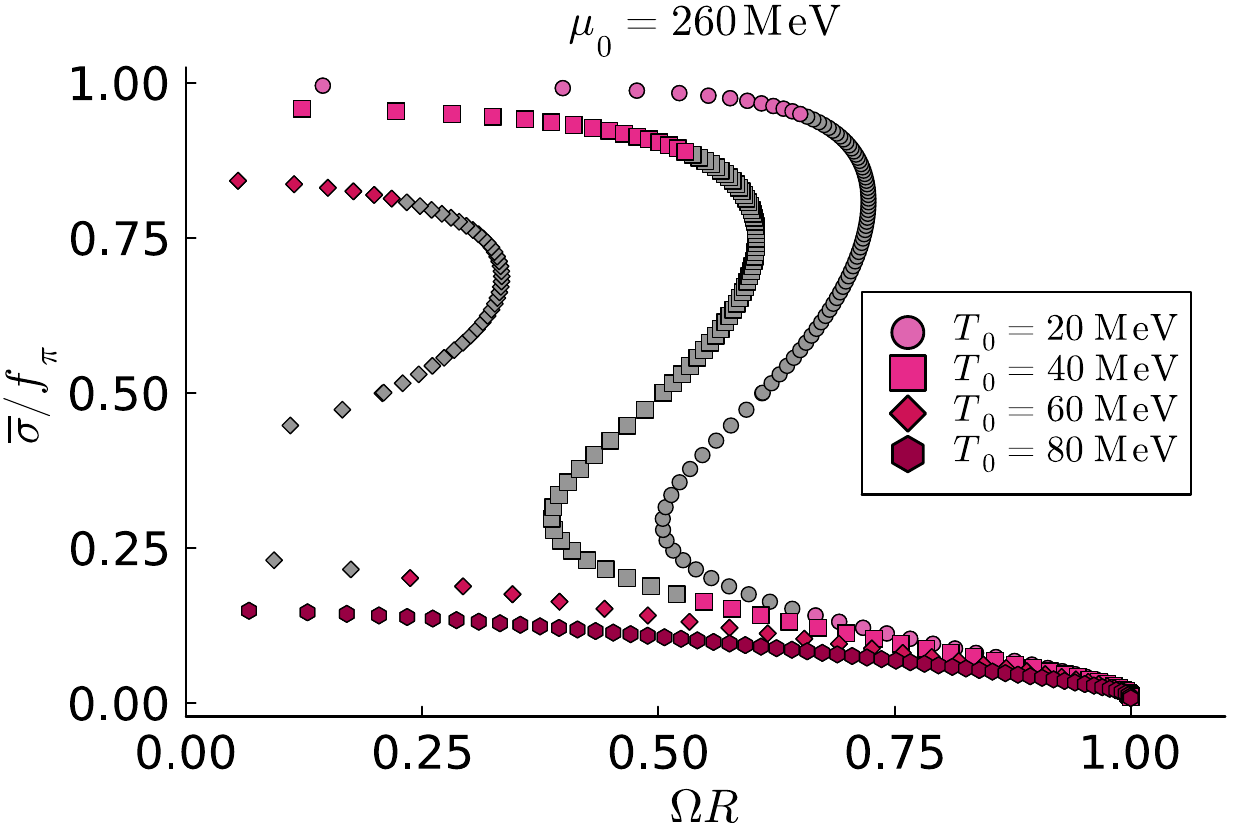}
\end{tabular}
    \caption{Uniform ground state: Normalized value of the condensate $\sigma$, obtained as a solution of the mass gap equation \eqref{eq_sbar_saddle}, for a set of on-axis temperatures $T_0$ at two values of the on-axis chemical potential $\mu_0$. At small chemical potential, the transition is crossover, whereas at large chemical potential, it is of first order.
    \label{fig_Averaged-sofR-mu}
    }
\end{figure}

\begin{figure}
\centering 
\begin{tabular}{c}
    \includegraphics[width=0.95\linewidth]{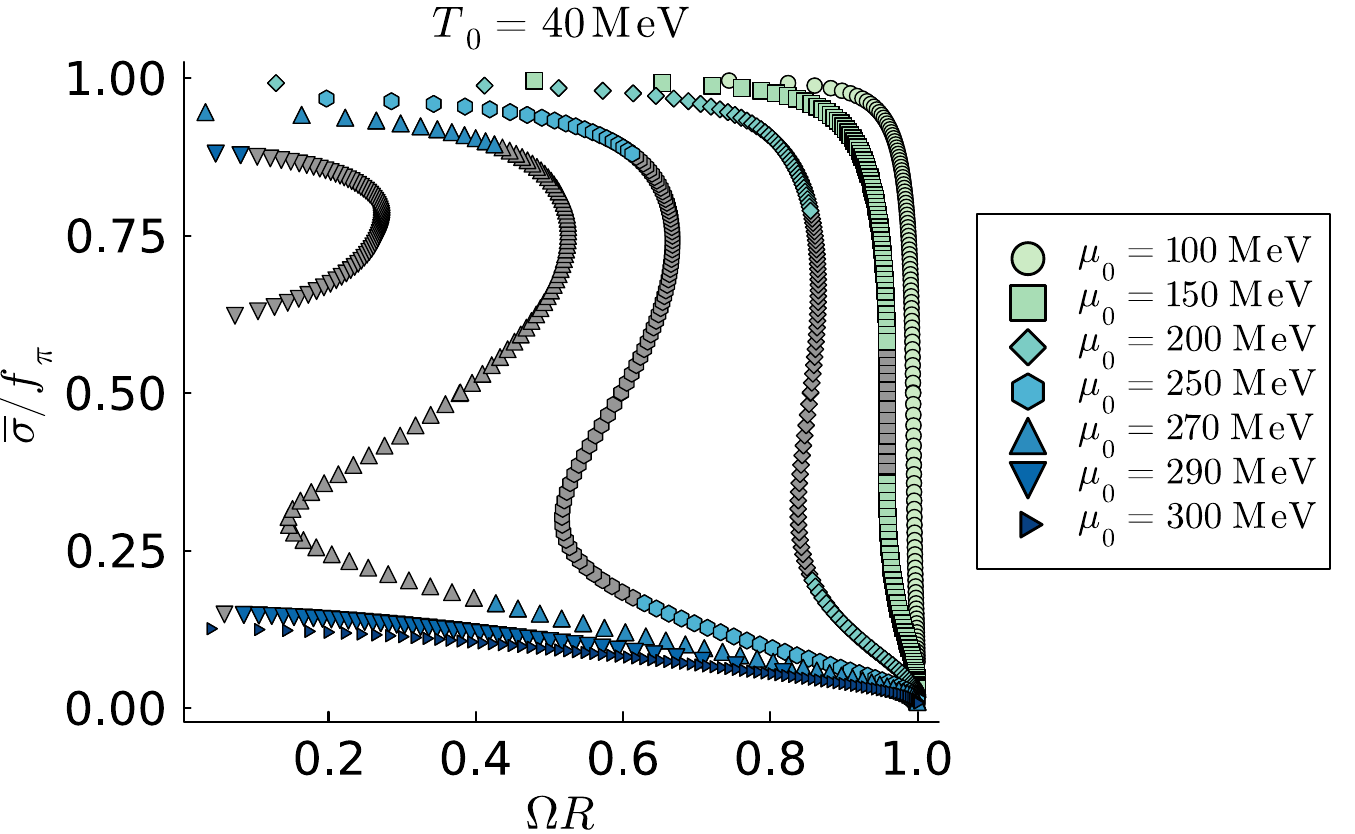} \\
    \includegraphics[width=0.95\linewidth]{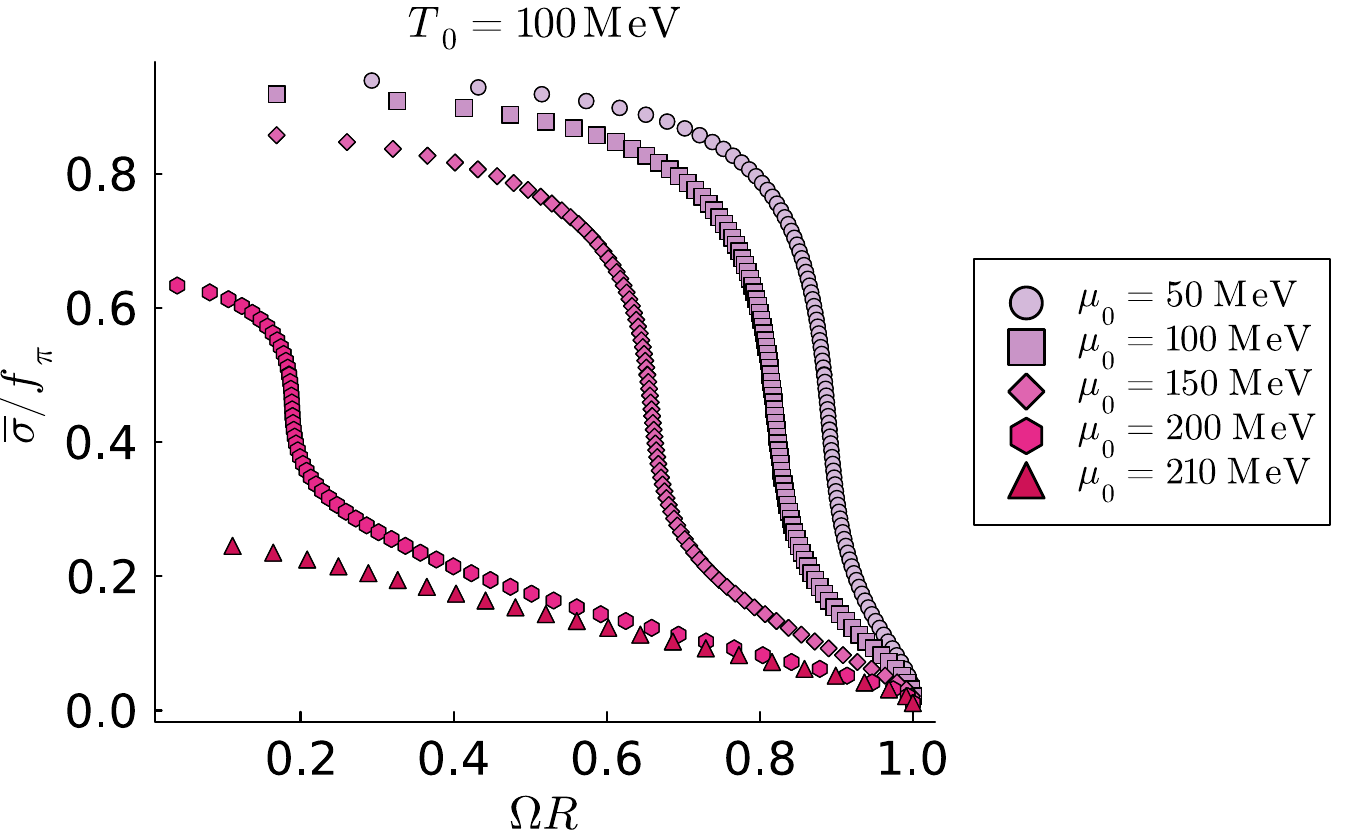}
\end{tabular}
    \caption{Uniform ground state: The same as in Fig.~\ref{fig_Averaged-sofR-mu} for fixed temperature and varying chemical potential. At small temperatures, the transition goes from crossover to first order as the chemical potential increases. At larger temperatures, the transition is always crossover.
    \label{fig_Averaged-sofR-T}
    }
\end{figure}

In the first approach, we consider that $\sigma \rightarrow \bar{\sigma}$ is a constant, position-independent field, as described in Sec.~\ref{sec:model:model1}. With this single degree of freedom, we can only minimize the grand canonical potential in a global (domain-integrated) way. We therefore consider a fictitious cylinder of radius $R$ and arrive at the mass gap equation~\eqref{eq:gap_averaged}: 
\begin{equation}
 \lambda(\bar{\sigma}^2 - v^2) \bar{\sigma} = h - \frac{2g}{R^2} \int_0^R d\rho\,\rho \langle \hat{\bar{\psi}} \hat{\psi} \rangle.
 \label{eq_sbar_saddle}
\end{equation}
In the TE approach, the fermion condensate depends on the radial coordinate $\rho$ only via the local temperature, $T(\rho) = T_0 \Gamma_\rho$, which is proportional to the Lorentz factor~\eqref{eq_Gamma_rho}. The integral over $\rho$ can be performed analytically. Changing variables to integrate over $\beta(\rho)$, we find
\begin{equation}
 \int_0^R d\rho\,\rho \langle \bar{\psi} \psi \rangle_{\rm TE} 
 = \frac{g\bar{\sigma} N_f N_c}{2\beta_0^2 \Omega^2} \sum_{\varsigma,\lambda} \int dP \int_{\beta_R}^{\beta_0} \frac{\beta_\rho\,d\beta_\rho}{e^{\beta_\rho E - \varsigma \alpha} + 1}\,,
 \label{eq_sbar_saddle_2} 
\end{equation}
where we have used that $\rho d\rho = -\beta_\rho d\beta_\rho / (\beta_0 \Omega)^2$, and we have substituted the expectation value of the fermion condensate given in Eq. \eqref{eq:tevs}.  The quantities $\beta_0$ and $\beta_R$ represent the local inverse temperature on the rotation axis and at a distance $R$, respectively.
The $\beta$ integral can be performed in terms of the polylogarithm function, defined as $\textrm{Li}_n(z)=\sum_{j=1}^\infty z^j/j^n$:
\begin{multline}
 \int_{\beta_R}^{\beta_0} \frac{\beta_\rho\,d\beta_\rho}{e^{\beta_\rho E - \varsigma \alpha} + 1} \\= 
 \left(\frac{1}{E^2}[{\rm Li}_2(-e^{\varsigma\alpha - \beta_0 E}) - {\rm Li}_2(-e^{\varsigma\alpha - \beta_R E})]\right.\\\left.
 + \frac{\beta_R}{E} \ln(1 + e^{\varsigma \alpha - \beta_R E}) - \frac{\beta_0}{E} \ln(1 + e^{\varsigma \alpha - \beta_0 E})\right). 
\end{multline}
 
\begin{figure}
\centering 
\begin{tabular}{c}
    \includegraphics[width=0.95\linewidth]{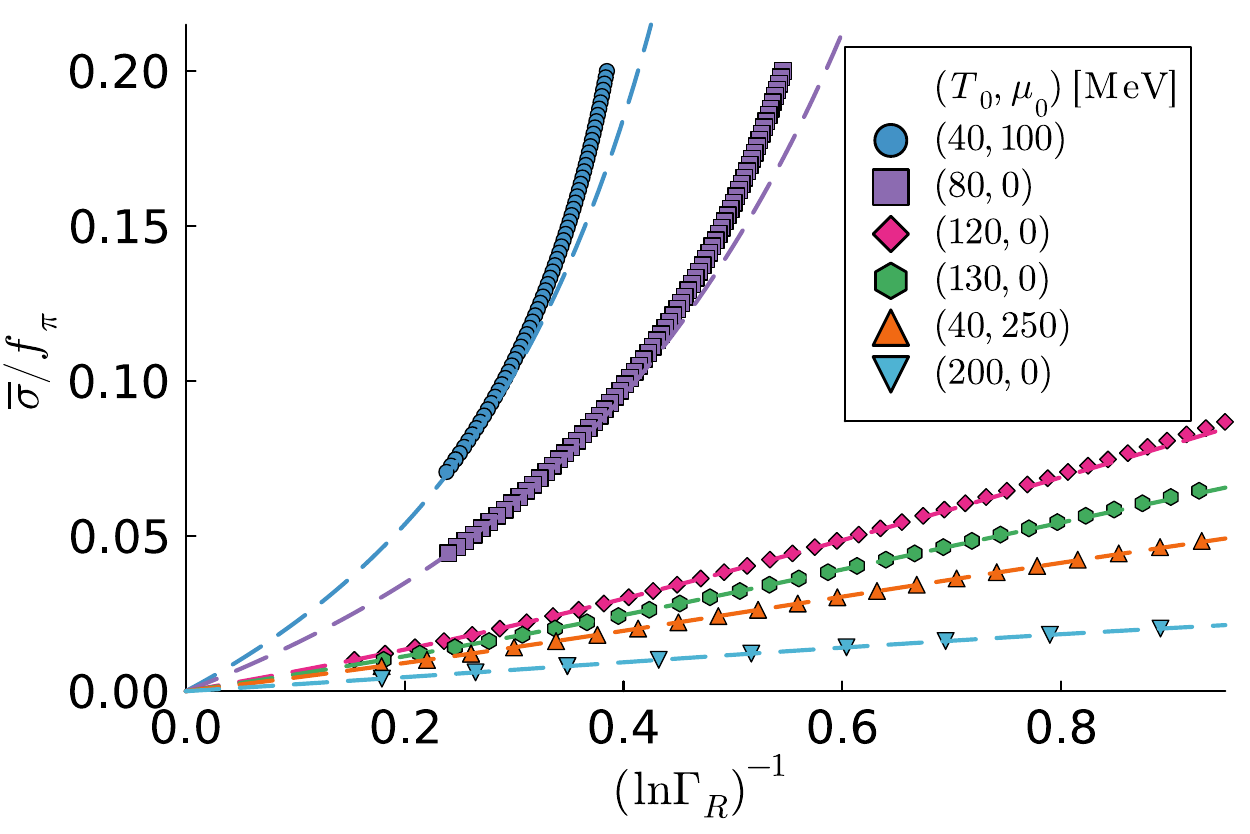} 
\end{tabular}
    \caption{Uniform ground state: Numerical confirmation of the asymptotic behaviour of the condensate plotted as the logarithmic function of the Lorentz factor at the boundary. Dashed lines correspond to the analytical expression \eqref{eq_averaged-near-firewall} while the points are the numerical solution of the mass gap equation \eqref{eq_sbar_saddle}. 
    \label{fig_Averaged-log}
    }
\end{figure}

In Figs. \ref{fig_Averaged-sofR-mu} and \ref{fig_Averaged-sofR-T} we show the value of $\overline\sigma$ that solves the averaged gap equation~\eqref{eq_sbar_saddle}, normalized to $f_{\pi}$, as a function of the dimensionless size of the system, $\Omega R$, for different equilibrium states, labelled by the on-axis temperature $T_0$ and the chemical potential $\mu_0$. As we allow the fictitious cylinder of radius $R$ to approach the light cylinder $R \to 1/\Omega$, we see that $\bar{\sigma}$ is pushed towards lower values, and the chiral symmetry of the system is restored. Indeed, close to the light cylinder, the local temperature and chemical potential become large and the chiral condensate $\langle \bar{\psi} \psi \rangle$ is approximately given by \cite{Ambrus:2014uqa,Ambrus:2019ayb,Ambrus:2019cvr}:
\begin{equation}\label{eq_FC_smallM}
 \langle \bar{\psi} \psi \rangle \simeq g\bar{\sigma} N_f N_c \Gamma^2_R \left(\frac{T_0^2}{6} + \frac{\mu^2_0}{2\pi^2}\right),
\end{equation}
where we neglected terms of higher order with respect to the effective mass, $g\bar{\sigma}$. Eventually, these high-temperature terms will dominate the volume average of the fermion condensate, leading to an overall decrease of the effective fermionic mass, $M=g\overline\sigma$. In this case, we can extract the limiting behavior:
\begin{equation}
 \frac{2}{R^2} \int_0^R d\rho\, \rho \langle \bar{\psi} \psi \rangle \simeq g \bar{\sigma} N_f N_c \left(\frac{T_0^2}{6} + \frac{\mu_0^2}{2\pi^2}\right) 
 \frac{2\ln \Gamma_R}{R^2 \Omega^2}.
\end{equation}
Imposing now Eq.~\eqref{eq_sbar_saddle}, in the limit when the fictitious boundary $R$ approaches the light cylinder, $R \rightarrow \Omega^{-1}$, we have 
\begin{equation}\label{eq_averaged-near-firewall}
 \bar{\sigma} \simeq h \left[\frac{2g^2 N_f N_c}{ R^2 \Omega^2} \left(\frac{T_0^2}{6} + \frac{\mu_0^2}{2\pi^2}\right) \ln \Gamma_R - v^2\lambda\right]^{-1}.
\end{equation}
In Fig. \ref{fig_Averaged-log}, we verify explicitly the validity of Eq.~\eqref{eq_averaged-near-firewall} for several pairs of $(T_0,\mu_0)$.

Coming back to Figs. \ref{fig_Averaged-sofR-mu} and \ref{fig_Averaged-sofR-T}, we observe that, when $R$ is sufficiently far from the light cylinder, the system can be in a chirally-broken phase, provided the temperature and chemical potential on the rotation axis lie below the transition line. In such cases, we can define a system size $\Omega R_c$ where the system transitions from a chirally-broken to a chirally-restored phase. The general trend is that increasing either the temperature or the chemical potential results in a decrease of  $\Omega R_c$. Furthermore, we observe that the phase transition can be either a crossover or first-order, depending on the thermodynamic state given by the pair $(T_0,\mu_0)$. 

In Figs. \ref{fig_Averaged-sofR-mu} and \ref{fig_Averaged-sofR-T}, the presence of a first-order phase transition is unambiguously signaled by the presence of simultaneous solutions to the averaged gap equation at a given temperature and chemical potential. For example, at $T_0=40$ MeV, $\mu_0=260$ MeV, and $\Omega R = 0.5$ (see lower panel of Fig. \ref{fig_Averaged-sofR-mu}) there are three coexisting solutions, which correspond to three local extrema of the grand canonical potential $\Phi_{\rm m.f.}$. The colored data points give the trajectory of the global minimum of $\Phi_{\rm m.f.}$, while the grey points are thermodynamically disfavored. As the size of the system is increased, there is an abrupt jump between two solution branches for ${\bar \sigma}$, indicating that the system undergoes a first-order phase transition in the region when the solution of Eq.~\eqref{eq_sbar_saddle} ${\bar \sigma}(R)$ is not a single-valued function. 

It is interesting to test how the transition line is crossed, given a pair of values $(T_0, \mu_0)$  on the rotation axis. Consider now a static system ($\Omega=0$) and denote  by $(T_c^{\Omega=0},\mu_c^{\Omega=0})$ the parameters corresponding to a phase transition. By the Tolman-Ehrenfest law, the local temperature and local chemical potential increase proportionally to the Lorentz factor \eqref{eq_Gamma_rho}. The trivial dependence on the Lorentz factor suggests that the transition points of the rotating system can be obtained from the non-rotating transition points by a simple transformation. In addition, we are solving the gap equation by averaging the fermion condensate over the radius of the fictitious cylinder $R$. Accordingly, we \textit{assume} that there is an averaged notion of the temperature $\overline T$ and chemical potential $\overline \mu$ that can serve to characterize the phase transition of the system under rotation. In particular, we define $\overline{T} \equiv \langle T^2 \rangle^{1/2}$ and $\overline{\mu} = \langle \mu^2 \rangle^{1/2}$ as the measure of averaged thermodynamic quantities, such that $\overline T  = \overline{\kappa}(R) T_0$ and $\overline \mu  = \overline{\kappa}(R) \mu_0$. Explicitly,
\begin{multline}\label{eq_effective_thermo}
 \begin{pmatrix}
  \overline{T} \\ \bar{\mu}
 \end{pmatrix} = \left[\frac{2}{R^2} 
 \begin{pmatrix}
  T_0^2 \\ \mu_0^2
 \end{pmatrix}
 \int_0^R \frac{\rho d\rho}{(1 - \rho^2 \Omega^2)}\right]^{1/2} \\
 = \frac{1}{R \Omega } 
 \begin{pmatrix}
  T_0 \\ \mu_0 
 \end{pmatrix}
 \sqrt{-\ln(1-R^2\Omega^2)} =\frac{\sqrt{2\ln \Gamma_R}}{R \Omega } 
 \begin{pmatrix}
  T_0 \\ \mu_0 
 \end{pmatrix} \,,
\end{multline}
from which we read off 
\begin{equation}
    \overline{\kappa} = \frac{\sqrt{2\ln \Gamma_R}}{R \Omega }\,. 
\end{equation}
Through numerical simulations, we measure the radius $R_c$ at which the system undergoes the phase transition at the level of $\bar{\sigma}$. We test the hypothesis that this point is determined by the requirement $T_c^{\Omega=0}  = \overline{\kappa}(R_c) T_0$ and $\mu_c^{\Omega=0} = \overline{\kappa}(R_c) \mu_0$.

\begin{figure}
\centering 
\begin{tabular}{c}
    \includegraphics[width=0.95\linewidth]{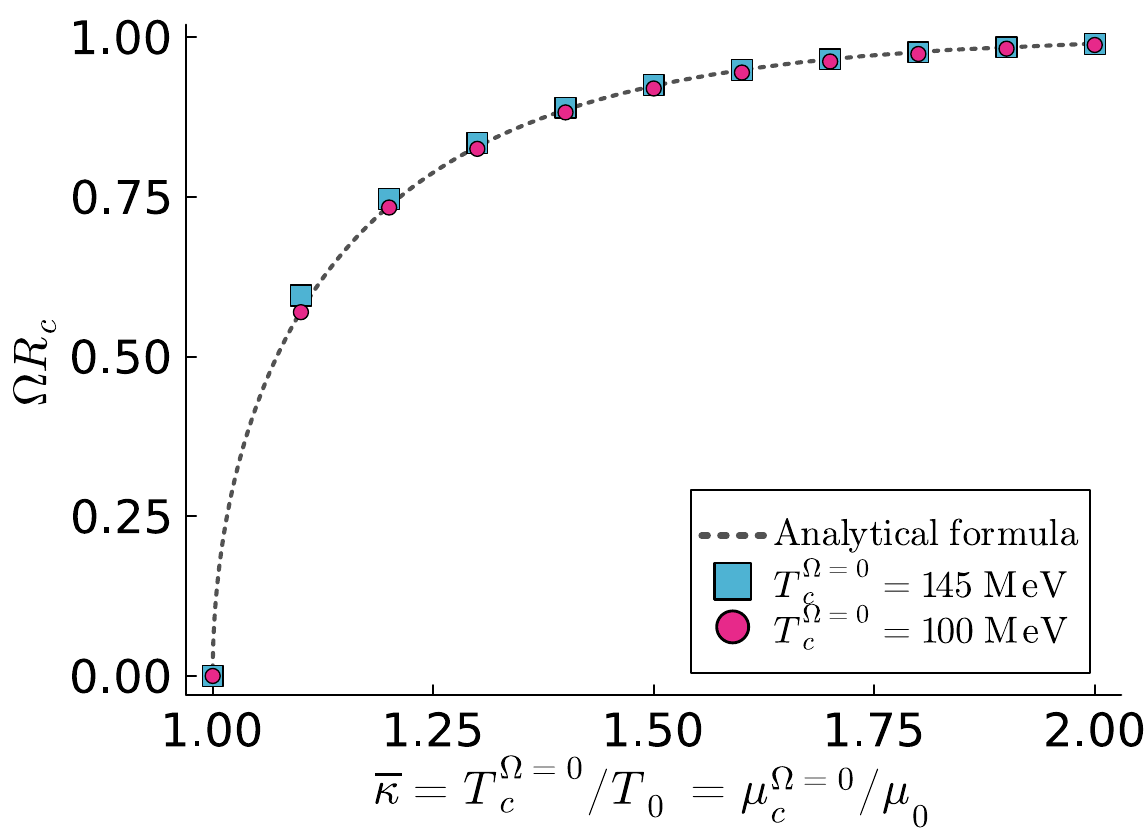} 
\end{tabular}
    \caption{Uniform ground state: Dimensionless critical radius $\Omega R_c$ of the system in a cylinder of radius $R$ that rotates with the angular velocity $\Omega$. The critical radius is shown as a function of the ratio of the critical temperature in the absence of rotation, $T_c^{\Omega = 0}$, and the on-axis temperature $T_0$ in a rotating system (this ratio of temperatures is equal to the ratio of the corresponding chemical potentials). The system resides in the chirally-broken (-restored) phase at $R < R_c$ ($R > R_c$). Dashed lines correspond to the analytical formula \eqref{eq_effective_thermo}, while data points correspond to the critical value obtained by solving the mass gap equation \eqref{eq_sbar_saddle}. 
    \label{fig_Averaged-crits}
    }
\end{figure}

The critical system size $\Omega R_c$ as a function of $\overline\kappa$ is shown in Fig. \ref{fig_Averaged-crits} with remarkable agreement between the numerical data and the estimate proposed in Eq. \eqref{eq_effective_thermo}. In general, the system size corresponding to the phase transition decreases as we increase the value of temperature and chemical potential on the axis. A good approximation of the phase diagram can be obtained from the non-rotating $T$-$\mu$ phase diagram, along with Eq.~\eqref{eq_effective_thermo}. The phase diagram with respect to the system size $\Omega R$ is shown in the top panel of Fig.~\ref{fig:phase_diagrams_12}. Dotted lines correspond to a crossover phase transition, while solid lines signal a first-order phase transition. In the inner (outer) regions, the system is in the chirally-broken (-restored) phase. We observe that, if the system extends to the light-cylinder ($\Omega R=1$), model 1 predicts that the system is in the chirally-restored phase regardless of the values of temperature and chemical potential.

\section{Model 2: Slowly-varying condensate}\label{sec:model2}

\begin{figure}
\centering 
\begin{tabular}{c}
    \includegraphics[width=0.95\linewidth]{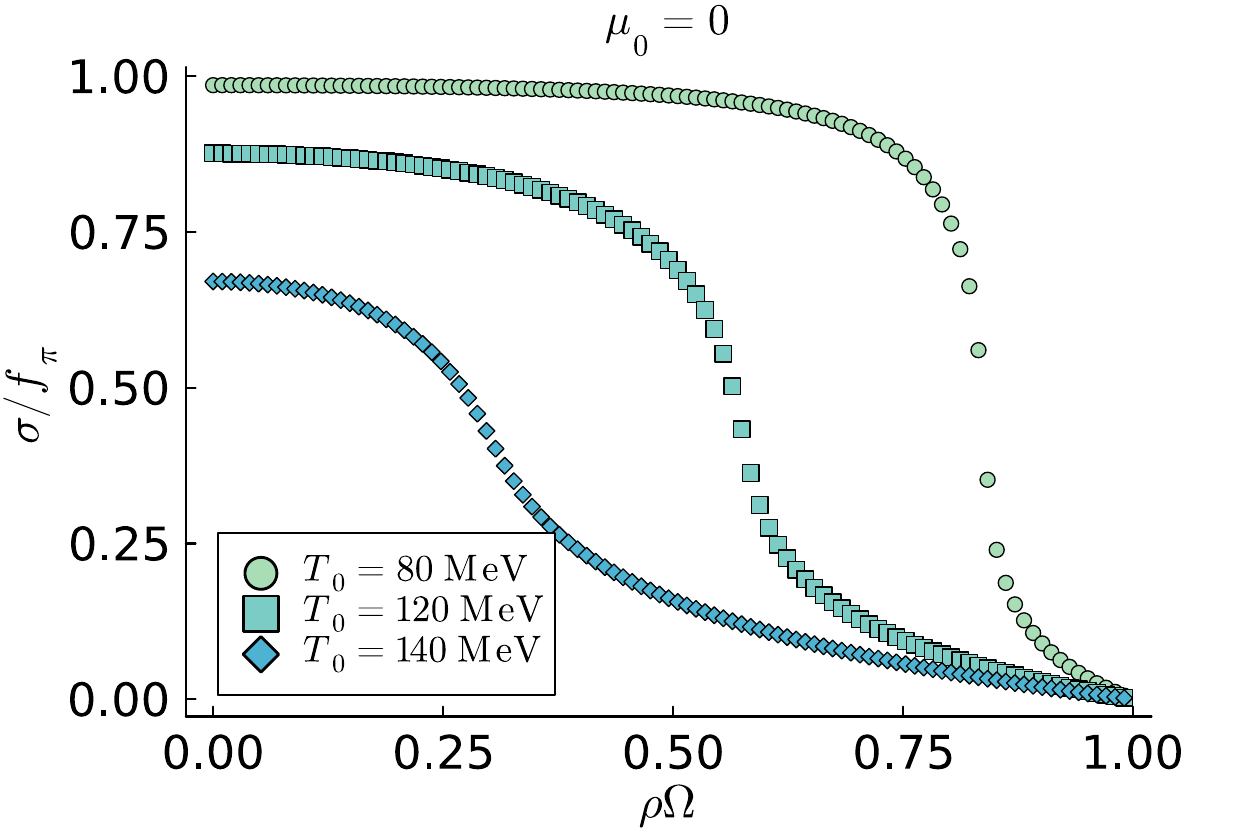} \\
    \includegraphics[width=0.95\linewidth]{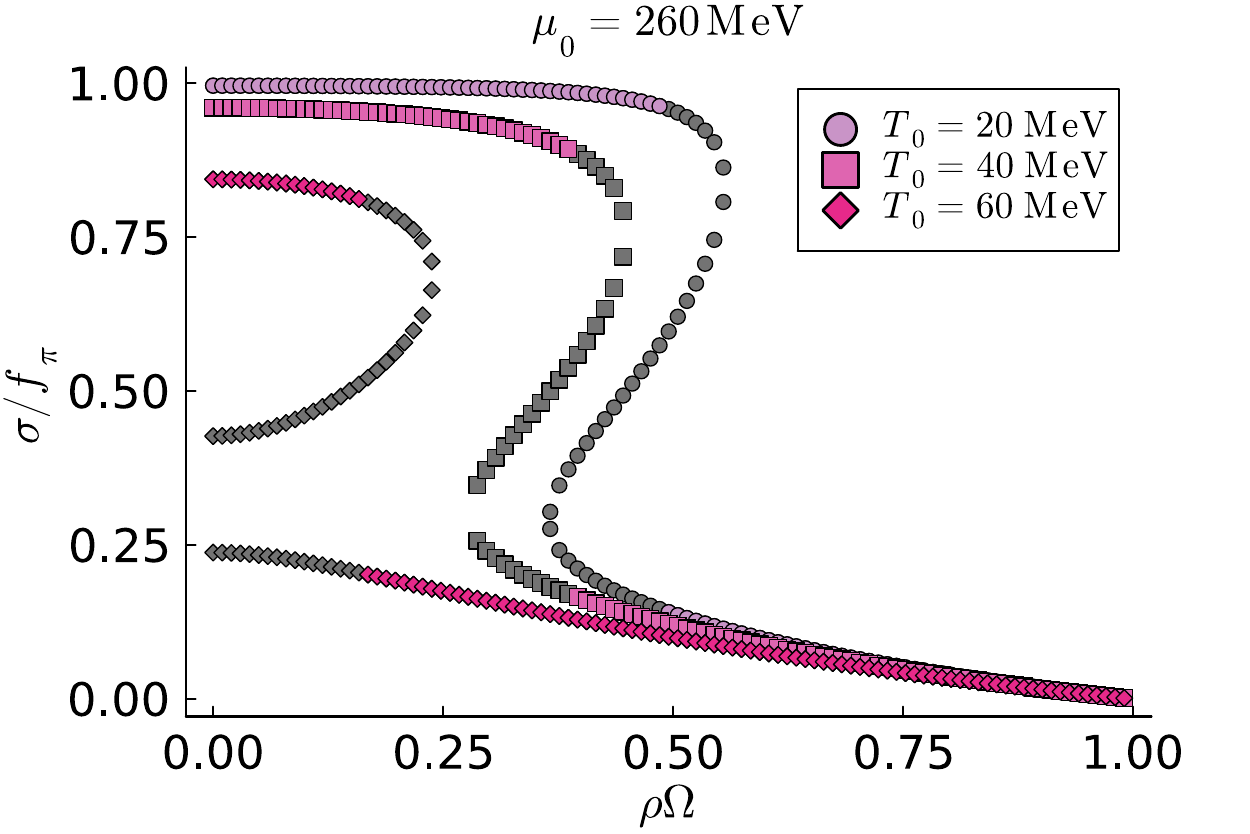}
\end{tabular}
    \caption{Slowly varying condensate: Normalized value of the inhomogeneous condensate $\sigma$, shown as a function of the rescaled radial distance $\rho \Omega$.
    The condensate is a solution of the mass gap equation \eqref{eq_massgap_slow}, shown for a set of on-axis temperatures $T_0$ at two values of the on-axis chemical potential $\mu_0$. The colored (black) points represent the thermodynamically (un)favored solutions.
    \label{fig_slow-sofR-mu}
    }
\end{figure}

In this section, we consider a point-dependent $\sigma$ while neglecting its spatial variations. The local gap equation is given by Eq. \eqref{eq:gap_local}:
\begin{equation}\label{eq_massgap_slow}
 \lambda[\sigma^2(\rho) - v^2] \sigma(\rho) = h - g \langle \bar{\psi} \psi \rangle,
\end{equation}
where 
\begin{equation}
 \langle \bar{\psi} \psi \rangle = g \sigma(\rho) N_f N_c \sum_{\varsigma,\lambda} \int \frac{dP}{e^{\beta_\rho E - \varsigma \alpha} + 1}.
\end{equation}
It is clear that in this approach, the system can develop local, inhomogeneous phases, depending on the local value of the condensate $\sigma$. This is exemplified in the solution to the local gap equation as a function of the dimensionless radial distance $\rho\Omega$ displayed in Fig. \ref{fig_slow-sofR-mu} for different on-axis values of the temperature and chemical potential. Contrary to the case when the effective mass was given by the average value $g\bar{\sigma}$, now the system is free to remain in the chirally-broken phase close to the rotation axis.

First, note that as the distance to the rotation axis increases and the light cylinder is approached, the local temperature and chemical potential increase according to the Tolman-Ehrenfest law, causing the effective mass to be reduced. Indeed, substituting the asymptotic value of the fermion condensate close to the light cylinder given in Eq. \eqref{eq_FC_smallM} into the local gap equation \eqref{eq_massgap_slow}, we find:
\begin{align}\label{eq_slow-near-firewall}
 \sigma(\rho) &\simeq\frac{ \sigma_\times }{\Gamma_\rho^2-\sigma_\times  h^{-1} \lambda v^2 }, \nonumber\\
 \sigma_\times &= \frac{h}{g^2 N_f N_c} \left(\frac{T_0^2}{6} + \frac{\mu_0^2}{2\pi^2} \right)^{-1},
\end{align}
where $\sigma_\times$ is a characteristic value that controls the behavior of the condensate $\sigma$ near the light-cylinder, where the Gamma factor becomes large, $\Gamma_\rho\to\infty$. We observe that in the limit $\rho \Omega \to 1$, the value of the $\sigma$ condensate near the boundary vanishes linearly:
\begin{equation}
    \sigma(\rho \Omega \to 1 ) \simeq 2 (1-\rho\Omega) \sigma_\times  . 
\end{equation}

\begin{figure}
\centering 
\begin{tabular}{c}
    \includegraphics[width=0.95\linewidth]{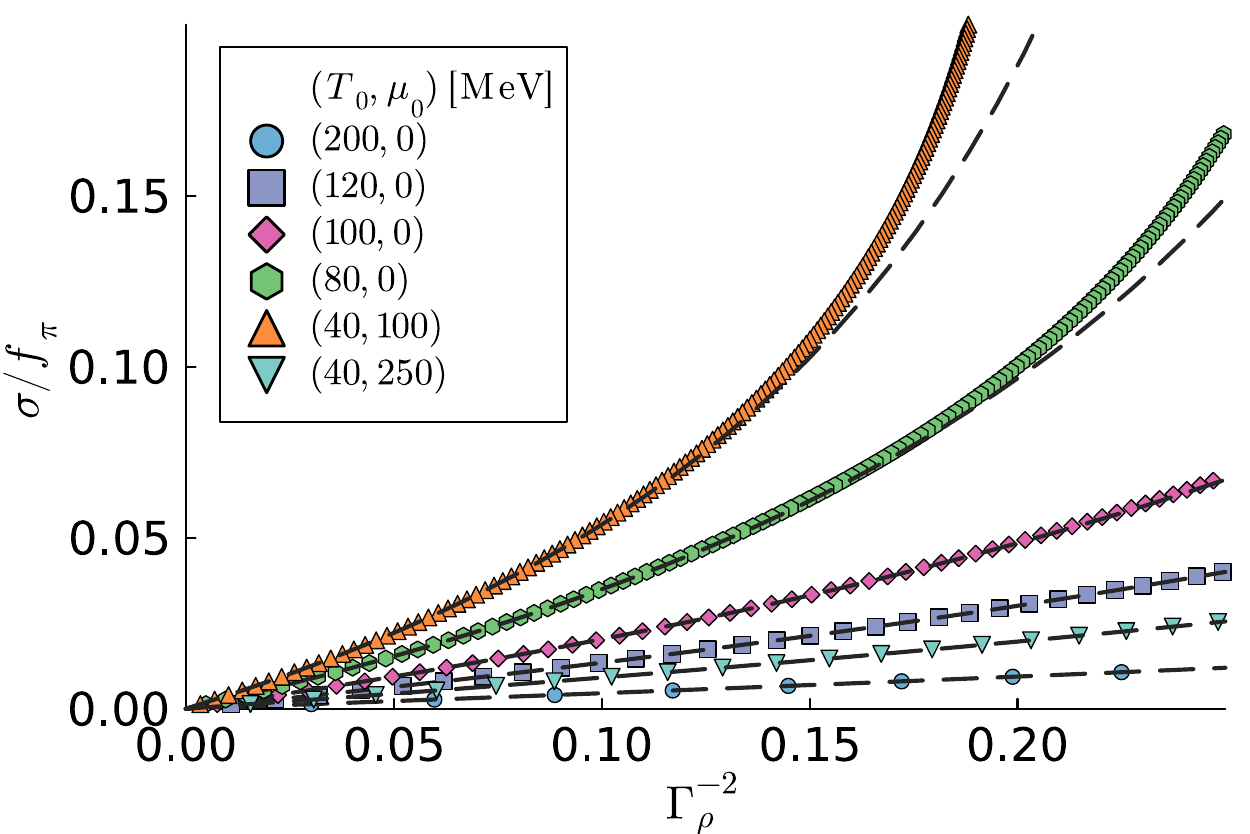} 
\end{tabular}
    \caption{Slowly varying condensate: Numerical confirmation of the asymptotic behavior of the condensate near the boundary, $\rho \to \Omega^{-1}$, plotted as a function of $\Gamma_\rho^{-2} = 1 - \rho^2 \Omega^2$. The dashed lines correspond to the analytical expression~\eqref{eq_slow-near-firewall}, while the points are obtained as numerical solutions of the mass gap equation \eqref{eq_massgap_slow}. The condensate is shown for several sets of temperatures and chemical potentials.
    \label{fig_slow-log}
    }
\end{figure}

In Fig. \ref{fig_slow-log} we compare the numerical solution of the gap equation close to the light cylinder with the analytical prediction given in Eq. \eqref{eq_slow-near-firewall}. The asymptotic formula agrees with the full solution provided that $\sigma/f_{\pi}< 0.1$ and deviates for higher values.

Similarly to Sec. \ref{sec:model1}, we are interested in the trajectory of the transition line in phase space. Considering, as in Sec.~\ref{sec:model1}, the values $(T_0, \mu_0)$ on the rotation axis, the phase transition will take place at the point where the effective temperature (chemical potential) equals the critical value for the transition in the absence of rotation: $[T(\rho_c), \mu(\rho_c)] = (T_c^{\Omega= 0}, \mu_c^{\Omega= 0})$ under the condition $T_c^{\Omega=0}/\mu_c^{\Omega=0}=T_0/\mu_0$. 
The distance at which the transition takes place, $\rho_c$, is a solution of the equation $\Gamma_{\rho_c} = \kappa$, namely:
\begin{equation}\label{eq_Slow-critomega}
 \rho_c = \frac{1}{\Omega} \sqrt{1 - \frac{1}{\kappa^2}}\,,
\end{equation}
where $\kappa = T_c^{\Omega= 0} / T_0 = \mu_c^{\Omega= 0} / \mu_0$ is the ratio between the (pseudo-)critical temperature (chemical potential) in the absence of rotation and the corresponding values on the rotation axis for the rotating system.

\begin{figure}
\centering 
\begin{tabular}{c}
    \includegraphics[width=0.95\linewidth]{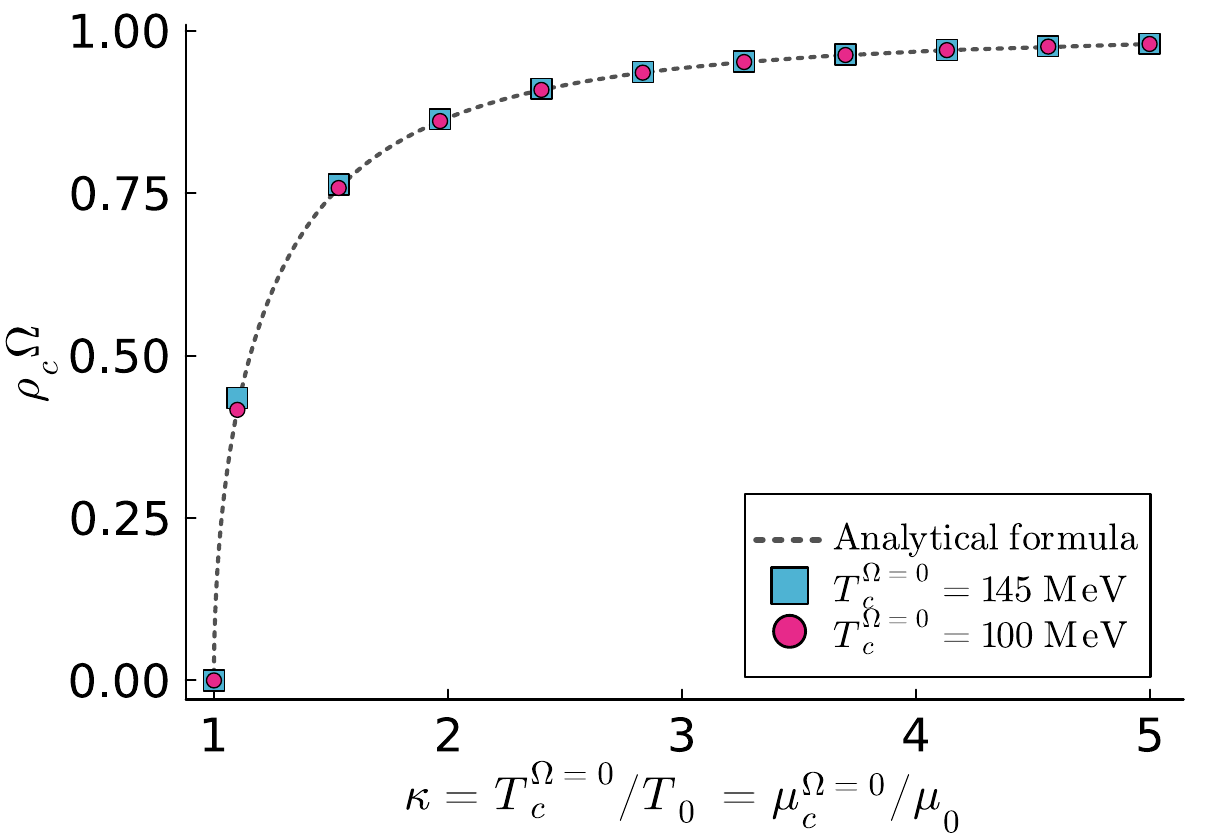} 
\end{tabular}
    \caption{Slowly varying condensate:  Dimensionless distance $\rho \Omega$ at which the phase transition takes place as a function of the ratio of the critical temperature (chemical potential) in the absence of rotation and the on-axis value of the temperature (chemical potential) for the rotating system. The dashed line corresponds to the analytical formula \eqref{eq_Slow-critomega}, while the data points correspond to the critical value obtained from solving the mass gap equation \eqref{eq_massgap_slow}.
    \label{fig_slow-crits}
    }
\end{figure}

We compute numerically the value of $\rho_c$ for a given ratio $\kappa$ and compare it with the analytical formula \eqref{eq_Slow-critomega}. The agreement between the numerical and analytical estimations is seen in Fig. \ref{fig_slow-crits}. Note that the analytical expression \eqref{eq_Slow-critomega} is exact, since it follows from the Tolman-Ehrenfest law applied to the transition points. This contrasts with the estimation of model 1 [c.f. Eq. \eqref{eq_effective_thermo}], which relies on the assumption that there exists an averaged measure of temperature and chemical potential encoding the response of the averaged system to rotation. 

\begin{figure}
    \centering
    \begin{tabular}{c}
     \includegraphics[width=0.95\linewidth]{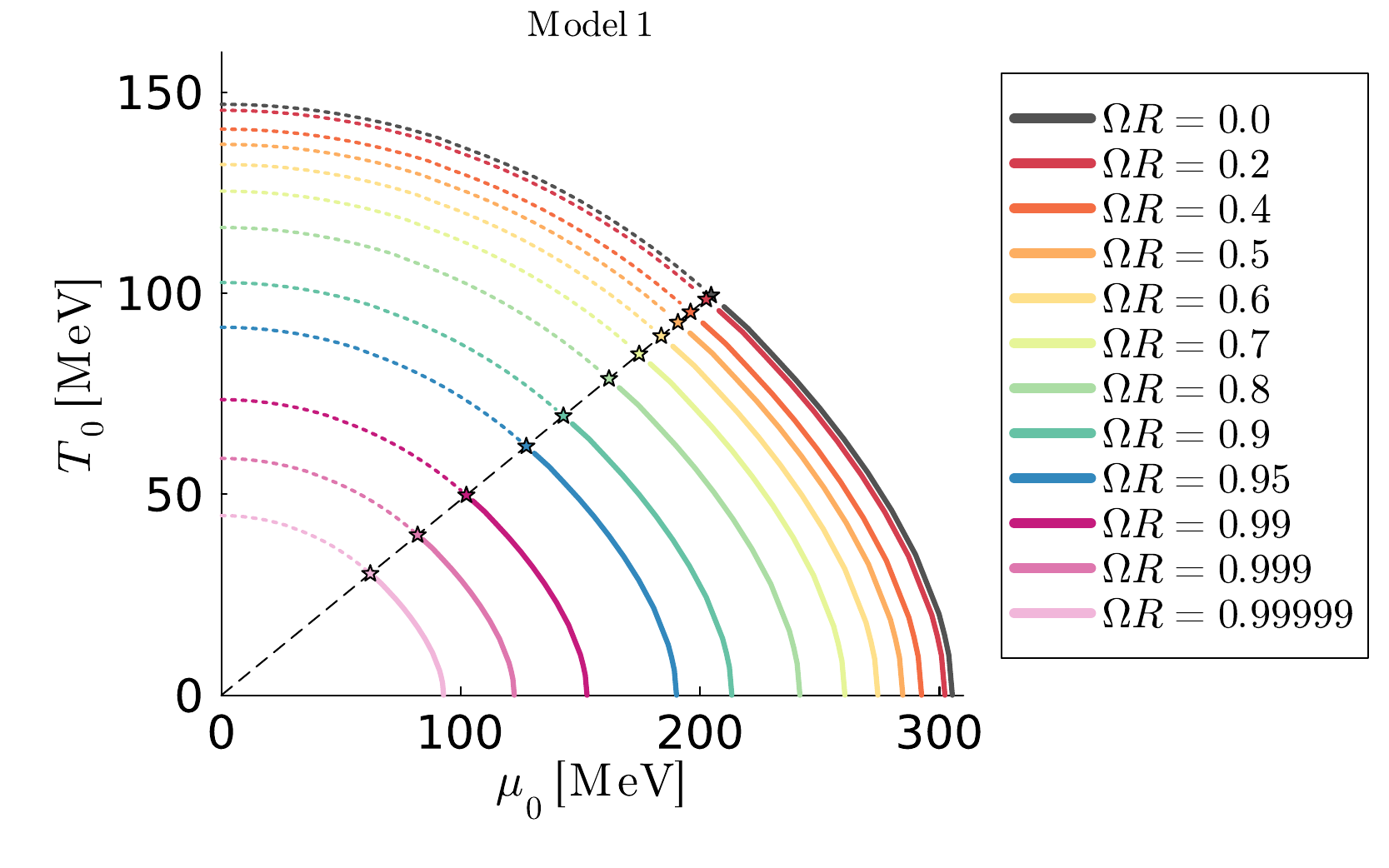}\\
     \includegraphics[width=0.95\linewidth]{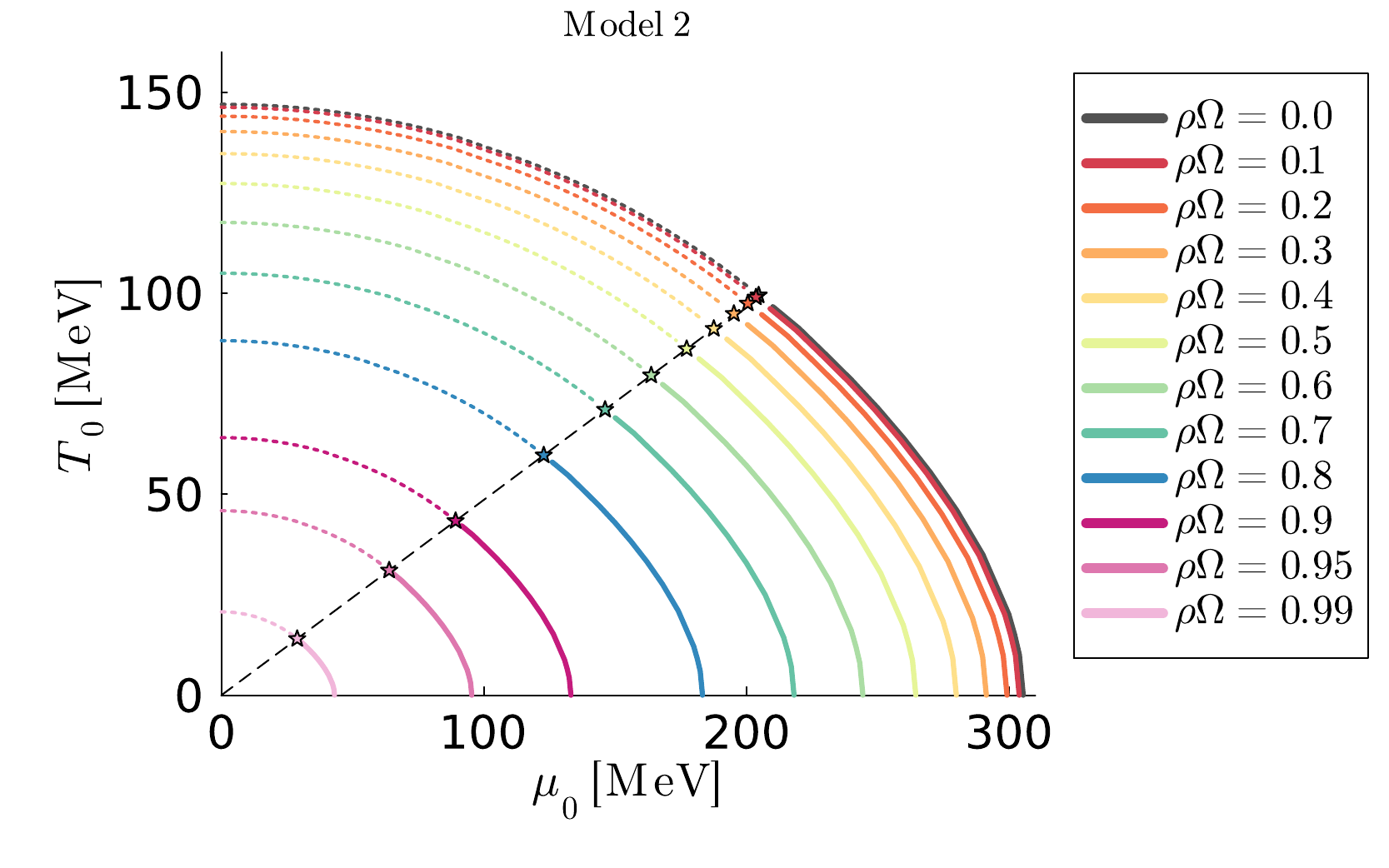}
    \end{tabular}
    \caption{(Top panel) $T$-$\mu$ phase diagram for model 1 (Sec. \ref{sec:model1}) at different values of the dimensionless system size $\Omega R$. Inner (outer) regions correspond to the chirally-broken (-restored) phase. (Bottom panel) $T$-$\mu$ phase diagram for model 2 (Sec. \ref{sec:model:model2}) at different values of the dimensionless radial distance $\rho\Omega$. 
    (Both panels) Crossover transitions are signaled with dotted lines, while first-order phase transitions are displayed with solid lines. }
    \label{fig:phase_diagrams_12}
\end{figure}

The phase diagram for model 2 can be obtained from the $T$-$\mu$ phase diagram at vanishing angular velocity, supplemented with Eq.~\eqref{eq_Slow-critomega}. The result is shown in the bottom panel of Fig.~\ref{fig:phase_diagrams_12}. Again, dotted and solid lines correspond to crossover and first-order phase transitions, respectively. In this case, the region enclosed by the outer black line $(\rho\Omega=0)$ corresponds to inhomogeneous phases where the chiral symmetry is broken close to the rotation axis and restored after some critical radial distance $\rho\Omega$. Each line shown in the plot gives the parameters $(T_0, \mu_0)$ at which the transition takes place at a given $\rho \Omega$. 

We point out that the interpretation of the phase diagrams for models 1 and 2 is physically very different, despite their visual similarity. On the one hand, model 1 asserts that the system as a whole is in a single phase, either broken or restored, and that for a system that extends to the light cylinder, the chiral symmetry is restored. On the other hand, model 2 allows for inhomogeneous phases, as explained above, and the nature of the phase close to the rotation axis is unaffected by the system properties close to the light-cylinder.

Another sensitive way to compare the global $\bar{\sigma}$ of Sec. \ref{sec:model1} and the slowly-varying local solution presented here is to consider the average value of $\sigma$ over the fictitious cylinder of radius~$R$:
\begin{equation}\label{eq:local_averaged}
 \bar{\sigma}(R) = \dfrac{2}{R^2} \int_0^{R} d\rho \, \rho \, \sigma(\rho)\,.
\end{equation}

\begin{figure}
\centering 
\begin{tabular}{c}
    \includegraphics[width=0.95\linewidth]{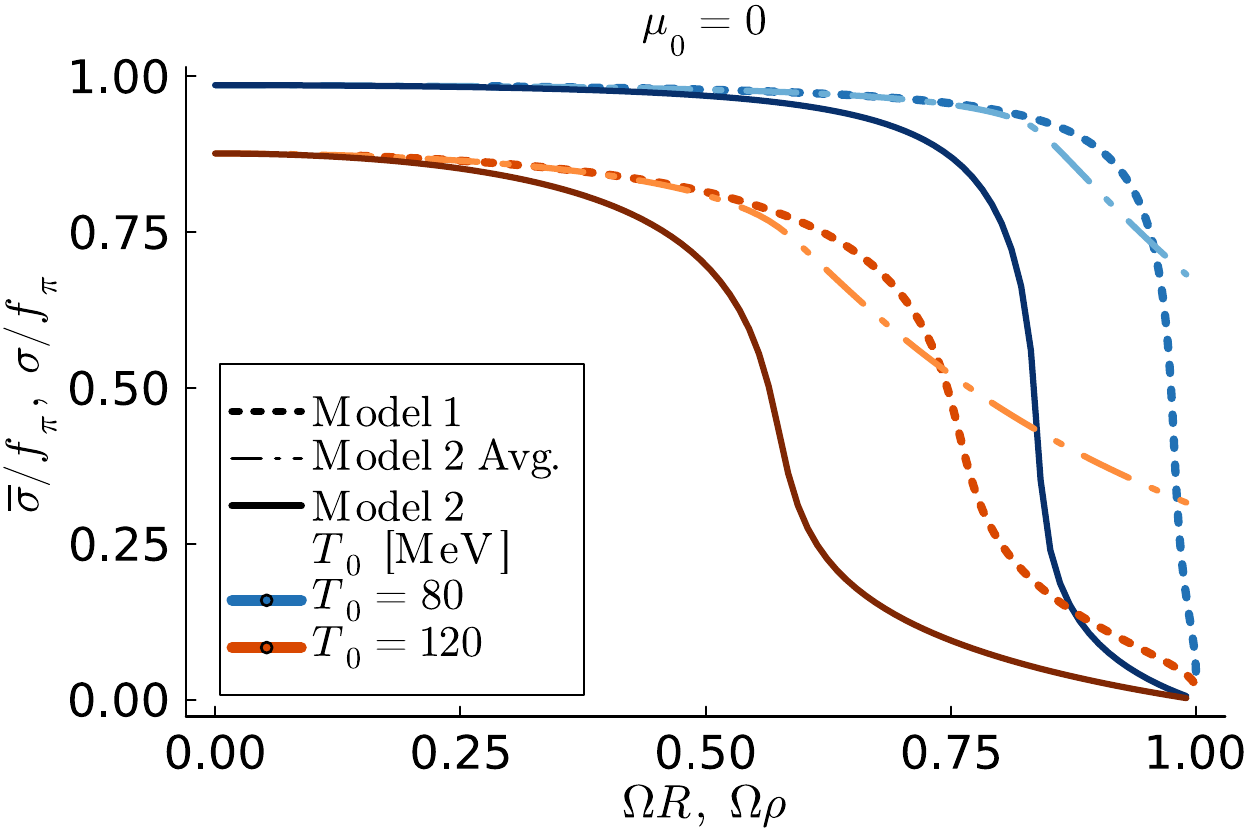} \\
    \includegraphics[width=0.95\linewidth]{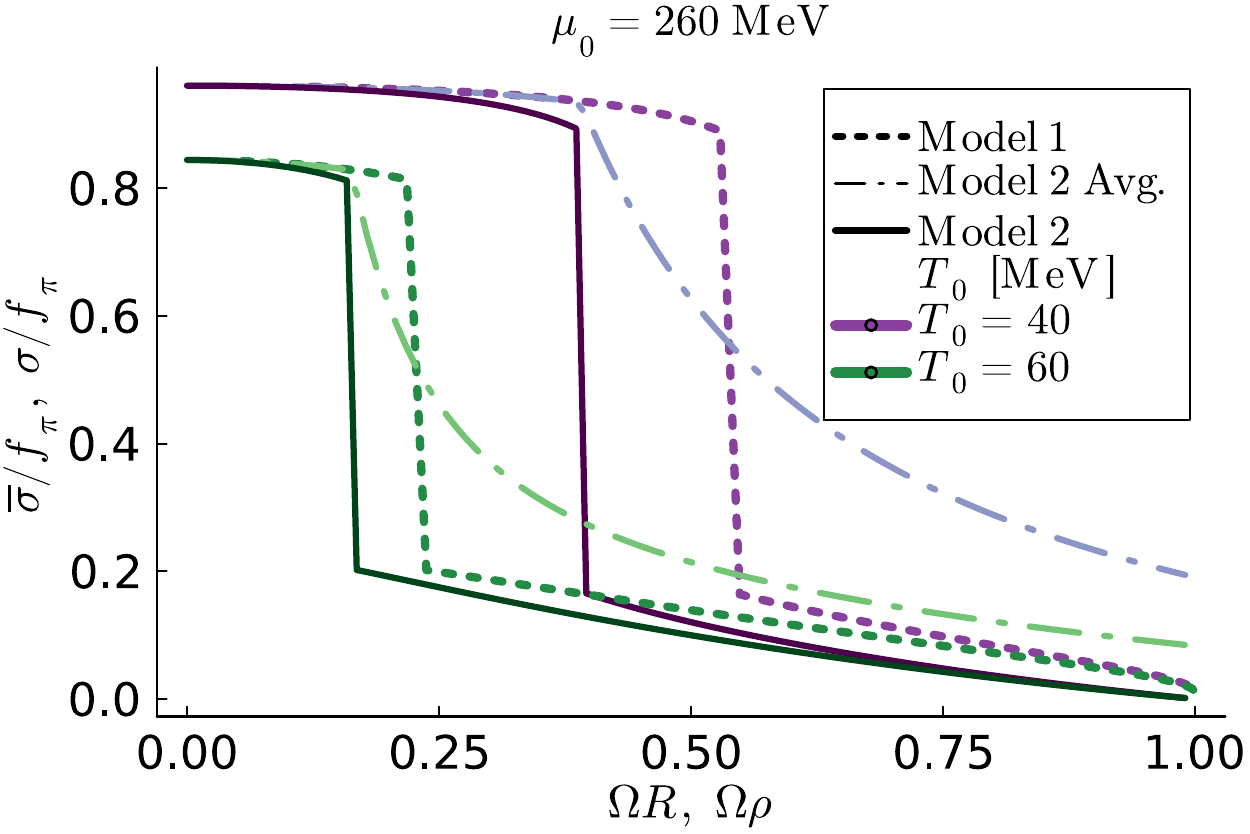}
\end{tabular}
    \caption{Comparison between the constant global value for condensate $\sigma$ (Model 1) and the local value of $\sigma$ neglecting the gradient contributions (Model 2). We include the averaged value $\overline\sigma(R)$ of the local $\sigma$ as given in Eq. \eqref{eq:local_averaged} (Model 2 Avg.). The transition happens systematically at a smaller value of $\Omega R$ in Model 2 (slowly varying condensate) compared to Model 1 (uniform condensate).
    \label{fig:Comparison12}
    }
\end{figure}

The comparison between the two approaches is shown in Fig. \ref{fig:Comparison12}. We first observe that the phase transition happens systematically at a smaller value of $\Omega R$ for the local $\sigma$ (Model~2) compared to the global ${\bar \sigma}$ (Model~1) of Sec. \ref{sec:model1}. Nevertheless, the curve corresponding to the averaged value $\overline{\sigma}(R)$ of the local condensate $\sigma$, given in Eq.~\eqref{eq:local_averaged}, is closer to the curve corresponding to the global condensate ${\bar \sigma}$ (Model~1), especially when the transition is not of first order type. However, note that Eq.~\eqref{eq:local_averaged} predicts a finite value of ${\bar \sigma}$ when the system extends to the light cylinder, while the global approach of Sec.~\ref{sec:model1} requires that the condensate ${\bar\sigma}$ vanishes as $R \to \Omega^{-1}$.

\section{Model 3: Fully inhomogeneous condensate}\label{sec:model3}

In the previous section, we determined the value of the condensate $\sigma$ locally, based on the expected Tolman-Ehrenfest law for the local temperature and chemical potential, in the approximation where the gradients of $\sigma$ are small and can, therefore, be neglected. We now move to the third approach, where the gradients of $\sigma$ are taken into account. It is important to note that now the gap equation \eqref{eq:model_diff} is a second-order differential equation with two integration constants. A priori, any pair of integration constants gives a solution to the extremization problem $\delta \Phi_{\rm m.f.} = 0$. After a careful analysis, we will show that, in fact, only 
a small number of solutions ensure the global minimization of the total grand potential of the system.

\subsection{Set-up of the boundary value problem}

\begin{figure}
\centering 
\begin{tabular}{c}
    \includegraphics[width=0.985\linewidth]{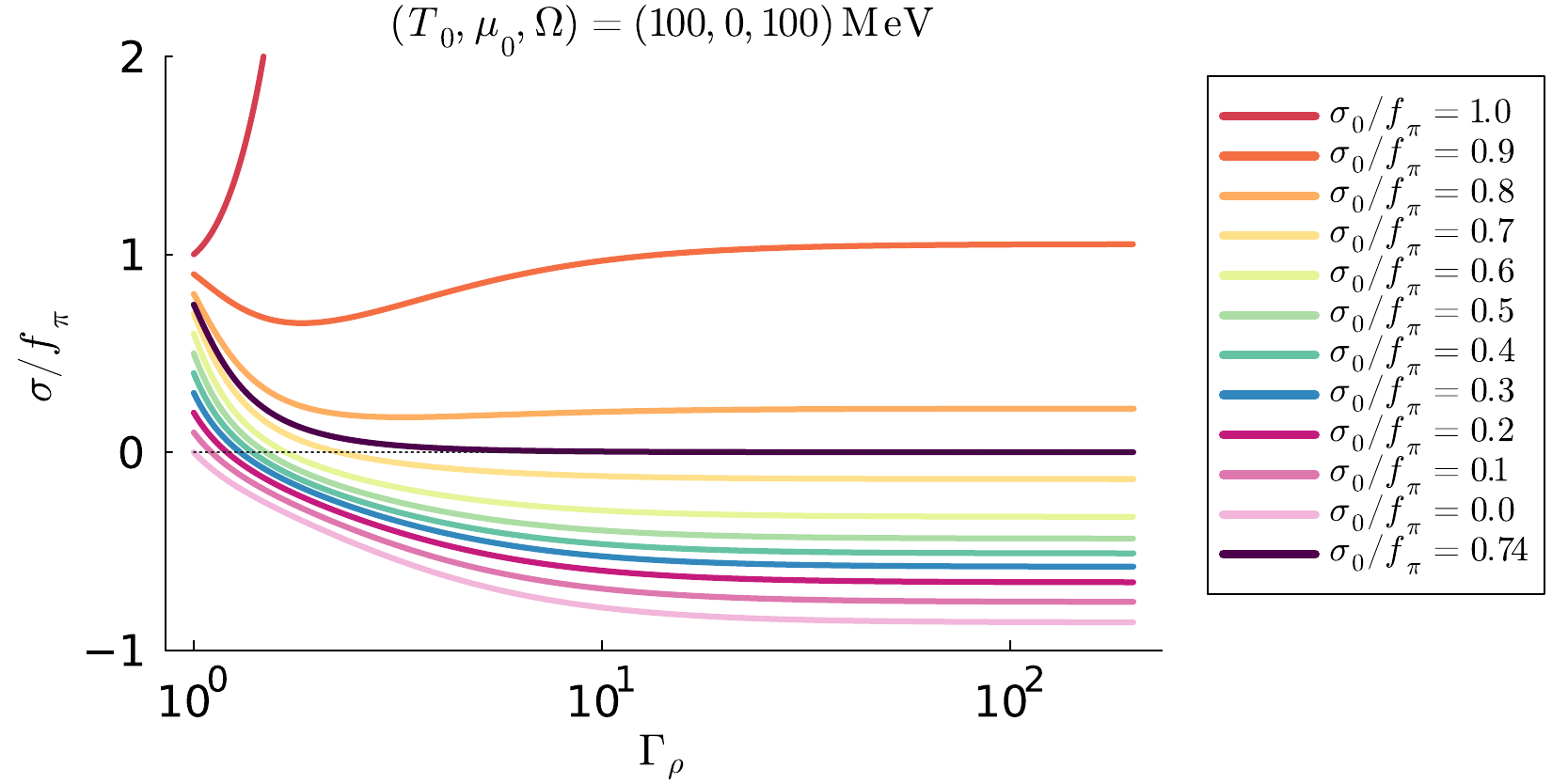} \\
    \includegraphics[width=0.985\linewidth]{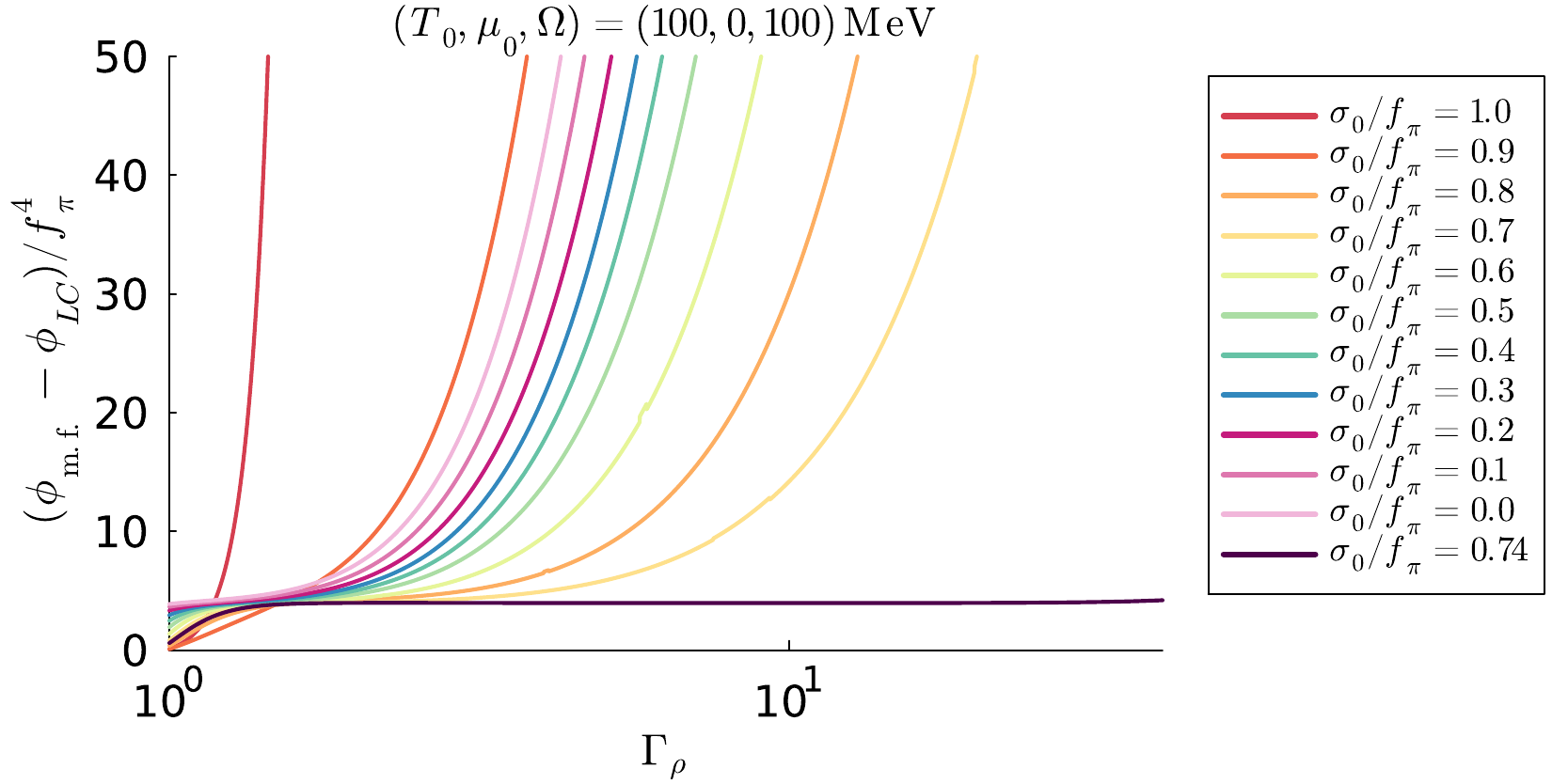}
\end{tabular}
    \caption{Full inhomogeneous solution: (Upper panel) Normalized value of the condensate $\sigma$ obtained from solving the differential equation \eqref{eq_Gap_diff_TE} as we vary its on-axis value. (Lower panel) Normalized grand canonical potential density, expressed relative to its asymptotic value, computed from the integrand of~\eqref{eq_grand_potential} for the corresponding solutions in the upper panel [cf. the definitions in Eqs.~\eqref{eq_PLC_1} and \eqref{eq_PLC_2}].}
    \label{fig:sigma_diff}
\end{figure}

We begin by rewriting Eq. \eqref{eq:model_diff} as a function of the Lorentz factor \eqref{eq_Gamma_rho}:
\begin{align}\label{eq_Gap_diff_TE}
    &\Omega^2\Gamma_\rho^4 (\Gamma_\rho^2 - 1)\sigma'' + \Omega^2\Gamma_\rho^3(3 \Gamma_\rho^2 - 1) \sigma'\nonumber \\& - \lambda \sigma(\sigma^2 -v^2) + h - g \langle \overline{\psi} \psi \rangle=0\,,
\end{align}
where the prime denotes derivatives with respect to the Lorentz factor $\Gamma_\rho$, i.e. $\sigma' = \partial \sigma / \partial \Gamma_\rho$.
The asymptotic solution for $\sigma$ near the rotation axis (located at $\Gamma_\rho = 1$) is given by
\begin{equation}\label{eq_sigma_diff_axis_TE}
    \sigma(\Gamma_\rho\to 1) = \sigma_{0} + \mathcal{C}_0 \log(1-\Gamma_\rho) + O(1-\Gamma_\rho)\,,
\end{equation}
where both $\sigma_{0}$ and $\mathcal{C}_0$ are integration constants. Note that the term proportional to $\mathcal{C}_0$ diverges on the rotation axis. Consequently, if $\mathcal{C}_0\neq 0$, the condensate $\sigma$ diverges inside the light cylinder, which, in turn, gives a finite contribution to the boundary term in Eq. \eqref{eq:variation_bdry} and implies that $\delta \Phi_{\rm m.f.}\neq0$, so this is not an acceptable solution. Thus, the regularity of the solution near the rotation axis enforces $\mathcal{C}_0=0$. We are left with a single integration constant, denoted by $\sigma_{0}$, which is equal to the value of $\sigma$ on the rotation axis. 
The top panel of Fig.~\ref{fig:sigma_diff} shows the condensate $\sigma / f_\pi$ as a function of the Lorentz factor $\Gamma_\rho$ for various on-axis values $\sigma_0$. It can be seen that $\sigma$ reaches a constant value close to the light cylinder, as $\Gamma_\rho \to \infty$, which affects the total grand potential of the system, as we discuss below.

Close to the light-cylinder, the fermion condensate is given in the Tolman-Ehrenfest approximation by Eq.~\eqref{eq_FC_smallM}. The asymptotic solution of Eq.~\eqref{eq_Gap_diff_TE} near the light-cylinder (where $\Gamma_\rho\to \infty$) is given by
\begin{multline}\label{eq_sigma_diff_LC_TE}
    \sigma(\Gamma_\rho\gg 1) = \sigma_{\rm LC} \left(1 + \frac{h}{2}\sigma_\times^{-1} \dfrac{\log\Gamma_\rho}{\Gamma_\rho^2}\right) + \dfrac{\mathcal{C}_{\rm LC}}{\Gamma_\rho^2} \\
    + O(\Gamma_\rho^{-4} \log\Gamma_\rho)\,,
\end{multline}
with $\sigma_{\rm LC}$ and $\mathcal{C}_{\rm LC}$ two integration constants, while
$\sigma_\times$ is defined in Eq. \eqref{eq_slow-near-firewall}. As opposed to models 1 and 2, the differential equation considered in model 3 allows for a finite (non-vanishing) value $\sigma_{\rm LC}$ of the $\sigma$ condensate on the light cylinder. Solving the differential equation for a given $\sigma_{0}$ will, in general, generate a finite value of $\sigma_{\rm LC}$. This is explicitly demonstrated in Fig.~\ref{fig:sigma_diff} (upper panel).

\begin{figure}
\centering 
\begin{tabular}{c}
    \includegraphics[width=0.985\linewidth]{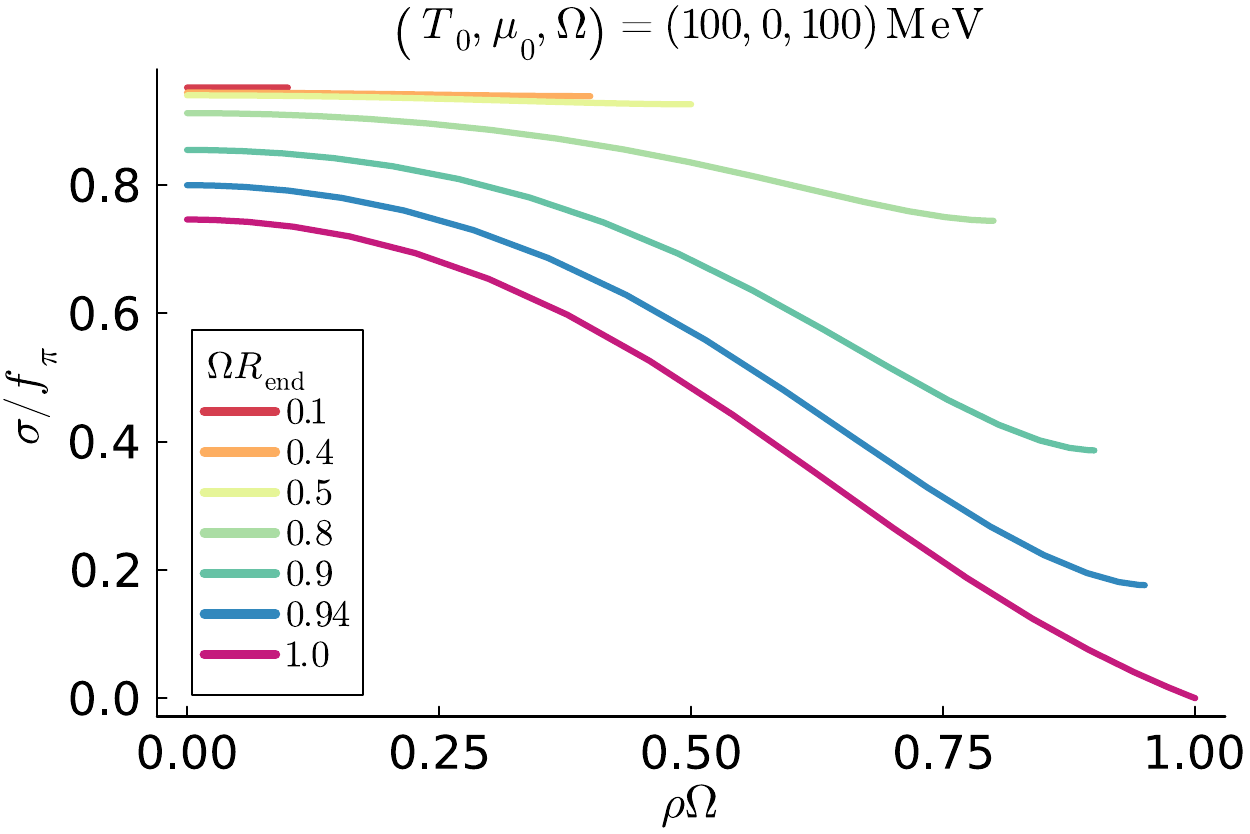} 
\end{tabular}
    \caption{    
    Normalized value of the condensate $\sigma$ obtained from solving the differential equation \eqref{eq_Gap_diff_TE} for finite-size systems extending to $\rho=R_{\rm end}$.
    }
    \label{fig:sigma_interm}
\end{figure}

Given a thermodynamic state, specified by the temperature $T$, the chemical potential $\mu$, and the angular velocity $\Omega$, the task is to find the value of the on-axis condensate $\sigma_{0}$ such that the grand canonical potential $\Phi_{\rm m.f.}$ is globally minimal. 
The result certainly depends on the size of the system: if we restrict the system to the rotation axis, we expect that the value of $\sigma_{0}$ coincides with that of the non-rotating system, whereas if we extend the size of the system up to the light-cylinder, the value of $\sigma_{0}$ can drastically change. For intermediate sizes, one expects an interpolation between these two values of $\sigma_{0}$ as exemplified in Fig. \ref{fig:sigma_interm}. For the remainder of this section, we will restrict ourselves to the situation where the system extends all the way up to the light-cylinder, since this is the situation where the presence/absence of boundary conditions becomes more relevant.
In order to determine the value of $\sigma_{0}$ giving a minimal $\Phi_{\rm m.f.}$, we first quote the value of the grand canonical potential density $\phi_{\rm m.f.}$, defined as the integrand of Eq. \eqref{eq_grand_potential}, as we approach the light-cylinder ($\Gamma_\rho\to \infty$):
\begin{equation}
    \phi_{\rm m.f.} =  \phi_{\rm LC} + \dfrac{N_f N_c}{12\pi^2} g^2 \sigma_{\rm LC}^2 \Gamma_\rho^2 (\pi^2 T_0^2 +3\mu_0^2) + O(g^4 \sigma_{\rm LC}^4 \Gamma_\rho^0)\,,
    \label{eq_PLC_1}
\end{equation}
where we have defined the leading contribution near the light cylinder as $\phi_{\rm LC} = \phi_{\rm LC}^0\Gamma_\rho^4$, with
\begin{equation}
    \phi_{\rm LC}^0= -\dfrac{N_f N_c}{180 \pi^2}  (7 \pi^4 T_0^4 + 30\pi^2 T_0^2 \mu_0^2 + 15 \mu_0^4 ) \,,
    \label{eq_PLC_2}
\end{equation}
and we have denoted the value of the condensate $\sigma$ at the light-cylinder as $\sigma(\rho\Omega=1)\equiv \sigma_{\rm LC}$.
The grand canonical potential diverges near the light cylinder. 

In order to determine the thermodynamically-favorable solution, we note that the leading contribution to the divergence depends only on the thermodynamic state, but not on the value of $\sigma$ at the boundary. Therefore, this contribution is irrelevant in determining the value of $\sigma_{0}$ that gives a minimal $\Phi_{\rm m.f.}$ and we can directly work with $\phi_{\rm m. f.}-\phi_{\rm LC}$. The previous quantity diverges as $\Gamma_\rho^2$ close to the light-cylinder, and the integrated version $\int d\rho \rho (\phi_{\rm m.f.}-\phi_{\rm LC}) \sim \int \frac{d\Gamma_\rho}{\Gamma_\rho^3} (\phi_{\rm m.f.}-\phi_{\rm LC})\sim \log \Gamma_\rho$ is also divergent unless the term proportional to $\Gamma_\rho^2$ in Eq. \eqref{eq_PLC_1} vanishes. Crucially, such a term is positive semi-definite and therefore any non-zero value of $\sigma$ at the light cylinder will give an infinite positive contribution to the grand canonical potential. As a result, the global solution for $\sigma$ given a thermodynamic state is such that $\sigma$ vanishes near the light cylinder. An explicit demonstration of the behavior of $\phi_{\rm m.f.}-\phi_{\rm LC}$ for vanishing and non-vanishing $\sigma$ on the light cylinder can be found in Fig.~\ref{fig:sigma_diff} (lower panel).

In summary, given a thermodynamic state, one needs to find the value of the condensate at the rotation axis $\sigma_{0}$ such that the condensate $\sigma$ vanishes at the light-cylinder. Thus, out of the continuous possible values for the two integration constants $\sigma_{0}$ and $\mathcal{C}_0$ in Eq. \eqref{eq_sigma_diff_axis_TE}, only a discrete set of values $(\sigma_{0}^{(i)},\mathcal{C}_0=0)$ gives an acceptable solution (here $i$ labels the independent solutions that extremize the grand potential $\Phi_{\rm m.f.}$).

\subsection{Physical solutions for the inhomogeneous condensate}

\begin{figure}
\centering 
\begin{tabular}{c}
    \includegraphics[width=0.95\linewidth]{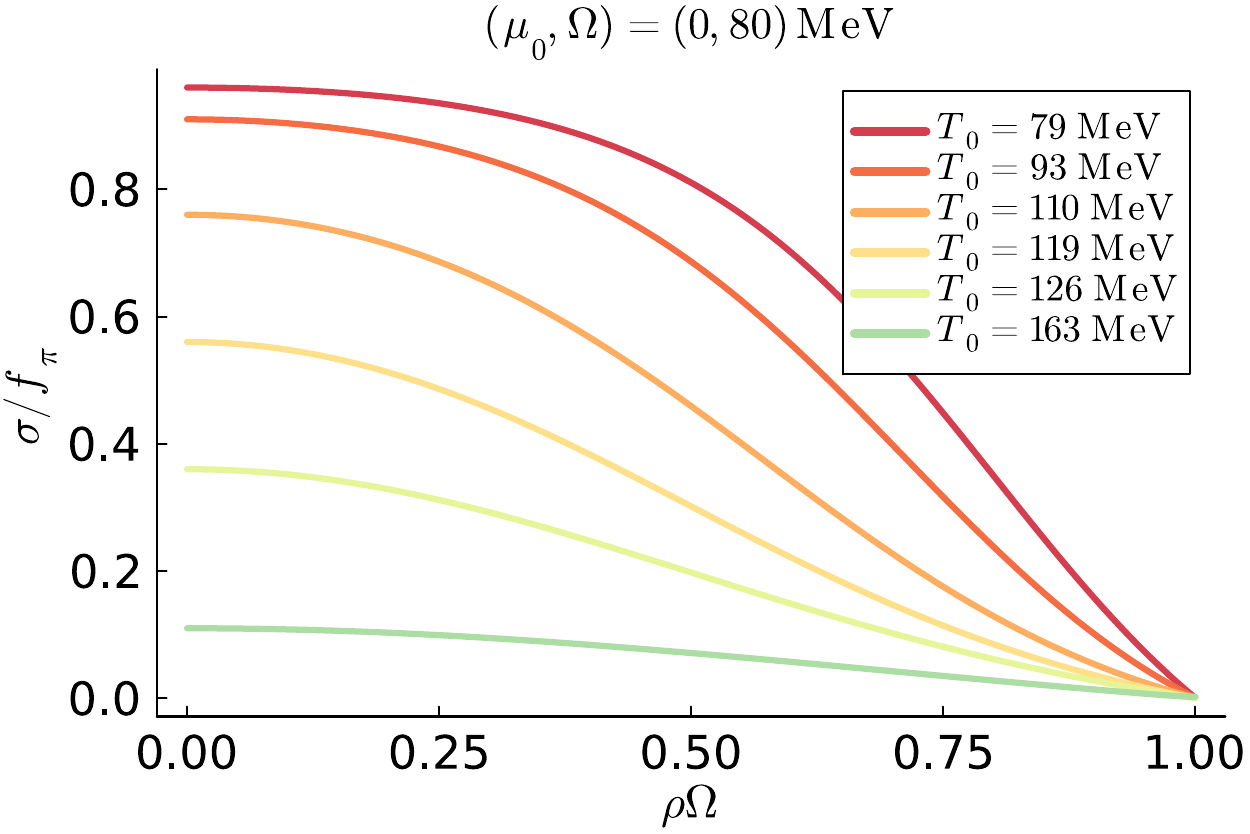} \\
    \includegraphics[width=0.95\linewidth]{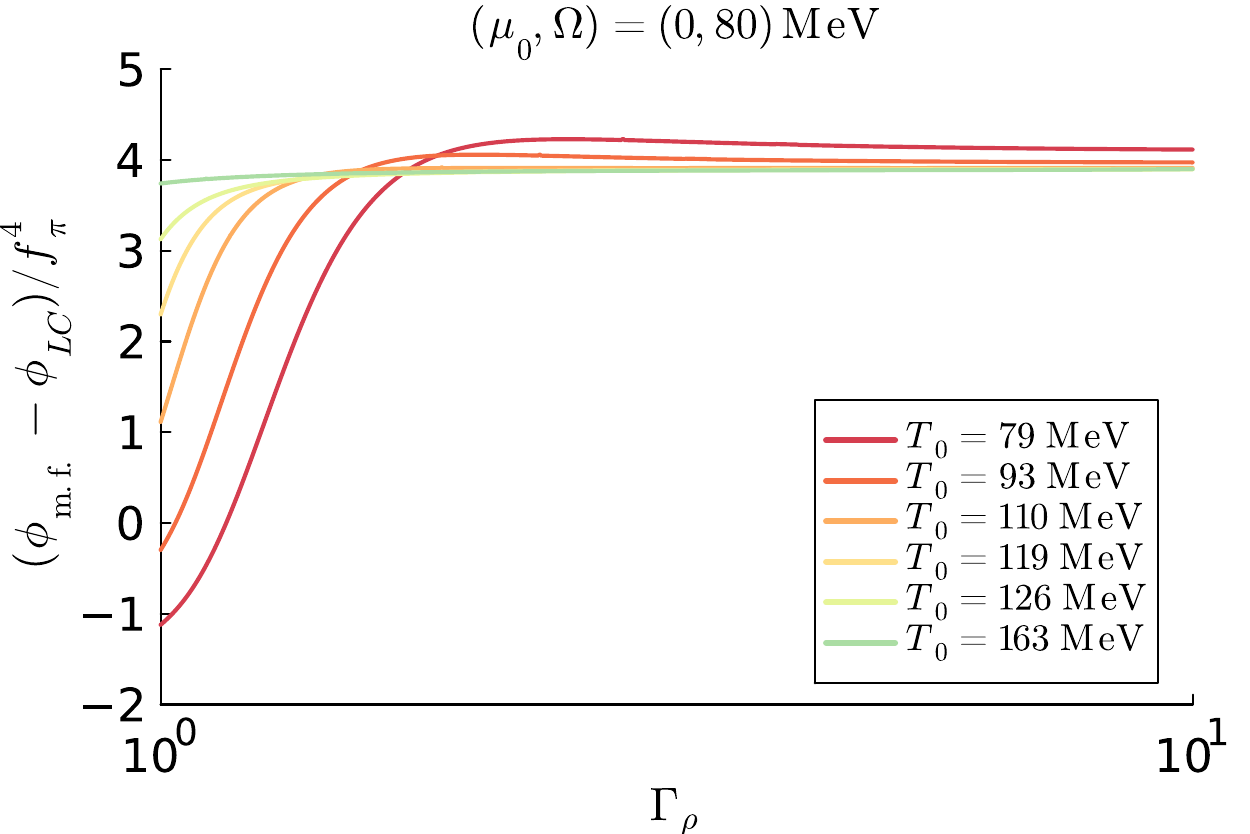}
\end{tabular}
    \caption{Full inhomogeneous solution: (Upper panel) Normalized value of the condensate $\sigma$ obtained from solving the differential equation \eqref{eq_Gap_diff_TE}, subjected to the boundary condition at the light cylinder $\sigma(\Gamma_\rho\to\infty)=0$, as we vary the on-axis temperature $T_0$. (Lower panel) Normalized subtracted grand canonical potential density computed as the integrand of Eq.~\eqref{eq_grand_potential} for the corresponding solutions shown in the upper panel.
    }\label{fig:sigma_good}
\end{figure}

In Fig.~\ref{fig:sigma_good}, we present the profile of $\sigma$ that minimizes the grand canonical potential $\Phi_{\rm m.f.}$ for zero chemical potential and finite angular velocity as we vary the temperature of the system. There is a discrete set of solutions of the differential equation for which $\sigma$ vanishes at the light cylinder and for which $\sigma$ is bounded. 
These conditions can be reformulated as a Dirichlet condition at the light cylinder and a Neumann condition at the rotation axis:
\begin{align}
	\frac{\partial \sigma(\rho)}{\partial \rho}{\biggl|}_{\rho=0}=0\,, \qquad \sigma(\rho) {\biggl|}_{\rho \Omega =1} = 0\,.
    \label{eq_conditions}
\end{align}
Generally, there is a unique solution $\sigma(\rho)$ satisfying Eqs. \eqref{eq_conditions}.
Later, we will see instances in which there are multiple solutions under the two conditions~\eqref{eq_conditions}.

We observe that $\sigma$ is monotonically decreasing, and therefore we typically have an inhomogeneous phase where the system is in the chirally-broken phase close to the axis of rotation but in the chirally-restored phase close to the light cylinder. This phenomenon can be understood from the fact that the effective temperature increases as we depart from the rotation axis, and thus the fermion condensate melts after some radial distance. The emergence of this phase structure is in agreement with the Tolman-Ehrenfest effect~\cite{Tolman:1930ona, Tolman:1930zza}, which implies that the local temperature increases as we move further from the central axis to the boundary of the system [see Eqs.~\eqref{eq_beta_rho} and \eqref{eq_Gamma_rho}].
For a high enough temperature, the whole system resides in the chirally-restored phase. 

This behavior of the phase structure is analogous to the results found in Sec.~\ref{sec:model2}, where the gradients of the field $\sigma$ are neglected. Nevertheless, the two approaches -- that neglect and take into account the derivatives-- are inequivalent. One of the major differences between them lies in the fact that the results of Sec.~\ref{sec:model2} depend on the dimensionless combination $\rho \Omega$, as a consequence of the Tolman-Ehrenfest approximation. The presence of the radial gradients breaks this degeneracy, and the system now depends separately on the radial distance $\rho$ and the angular velocity $\Omega$. 

One can naturally ask how these two approaches can be compared. To this end, we note that, after the change of variables\footnote{The same conclusion applies for a change of variables that depends on the combination $\rho\Omega$, i.e. $\rho \to f(\rho\Omega)$ for any function~$f$.} $\rho \to \Gamma_\rho$ in the differential equation~\eqref{eq_Gap_diff_TE}, the gradients of the condensate $\sigma$ are proportional to the angular velocity $\Omega$. As a result, the limit $\Omega\to 0 $ in the differential gap equation~\eqref{eq_Gap_diff_TE} reduces to the local gap equation~\eqref{eq:gap_local}. Accordingly, the approximation of a slowly varying $\sigma$ is valid for a sufficiently small angular velocity $\Omega/f_\pi\ll 1$. 

\begin{figure}
\centering 
\begin{tabular}{c}
    \includegraphics[width=0.95\linewidth]{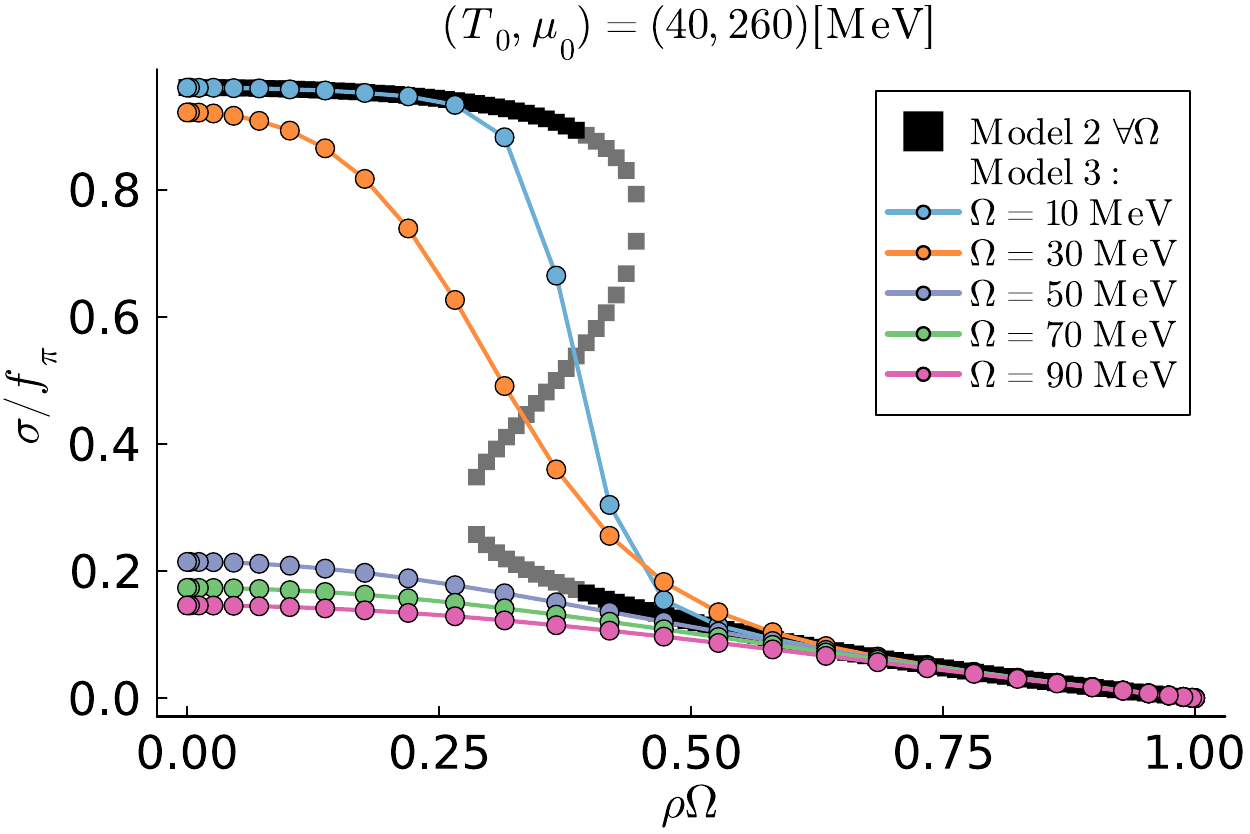} 
\end{tabular}
    \caption{Slowly-varying condensate vs. fully-inhomogeneous solution: Comparison between (i) the local approach that neglects the gradients (Model~2) [the condensate $\sigma$ is obtained from the mass gap equation \eqref{eq_massgap_slow}], and (ii) the fully-inhomogeneous approach, which takes them into account (Model 3) [the condensate $\sigma$ is obtained as a solution of Eq. \eqref{eq_Gap_diff_TE}]. Interestingly, Model~2 corresponds precisely to the vanishing rotation limit $\Omega\to 0$ of Model~3. The gray squares correspond to solutions of Model 2 that are not thermodynamically favored.
    \label{fig:Comparison23}
    }
\end{figure}

In Fig. \ref{fig:Comparison23}, we show the radial dependence of $\sigma$ for different angular velocities, obtained from solving the differential equation~\eqref{eq_Gap_diff_TE} that takes into account full inhomogeneities (Model 3), and the radial dependence of $\sigma$ for the same temperature and chemical potentials as obtained from Sec.~\ref{sec:model2}, which ignores the derivatives in the assumption of a slowly-varying condensate (Model 2). Indeed, we observe that, as the angular velocity decreases, the results from Model~3 approach those of Model~2. We note that the differential equation considered in Model~3 precludes the development of sharp jumps in the system. Therefore, the system can go from the chirally-broken to the chirally-restored phase only via a crossover transition.

\begin{figure}
\centering 
\begin{tabular}{c}
    \includegraphics[width=0.95\linewidth]{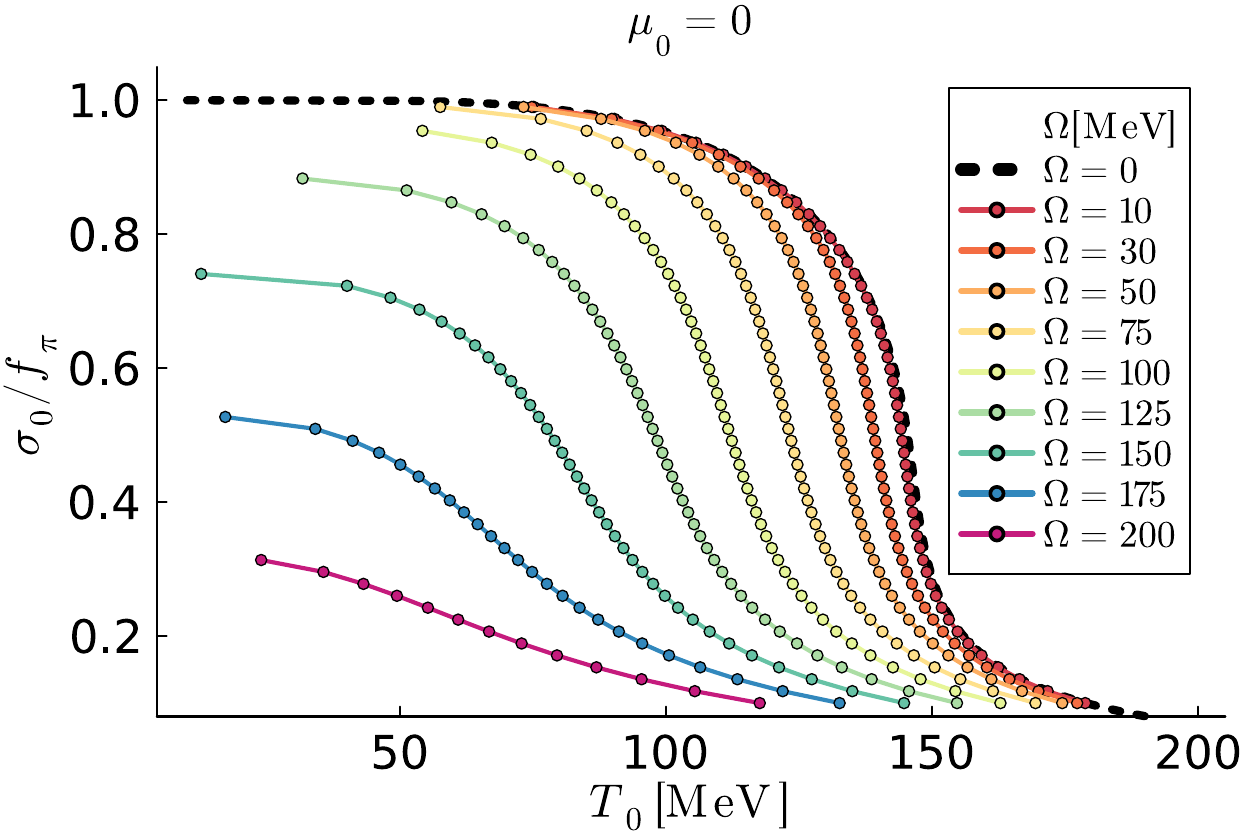} 
\end{tabular}
    \caption{Fully-inhomogeneous solution:  Normalized on-axis value of the condensate $\sigma$, obtained as a solution of the differential equation \eqref{eq:model_diff}, subjected to the boundary conditions~\eqref{eq_conditions}. The solutions are shown as functions of the on-axis temperature $T_0$ for different angular velocities $\Omega$ and a vanishing chemical potential $\mu=0$.
    \label{fig:saxis_mu0}
    }
\end{figure}

In the following, we will display only the on-axis value of the condensate $\sigma$ that gives a minimal value of the thermodynamic potential $\Phi_{\rm m.f.}$ for the given thermodynamic variables. In Fig.~\ref{fig:saxis_mu0}, we see how the on-axis value of the condensate $\sigma$ changes for zero chemical potential as we vary both the angular velocity and the temperature. It is clearly seen that as we increase rotation, both the condensate $\sigma$ and the critical temperature decrease. 

\begin{figure}
\centering 
\begin{tabular}{c}
    \includegraphics[width=0.95\linewidth]{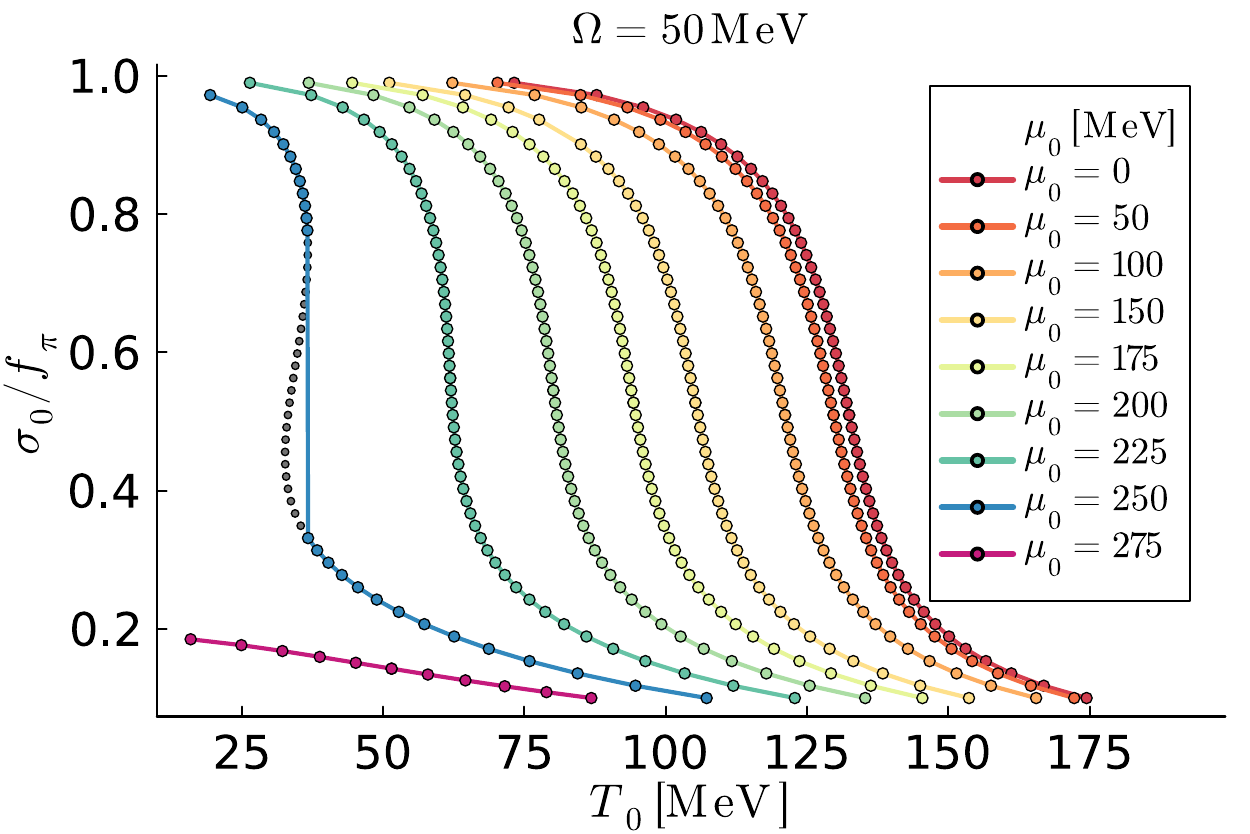} \\
    \includegraphics[width=0.95\linewidth]{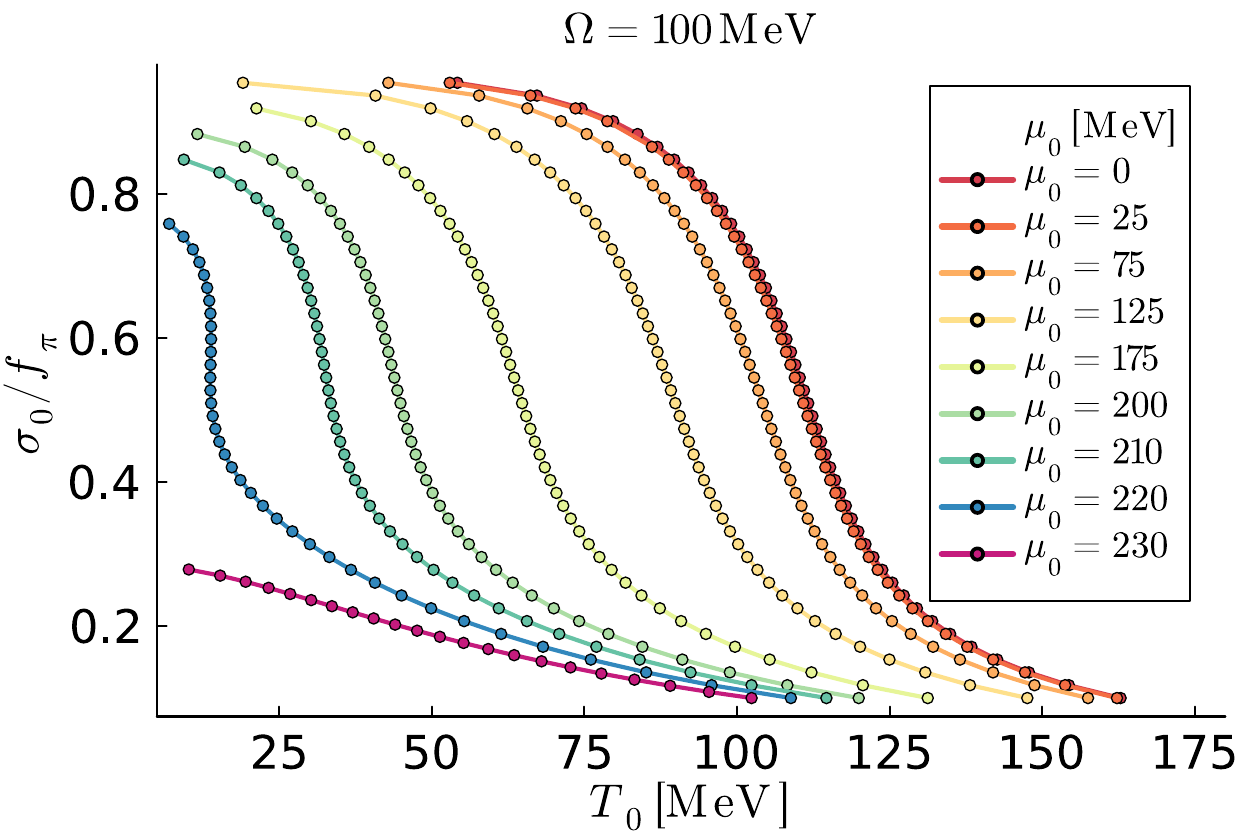}
\end{tabular}
    \caption{Fully-inhomogeneous solution: The same as in Fig.~\ref{fig:saxis_mu0}, but for different chemical potentials $\mu_0$ and two values of the angular velocity.}
    \label{fig:saxis_omegas}
\end{figure}

\begin{figure}
\centering 
\begin{tabular}{c}
    \includegraphics[width=0.95\linewidth]{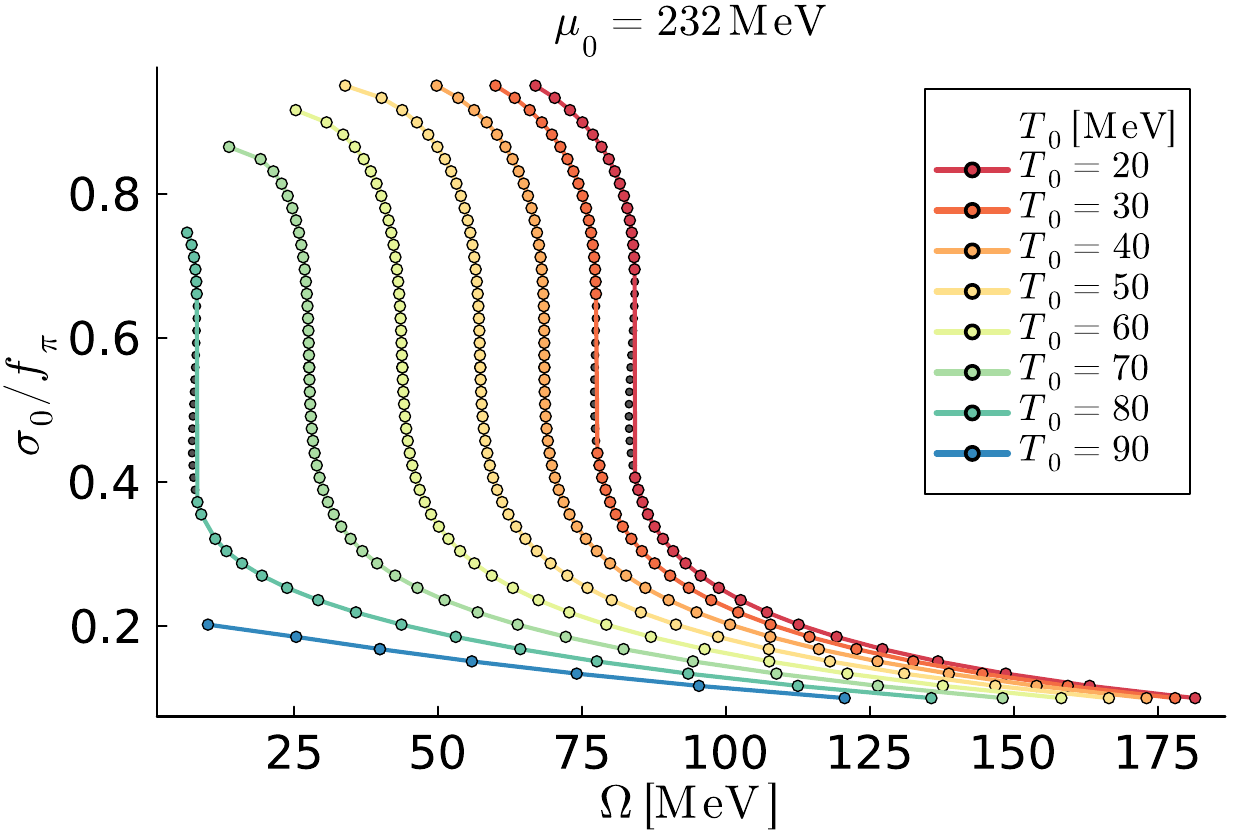} \\
    \includegraphics[width=0.95\linewidth]{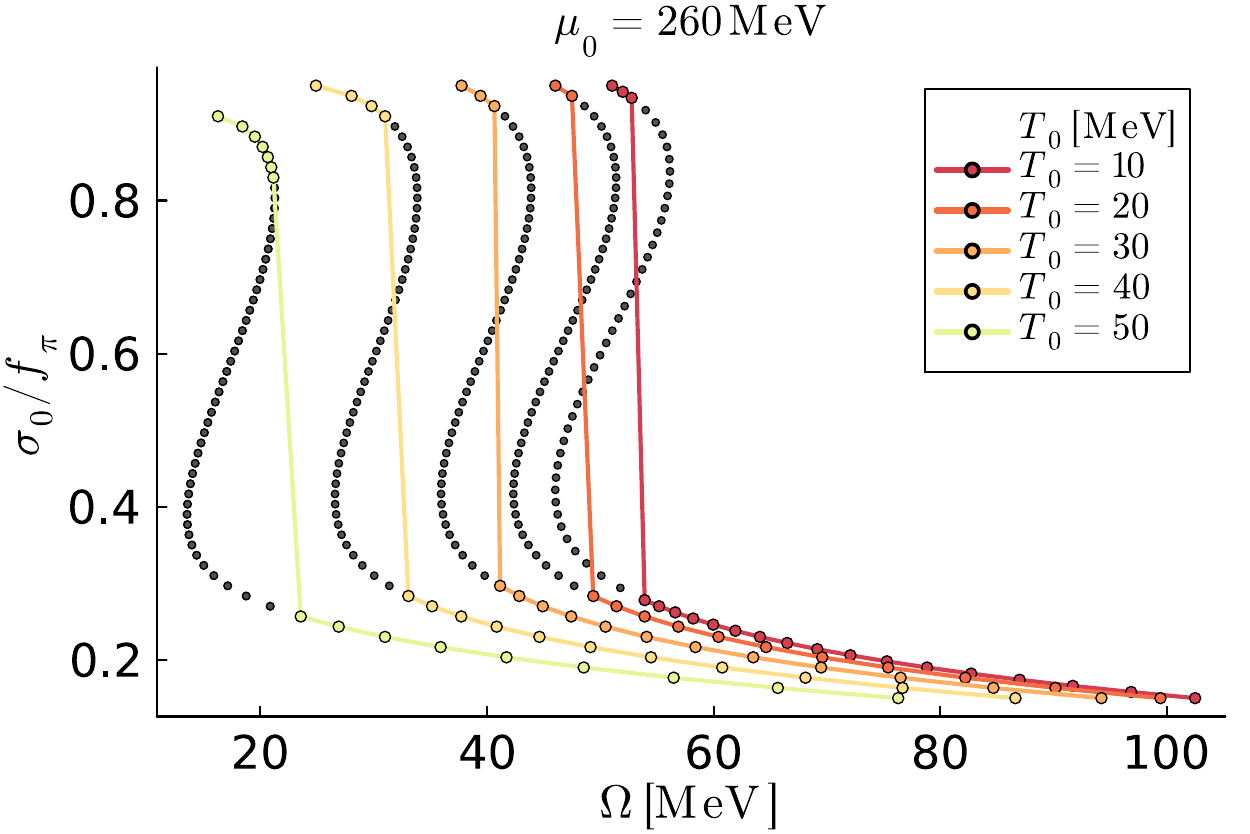}
\end{tabular}
    \caption{Full inhomogeneous solution: The same normalized on-axis value of the condensate $\sigma$, as in Fig.~\ref{fig:saxis_mu0}, but now shown as a function of the angular velocity $\Omega$ for different on-axis temperatures $T_0$ and two on-axis values of the chemical potential. The visible lines in panel (b) indicate a first-order phase transition.
    }\label{fig:saxis_omegasTs}
\end{figure}

In Fig.~\ref{fig:saxis_omegas}, we discuss the analogous problem for  fixed angular velocity while varying temperature and the chemical potential. Remarkably, we see that there are instances in which there are three different solutions of the differential equation for the same temperature, chemical potential, and angular velocity, which imply that a first-order phase transition takes place. Similar features appear in Fig.~\ref{fig:saxis_omegasTs}, where the value of the condensate $\sigma$ on the rotation axis is obtained as a function of the angular velocity $\Omega$. The results discussed above show that the system can undergo a first-order phase transition also in Model 3. This happens when the on-axis $\sigma$ condensate abruptly decreases, and the chiral symmetry is restored throughout the system as a whole.

\section{Phase diagram for Model 3}\label{sec:phasediagram}

We now proceed to discuss the phase diagram for the strongly inhomogeneous case considered in Sec.~\ref{sec:model3}. Recall that $\sigma(\rho)$ is monotonically decreasing and vanishes at the firewall. As a result, there are two possible phases: (1) an inhomogeneous phase where the system is in the chirally-broken phase close to the rotation axis while it is in the chirally-restored phase close to the light cylinder and (2) a global chirally-restored phase.

\begin{figure}
\centering 
\begin{tabular}{c}
    \includegraphics[width=0.99\linewidth]{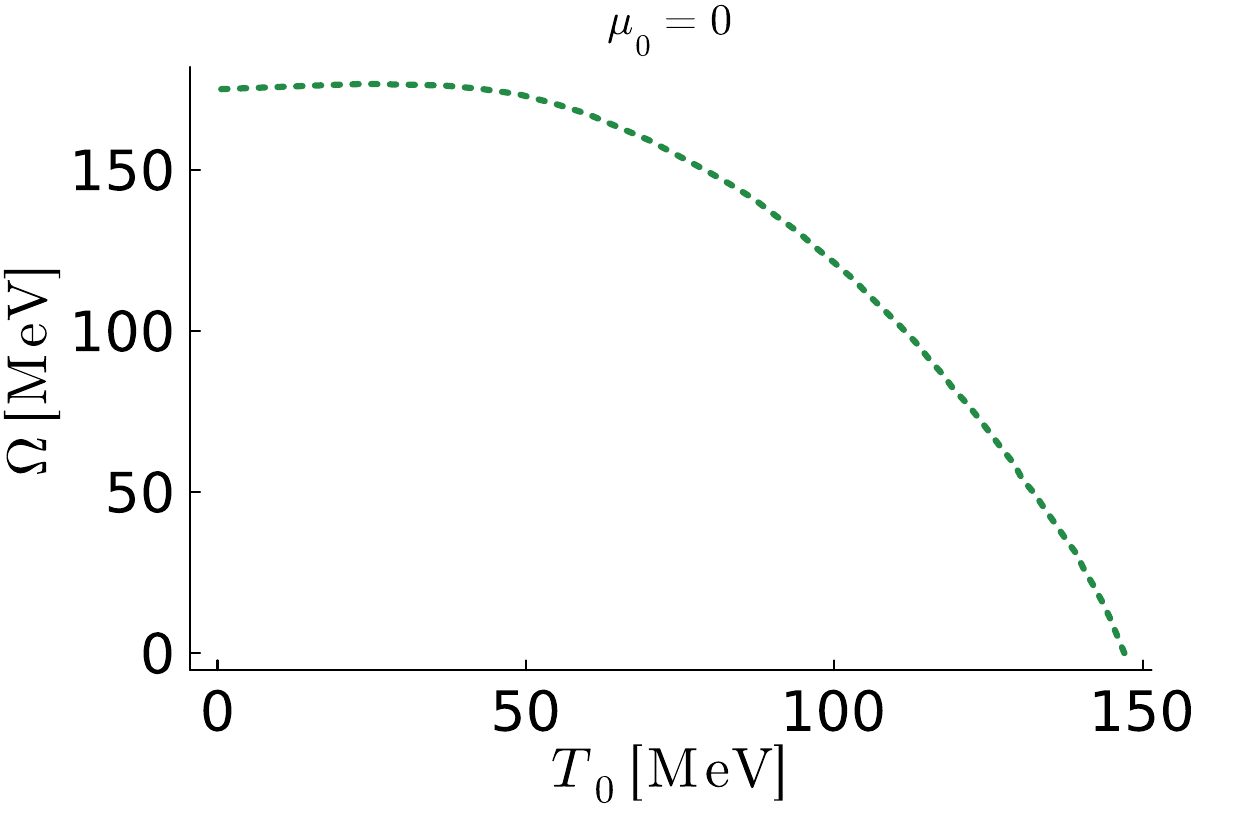} 
\end{tabular}
    \caption{The $\Omega-T_0$ phase diagram for the fully-inhomogeneous case at vanishing chemical potential $\mu_0=0$.
    \label{fig:OmTphase}
    }
\end{figure}

In order to understand the features of the phase diagram, it is useful to study separately the phase structure at vanishing chemical potential or at vanishing temperature.  We point out that the simultaneous limit $T \to 0$ and $\mu \to 0$ with finite angular velocity $\Omega$ is merely academic in this setup. In particular, the considerations that lead to the boundary condition that $\sigma$ vanishes at the light cylinder $\sigma(\rho \Omega=1)=0$ no longer apply. In Fig.~\ref{fig:OmTphase}, we show the phase diagram for $\mu=0$. The transition is everywhere crossover, with the inhomogeneous phase located in the inner region and the restored phase in the outer region. Notably, there exists a maximum value of angular velocity, $\Omega\simeq 175$ MeV, beyond which the system is always in the chirally-restored phase regardless of the values of temperature and chemical potential. 

\begin{figure}
\centering 
\begin{tabular}{c}
    \includegraphics[width=0.99\linewidth]{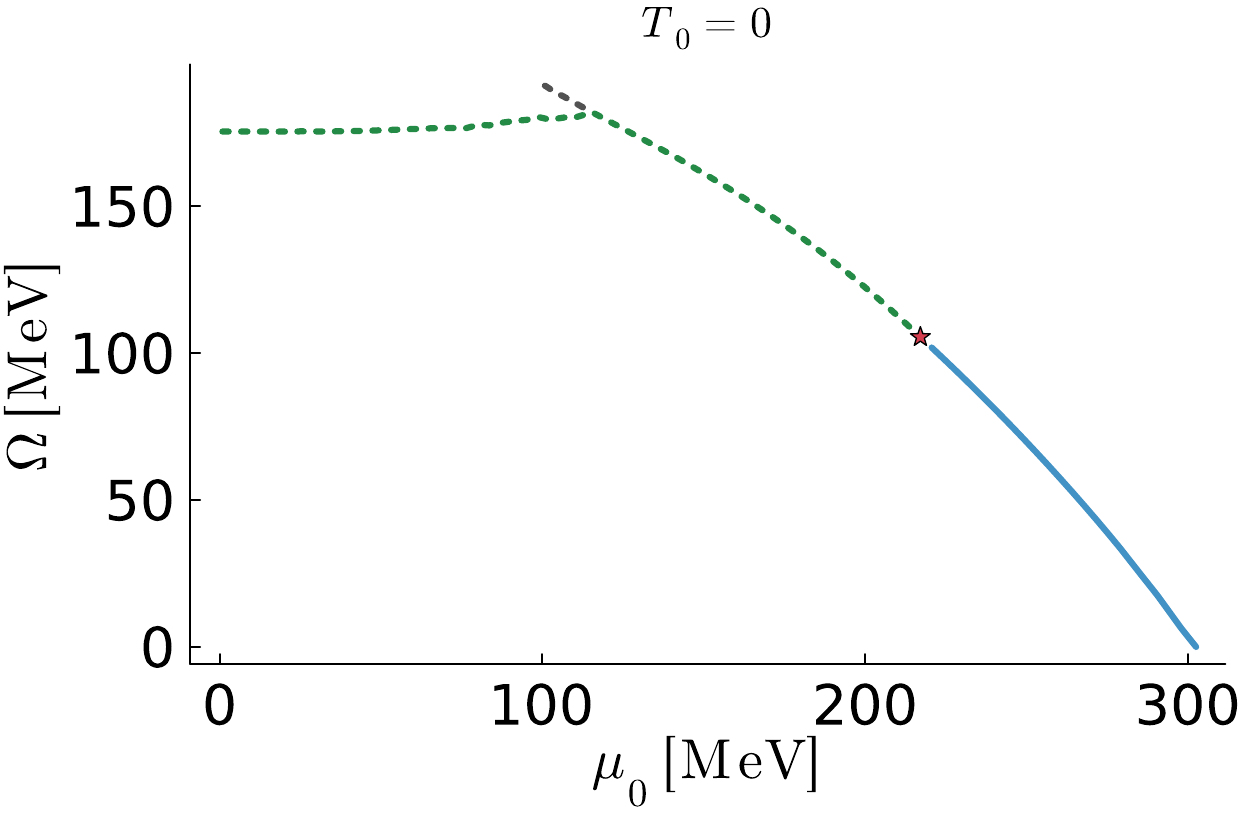} 
\end{tabular}
    \caption{The $\Omega$-$\mu_0$ phase diagram for the fully-inhomogeneous case at vanishing temperature $T_0=0$.
    }\label{fig:Ommuphase}
\end{figure}

\begin{figure}
\centering 
\begin{tabular}{c}
    \includegraphics[width=0.99\linewidth]{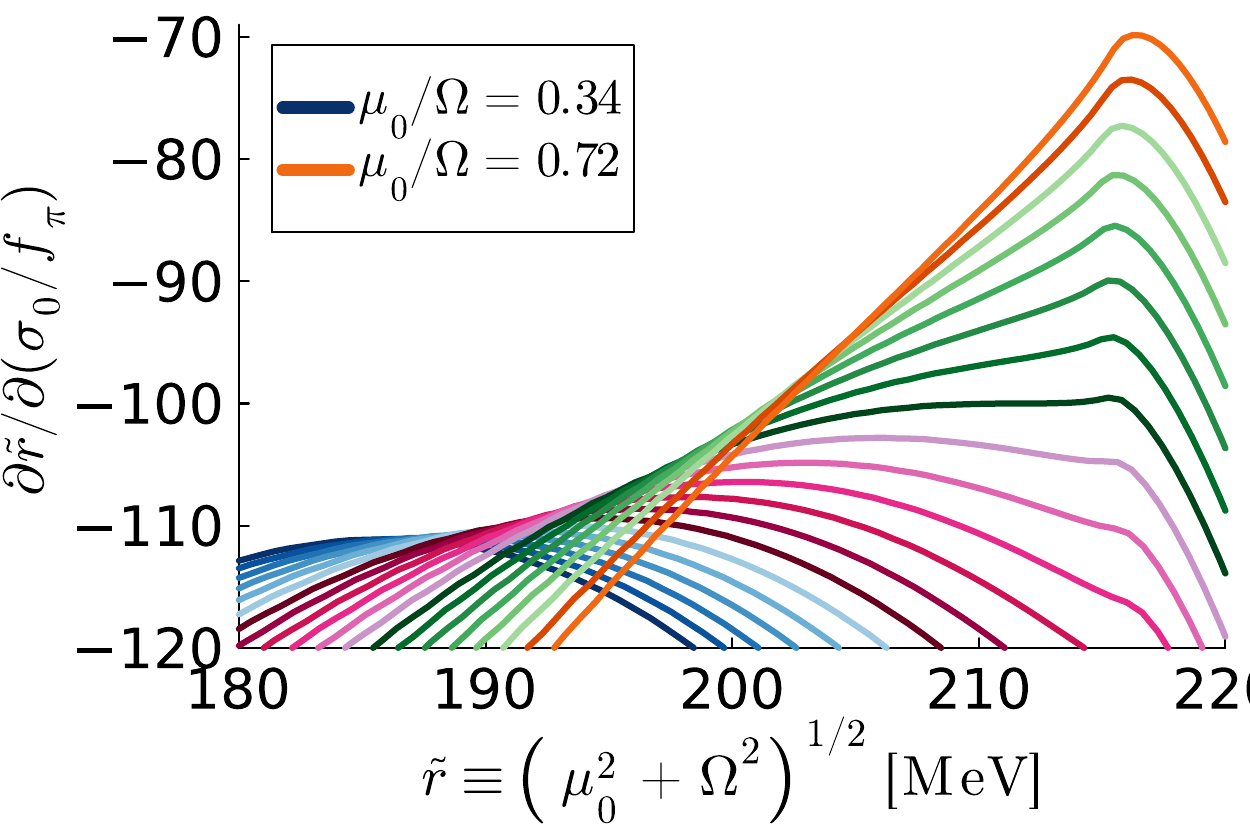} 
\end{tabular}
    \caption{Fully-inhomogeneous solution: (Inverse) derivative of the on-axis value of $\sigma$ with respect to the parameter $\tilde{r}=(\mu_0^2+\Omega^2)^{1/2}$ used to construct the phase diagram at zero temperature for a set of fixed ratios $\mu_0/\Omega$ that are evenly spaced between the minimal and maximal values of $\mu_0/\Omega$ (shown in the caption). The local maxima correspond to the points where the crossover phase transition takes place. There is a small range of the ratios $\mu_0/\Omega$ for which two extrema coexist, giving rise to the branching of the phase diagram of Fig. \ref{fig:Ommuphase}. }\label{fig:doublepeak}
\end{figure}

In Fig. \ref{fig:Ommuphase}, we display the phase diagram in the $\Omega$-$\mu$ plane at zero temperature.  Dotted lines are used to represent a crossover phase transition, while solid lines correspond to a first order phase transition. Similarly to the previous case, the outer region corresponds to the chirally-restored phase while the inner region corresponds to the inhomogeneous phase. Several comments are in order. Firstly, when rotation is small, we recover the known result of a first-order phase transition slightly above $\mu_0\sim 300$ MeV. As we increase the angular velocity $\Omega$, the first-order phase transition turns into a crossover phase transition, with the critical point located at $(\mu_c,\Omega_c)_{T=0} \simeq (216,107)$ MeV. Secondly, notice that there is a point where two branches appear for the crossover phase transition. This corresponds to a situation in which the condensate $\sigma$ features two (locally) steepest points, as shown in Fig. \ref{fig:doublepeak}, the (inverse) derivative of $\sigma_0$ can have two local extrema. Finally, we observe that there is a maximum value of angular velocity $\Omega\simeq 175$ MeV beyond which the system is in the chirally-restored phase regardless of the chemical potential. This value agrees with the one obtained in the $\mu=0$ case considered in Fig.~\ref{fig:OmTphase}. 

\begin{figure}
\centering 
\begin{tabular}{c}
    \includegraphics[width=0.99\linewidth]{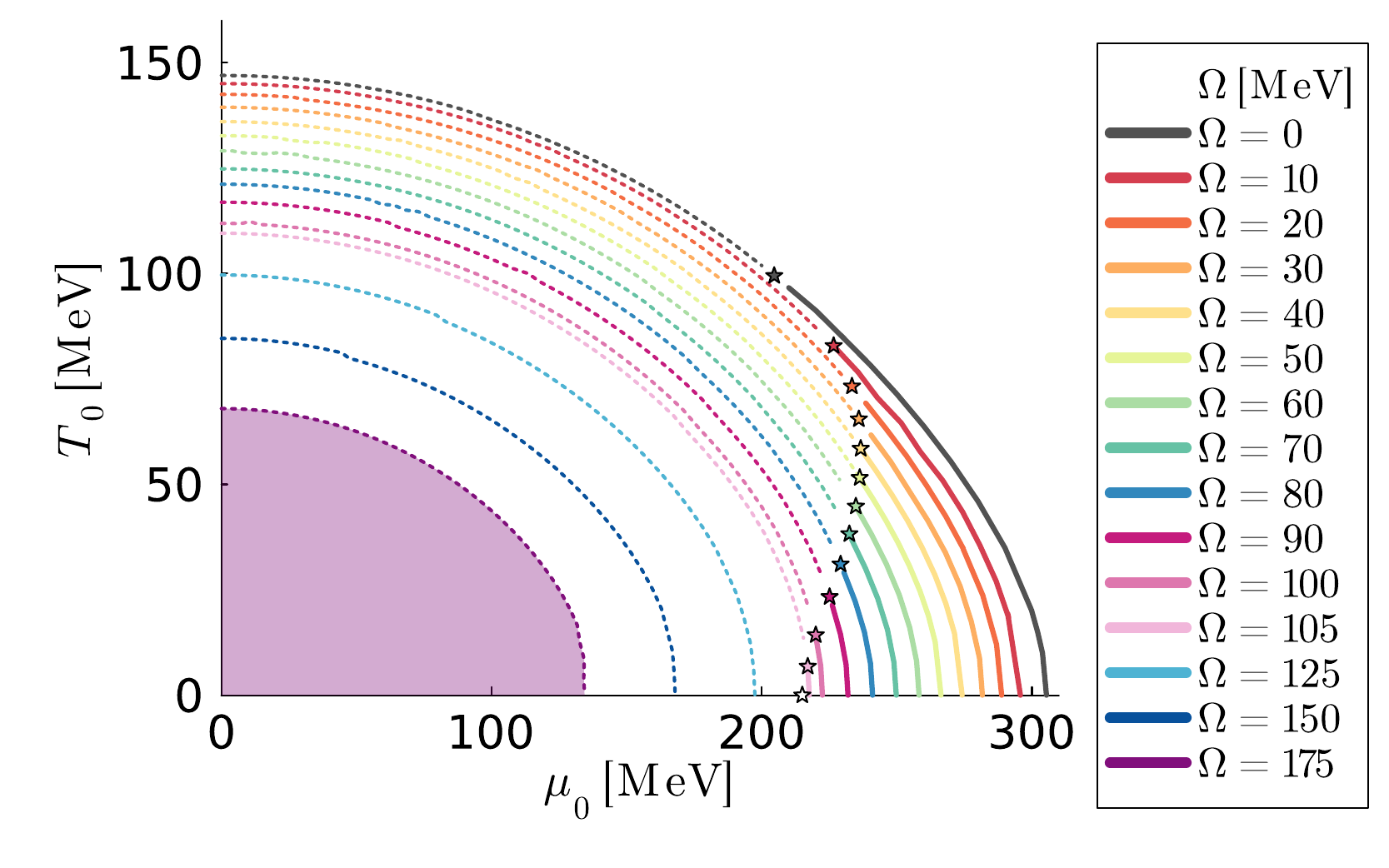} 
\end{tabular}
    \caption{Phase diagram of the fully-inhomogeneous system, shown in the $T$-$\mu$ plane at different cuts of fixed angular frequency $\Omega$. The system size extends to the light cylinder, i.e., $R = \Omega^{-1}$. 
    For $\Omega > \Omega_c = 175$ MeV, the system is in the chirally-restored phase, regardless of its temperature or chemical potential. The purple region marks the region where no phase transition can take place.
    }\label{fig:Tmuphase}
\end{figure}

In Fig. \ref{fig:Tmuphase} we show sections of the phase diagram in the $T$-$\mu$ plane at different constant angular velocities $\Omega$. The inner regions correspond to the inhomogeneous phase, while in the outer regions the system is in the chirally-restored phase. Interestingly, as we increase the angular velocity $\Omega$, the critical point separating the crossover and first-order phase transition, as a function of the chemical potential, follows a non-monotonic trajectory in the phase diagram. At $\Omega\simeq107$ MeV, the critical point approaches the zero-temperature axis, in agreement with the phase structure at vanishing temperature from Fig. \ref{fig:Ommuphase}. At higher angular velocities, there is no first-order phase transition. More specifically, in the range $107\ \text{MeV}\leq \Omega\leq 175 \ \text{MeV}$ the phase transition is a crossover. Above $\Omega\simeq 175$ MeV, the phase structure at vanishing temperature and at vanishing chemical potential shows that the whole system resides in the chirally-restored phase. In addition, the on-axis value of $\sigma$ is relatively small (see Figs. \ref{fig:saxis_mu0} and \ref{fig:saxis_omegasTs}) and continues to decrease as a function of the radial coordinate $\rho$. These two observations suggest the system should be in the (approximately) chirally-restored phase beyond $\Omega>175$ MeV. We included a shaded area in Fig.~\ref{fig:Tmuphase} to emphasize that the phase diagram becomes trivial above the quoted maximum angular velocity $\Omega\simeq175$ MeV.

We conclude this section discussing the analogous phase diagram for the approaches followed in Secs.~\ref{sec:model1} and~\ref{sec:model2}. Prior to the comparison of the results of different approaches, we emphasize that the phase diagram in Fig.~\ref{fig:Tmuphase} describes the medium of interacting quarks and mesons in the volume that extends all the way to the light-cylinder: $R=\Omega^{-1}$. In the approach of the uniform ground state discussed in Sec.~\ref{sec:model1}, a rotating medium that extends over the entire light cylinder, $R = \Omega^{-1}$, should possess a vanishing condensate everywhere $\overline{\sigma}=0$ [c.f. Eq. \eqref{eq_averaged-near-firewall}], and the system is trivially in the chirally-restored phase regardless of the values of temperature and chemical potential, as seen in Fig. \ref{fig:phase_diagrams_12} (top panel). Finally, the slowly varying approach of Sec.~\ref{sec:model2} is equivalent to the full-inhomogeneous case in the absence of rotation, $\Omega=0$, albeit subject to the boundary conditions \eqref{eq_conditions}, [c.f. Fig.~\ref{fig:Comparison23}] and the same holds true for the phase diagram, see Fig. \ref{fig:phase_diagrams_12} (bottom panel).

\section{Inhomogeneous rotation $\Omega(\rho)$} \label{sec:inh}

In the previous sections we considered the global equilibrium state in which the angular velocity $\Omega$ is constant. In the local thermal equilibrium approach of Sec.~\ref{sec:model:therm}, we saw that such a configuration induces an effective local temperature $T_{\rho}$, defined in Eq.~\eqref{eq_beta_rho}, together with an effective local chemical potential $\mu_\rho$, such that $\mu_\rho/T_\rho$ is constant. In this section, we consider configurations with inhomogeneous angular velocity, $\Omega \to \Omega(\rho)$. We investigate the properties of this out-of-equilibrium configuration in the local thermal equilibrium approximation, by which we assume that the system is stationary and dissipative processes are absent. We show that hydrodynamic consistency leads again to inhomogeneous local temperature $T_\rho$ and chemical potential $\mu_\rho$, however their ratio $\mu_\rho / T_\rho$ remains constant in thermal equilibrium \cite{Cercignani:2002}.

We start with a general discussion of local thermal equilibrium configurations with inhomogeneous angular velocity in Sec.~\ref{sec:inh:LTE}. We discuss the expected dissipative effects on the basis of the gradients of the angular velocity $\Omega(\rho)$ in Sec.~\ref{sec:inh:diss}. We consider the simplest non-trivial model, where $\Omega(\rho)$ is continuous while the vorticity $\omega$ vanishes outside a rotating core of radius $R_c$, known as the Rankine vortex, in Sec.~\ref{sec:inh:Rankine}.

\subsection{Local thermal equilibrium: kinematic properties} \label{sec:inh:LTE}

Consider the four-velocity describing inhomogeneous rotation $\Omega_\rho \equiv \Omega(\rho)$ in cylindrical coordinates:
\begin{equation}\label{eq:fourvel}
    u^\mu \partial_\mu = \Gamma_\rho \left(\partial_t +\Omega_\rho \partial_\varphi  \right)\,,
\end{equation}
where the Lorentz factor $\Gamma_\rho=1/\sqrt{1-(\rho\Omega_\rho)^2}$ ensures the normalisation $u_\mu u^\mu =1$. We begin this section with a discussion of the kinematic properties of this flow. For the explicit computations, we employ the cylindrical coordinate system with line element $ds^2 = dt^2 - d\rho^2 - \rho^2 d\varphi^2 -dz^2$ and connection coefficients
\begin{equation}
 \Gamma^\rho{}_{\varphi\varphi} = -\rho, \qquad 
 \Gamma^\varphi{}_{\varphi\rho} = \Gamma^\varphi{}_{\rho\varphi} = \frac{1}{\rho}.
\end{equation}
In the following, we denote covariant differentiation using $\nabla_\mu$. The notation $\nabla^\perp_\mu = \Delta^\alpha_\mu  \nabla_\alpha$ represents the gradient in the directions transverse to the four-velocity $u^\mu$, with $\Delta^\alpha_\mu = \delta^\alpha_\mu - u^\alpha u_\mu$ being the projector orthogonal to $u^\mu$, while $D = u^\mu \nabla_\mu$ denotes the comoving (covariant) derivative. We will work under the assumption of axial symmetry and homogeneity with respect to the $z$ direction. In this section, we will consider the state to be stationary. We will point out the deviations from stationarity on the basis of the non-conservation of the energy-momentum tensor due to dissipative effects, in Subsec.~\ref{sec:inh:diss}. 

Under the assumptions discussed above, the expansion scalar vanishes, $\theta = \nabla_\mu u^\mu = 0$. The vorticity tensor $\omega_{\mu\nu} = \frac{1}{2} (\nabla^\perp_\mu u_\nu - \nabla^\perp_\nu u_\mu)$ has the following non-vanishing components:
\begin{equation}
 \omega_{t\rho} = \Omega_\rho \omega_{\rho\varphi} = -\rho\Omega_\rho \Gamma_\rho^3(\Omega_\rho + \tfrac{1}{2} \rho \partial_\rho \Omega_\rho).
 \label{eq:inh_omega_tens}
\end{equation}
The corresponding vorticity vector, $\omega^\mu = \dfrac{1}{2}\varepsilon^{\mu\nu\alpha\lambda} u_\nu \omega_{\alpha\beta}$, evaluates to
\begin{equation}
    \omega^\mu\partial_\mu = \dfrac{\Gamma_\rho^2}{2\rho} \left(\frac{\partial(\rho^2 \Omega_\rho)}{\partial \rho}\right) \partial_z,
    \label{eq:inh_omega_vec}
\end{equation}
where we used $\varepsilon^{t\rho\varphi z} = 1/\rho$.
Note that there exists a non-trivial angular velocity $\Omega^{\rm v.f.}_\rho$ corresponding to a vorticity-free velocity profile,
\begin{equation}\label{eq:omnovort}
    \Omega^{\rm v.f.}_\rho = \dfrac{\mathcal{C}_\Omega
    }{\rho^2}, \quad \textrm{with} \  \mathcal{C}_\Omega \ \textrm{constant},
\end{equation}
which will become useful later.
Finally, the shear tensor $\sigma_{\mu\nu} = \Delta_{\mu\nu}^{\alpha\beta} \nabla_\alpha u_\beta$, with $\Delta_{\mu\nu}^{\alpha\beta} = \frac{1}{2}(\Delta_\mu^\alpha \Delta_\nu^\beta + \Delta_\mu^\beta \Delta_\nu^\alpha) - \frac{1}{3} \Delta_{\mu\nu} \Delta^{\alpha\beta}$, has the following non-vanishing components: 
\begin{equation}
 \sigma_{t\rho} = -\Omega_\rho \sigma_{\rho\varphi} = \frac{1}{2}\rho^2\Gamma_\rho^3 \Omega_\rho \partial_\rho \Omega_\rho\,.
 \label{eq:inh_sigma}
\end{equation}

In local thermal equilibrium, the charge current and energy-momentum tensor take the perfect fluid form,
\begin{gather}
    J^{\mu} = Q u^\mu, \quad 
    T^{\mu\nu} = \epsilon u^\mu u^\nu - P \Delta^{\mu\nu},
    \label{eq:inh_JT}
\end{gather}
where the charge density $Q$, energy density $\epsilon$ and pressure $P$ are assumed to depend on the transverse radial coordinate $\rho$. The conservation equations for charge, energy and momentum read:
\begin{subequations}\label{eq:inh_cons}
\begin{align}
 DQ + Q\theta &= 0, \label{eq:inh_consj} \\
 D\epsilon + (\epsilon + P) \theta &= 0, \label{eq:inh_conse} \\
 (\epsilon + P) Du^\mu - \nabla_\perp^\mu P &= 0.\label{eq:inh_consu}
\end{align}
\end{subequations}
The first two equations are automatically satisfied, since $\theta = 0$ and $DQ = D\epsilon = 0$, by virtue of the assumptions of stationarity and axial symmetry. Noting that $Du^\mu = -\delta^\mu_\rho \rho \Omega_\rho^2 \Gamma_\rho^2$ and $\nabla^\mu_\perp P = -\delta^\mu_\rho \partial_\rho P$, the momentum Eq.~\eqref{eq:inh_consu} reduces to
\begin{equation}
 \partial_\rho P - (\epsilon + P ) \rho \Omega_\rho^2 \Gamma_\rho^2 = 0.
 \label{eq:inh_dP_aux}
\end{equation}
Taking into account that the pressure is a thermodynamic function of the temperature $T$ and chemical potential $\mu$, the radial derivative of the pressure can be expressed as 
\begin{equation}
    \frac{\partial P}{\partial \rho} = \dfrac{\partial P}{\partial T} \frac{\partial T}{\partial \rho} + \dfrac{\partial P}{\partial \mu} \frac{\partial \mu}{\partial \rho}.
\end{equation}
The derivatives of the pressure appearing above are standard:
\begin{equation}
 \dfrac{\partial P}{\partial T} = s, \quad 
 \dfrac{\partial P}{\partial \mu} = Q,
\end{equation}
where $Q$ is the charge density and $s = (\epsilon + P - \mu Q) / T$ is the entropy density. Taking into account that the ratio $\mu / T$ is constant in thermodynamic equilibrium \cite{Cercignani:2002},
\begin{equation}\label{eq:alpha}
 \frac{\mu}{T} = {\rm const},
\end{equation}
we arrive at
\begin{equation}
     \frac{\partial P}{\partial \rho} = \frac{\epsilon + P}{T} \frac{\partial T}{\partial \rho}.
\end{equation}
Substituting the above result into Eq.~\eqref{eq:inh_dP_aux} leads to
\begin{equation}\label{eq:inh_difT}
 \dfrac{\partial_\rho T}{T} -\rho \Omega_\rho^2 \Gamma_\rho^2 = 0\,,
\end{equation}
whose solution gives the effective temperature as a function of the radial distance $T(\rho)$. In the case of rigid rotation, $\Omega_\rho = \Omega_0$, we recover the Tolman-Ehrenfest dependence given in Eq.~\eqref{eq_beta_rho}.

\subsection{Dissipative effects}\label{sec:inh:diss}

Let us assume for a moment that the system is not in local thermal equilibrium. The charge current $J^\mu$ and energy-momentum tensor $T^{\mu\nu}$ receive viscous contributions. In the Landau frame, Eqs.~\eqref{eq:inh_JT} are extended to
\begin{gather}
    J^{\mu} = Q u^\mu + V^\mu,\nonumber\\
    T^{\mu\nu} = \epsilon u^\mu u^\nu - (P + \Pi) \Delta^{\mu\nu} + \pi^{\mu\nu},
    \label{eq:inh_diss_JT}
\end{gather}
where $\Pi$ is the viscous (dynamic) pressure,  $V^\mu$ is the diffusion current and $\pi^{\mu\nu}$ is the shear-stress tensor. These quantities encode the dissipative degrees of freedom. The conservation Eqs.~\eqref{eq:inh_cons} are modified to \cite{Denicol2022:Springer}
\begin{subequations}\label{eq:inh_diss_cons}
\begin{align}
 DQ+Q\theta + \nabla^\perp_\mu V^\mu &= 0, \label{eq:inh_diss_consj} \\
 D\epsilon + (\epsilon + P + \Pi) \theta - \pi^{\mu\nu} \sigma_{\mu\nu} &= 0, 
 \label{eq:inh_diss_conse} \\
 (\epsilon + P + \Pi) Du^\mu - \nabla_\perp^\mu(P + \Pi) + \Delta^\mu_\lambda \nabla_\nu \pi^{\lambda\nu} &= 0.\label{eq:inh_diss_consu}
\end{align}
\end{subequations}

To address Eqs.~\eqref{eq:inh_diss_cons}, the dissipative fluxes $\Pi$, $V^\mu$ and $\pi^{\mu\nu}$ must be specified via constitutive equations. In relativistic hydrodynamics, these fluxes must be promoted to dynamical variables obeying relaxation equations, to avoid faster-than-light signal propagation \cite{Israel:1979wp,Hiscock:1983zz,Hiscock:1985zz} (see Ref.~\cite{Bemfica:2017wps,Kovtun:2019hdm} for a notable alternative). 
A first approximation can be obtained by neglecting the relaxation process and instead employing the first-order Navier-Stokes constitutive relations \cite{Denicol2022:Springer},
\begin{equation}
 \Pi = -\zeta \theta, \quad    
 V^\mu = \kappa \nabla_\perp^\mu\left(\frac{\mu}{T}\right), \quad 
 \pi^{\mu\nu} = 2\eta \sigma^{\mu\nu},
\end{equation}
with $\zeta$, $\kappa$ and $\eta$ being the bulk viscosity, diffusion coefficient and shear viscosity. Under the assumption of stationarity, $\theta = \partial_t \Gamma_\rho \simeq 0$ implies $\Pi \simeq 0$. Moreover, when $\mu / T = {\rm const}$, $V^\mu$ also vanishes and the charge conservation Eq.~\eqref{eq:inh_diss_consj} is again automatically satisfied. 

The stationarity assumption is invalidated by the energy Eq.~\eqref{eq:inh_diss_conse}, since 
\begin{equation}
 \pi^{\mu\nu} \sigma_{\mu\nu} 
 \simeq -\pi^{\rho\varphi} \Gamma_\rho \rho^2\partial_\rho \Omega
 \simeq \eta (\Gamma_\rho^2 \rho \partial_\rho \Omega_\rho)^2,
\end{equation}
where the first approximation sign indicates that we ignored the time derivative of $\Omega_\rho$, while the second approximation sign reminds us that $\pi^{\mu\nu}$ is replaced by its Navier-Stokes value, $2 \eta \sigma^{\mu\nu}$. 
The relation $\Delta^\mu_\lambda \nabla_\nu \pi^{\lambda\nu} = \nabla_\nu \pi^{\mu\nu} + u^\mu \pi^{\lambda\nu} \sigma_{\nu\lambda}$ highlights the role of this same dissipative term at the level of the momentum Eq.~\eqref{eq:inh_diss_consu}. Since $u^\rho = \pi^{\rho\rho} = \pi^{\varphi\varphi} = 0$, we have $\Delta^\rho_\lambda \nabla_\nu \pi^{\lambda\nu} = 0$ and the radial Eq.~\eqref{eq:inh_dP_aux} remains unaffected. In the case of the azimuthal component, we obtain
\begin{multline}
 (\epsilon + P + \Pi) D u^\varphi + u^\varphi D(P + \Pi) \\
 + u^\varphi \pi^{\lambda\nu} \sigma_{\lambda\nu} + \nabla_\nu \pi^{\varphi \nu} = 0.
 \label{eq:inh_diss_consuphi}
\end{multline}
The above relation can be simplified after the addition of Eq.~\eqref{eq:inh_diss_conse} multiplied by $u^\varphi$:
\begin{equation}
 \partial_t[(\epsilon + P + \Pi) \Gamma_\rho^2 \Omega_\rho] + \nabla_\nu \pi^{\varphi\nu} = 0,
\end{equation}
where we took into account that $\theta = \partial_t \Gamma_\rho$. Noting that $\pi^{\lambda\nu}\sigma_{\lambda\nu} = -u_\lambda \nabla_\nu \pi^{\lambda\nu}$, it is clear that the dissipative terms in Eqs.~\eqref{eq:inh_diss_conse}--\eqref{eq:inh_diss_consu} vanish only if $\nabla_\nu \pi^{t\nu}$ and $\nabla_\nu \pi^{\varphi \nu}$ vanish simultaneously. Assuming that $\pi^{\mu\nu}$ follows the structure of the shear tensor $\sigma_{\mu\nu}$ shown in Eq.~\eqref{eq:inh_sigma}, we find
\begin{equation}
 \nabla_\nu \pi^{\varphi \nu} = \frac{1}{\rho^3} \frac{\partial (\rho^3 \pi^{\varphi\rho})}{\partial \rho}, \quad 
 \nabla_\nu \pi^{t \nu} = \frac{1}{\rho} \frac{\partial (\rho \pi^{t\rho})}{\partial \rho}.
\end{equation}
Since $u_\mu \pi^{\mu\nu} = 0$, it can be seen that $\pi^{t\nu} = \rho^2 \Omega_\rho \pi^{\varphi\nu}$. Thus, the above relations vanish simultaneously only in the case when $\Omega_\rho = {\rm const}$, i.e., in the case of rigid-body rotation. In all other cases, dissipation will lead to the damping of the velocity profile. It is easily seen that the timescale for this damping process is related to the magnitude of the shear viscosity, $\eta$, and to the gradients $\partial_\rho \Omega_\rho$ of the angular velocity.

In order to provide simple explicit examples, we consider now a slowly rotating system, $\rho \Omega_\rho\ll1$, neglecting quadratic and higher order terms. As a first example, take $\Omega_\rho(t) = C \rho^{-\nu}$ outside of a rotating core of radius $R_c$, i.e. $\rho>R_c$, with $C$ and $\nu\geq0$ being $\rho$-independent functions of time. Then,
Eq.~\eqref{eq:inh_diss_consuphi} leads to
\begin{equation}
 \partial_t \Omega_\rho(t) \simeq -\frac{\nu(2 - \nu)}{\rho^2} \frac{\eta \Omega_\rho(t)}{\epsilon + P}.
 \label{eq:inh_diss_dOmega_poly}
\end{equation}
The case $\nu = 0$ corresponds to rigid rotation, while $\nu = 2$ corresponds to the vorticity-free profile discussed in Eq.~\eqref{eq:omnovort}. If $0 < \nu < 2$, shearing leads to a local damping of the flow, while for $\nu > 2$, the flow is enhanced. 
We can estimate the lifetime $\tau$ of such a configuration assuming that $\Omega_\rho(t) \propto e^{-t / \tau}$. In the case when $\eta / s = {\rm const}$, and considering a fluid close to neutrality, we can replace $\eta / (\epsilon + P) \simeq T^{-1} (\eta / s)$. Altogether, from Eq. \eqref{eq:inh_diss_dOmega_poly} we find 
\begin{align}
 \tau &\simeq \left(\frac{\rho}{5\ {\rm fm}}\right)^2 \left(\frac{T}{300\ {\rm MeV}}\right) \left(\frac{1/4\pi}{\eta / s}\right) \times 
 \frac{3\ {\rm fm}}{\nu(2-\nu)},
\end{align}
which is bounded below for $\rho=R_c$. As a second example, we take an
 exponentially-suppressed angular velocity, $\Omega_\rho(t) = C e^{-\alpha \rho}$, with $C$ and $\alpha$ being $\rho$-independent functions of time, we find 
\begin{equation}
 \partial_t \Omega_\rho(t) = \alpha\left(\alpha - \frac{3}{\rho}\right) \frac{\eta \Omega_\rho(t)}{\epsilon + P}.
 \label{eq:inh_diss_dOmega_exp}
\end{equation}
which is damped as
\begin{equation}
 \tau \simeq \left(\frac{T}{300\ {\rm MeV}}\right) \left(\frac{1/4\pi}{\eta / s}\right) \times 
 \frac{1}{\alpha^2} \left(1 - \frac{3}{\alpha\rho}\right)^{-1}.
\end{equation}
The exponential profile
is in general enhanced by shearing at sufficiently-large values of $\rho$. Hence, we are led to conclude that any inhomogeneous vortical configuration centered on the rotation axis decays at large distances as $\Omega_\rho \propto \rho^{-2}$. 

\subsection{Relativistic Rankine vortex}\label{sec:inh:Rankine}

To illustrate further our approach in the case of inhomogeneous backgrounds, in this section we consider a relativistic generalization of the Rankine vortex, which represents an idealized model of a spatially-compact rotating fluid flow that combines a solid-body rotation of angular velocity $\Omega_0$, within a core of radius $R_c$, with an irrotational (potential) flow outside this core.

There are two types of cylindrically-symmetric flow profiles with vanishing vorticity: the trivial profile, with a vanishing angular velocity $\Omega=0$, and the profile given by Eq.~\eqref{eq:omnovort}. In the first case, we would have 
\begin{equation}\label{eq:omega_discont}
    \Omega_\rho  = \Omega_0 \Theta(R_c-\rho)\,,
\end{equation}
which is a discontinuous function of the radial coordinate~$\rho$. As a consequence, the dissipative contributions $\nabla_\mu \pi^{\mu \nu}$ diverge at $R_c$, and dissipation will quickly erase such a configuration. Even if one regularizes the discontinuous step function in the angular velocity~\eqref{eq:omega_discont}, the large gradients of $\Omega(\rho)$ will favor the dissipative damping of such a profile. As shown in Eqs.~\eqref{eq:inh_diss_dOmega_poly}--\eqref{eq:inh_diss_dOmega_exp}, the polynomial $\Omega_\rho = C \rho^{-\nu}$ or exponential $\Omega_\rho = C e^{-\alpha \rho}$ ans\"atze would lead to fast dissipation, as the parameters $\nu, \alpha \to \infty$ to ensure the rapid decrease $\Omega \to 0$ for $\rho > R_c$ implied by Eq.~\eqref{eq:omega_discont}, thus making the example of Eq.~\eqref{eq:omega_discont} a short-live transient.

\begin{figure}
    \centering
    \begin{tabular}{c}
    \includegraphics[width=0.95\linewidth]{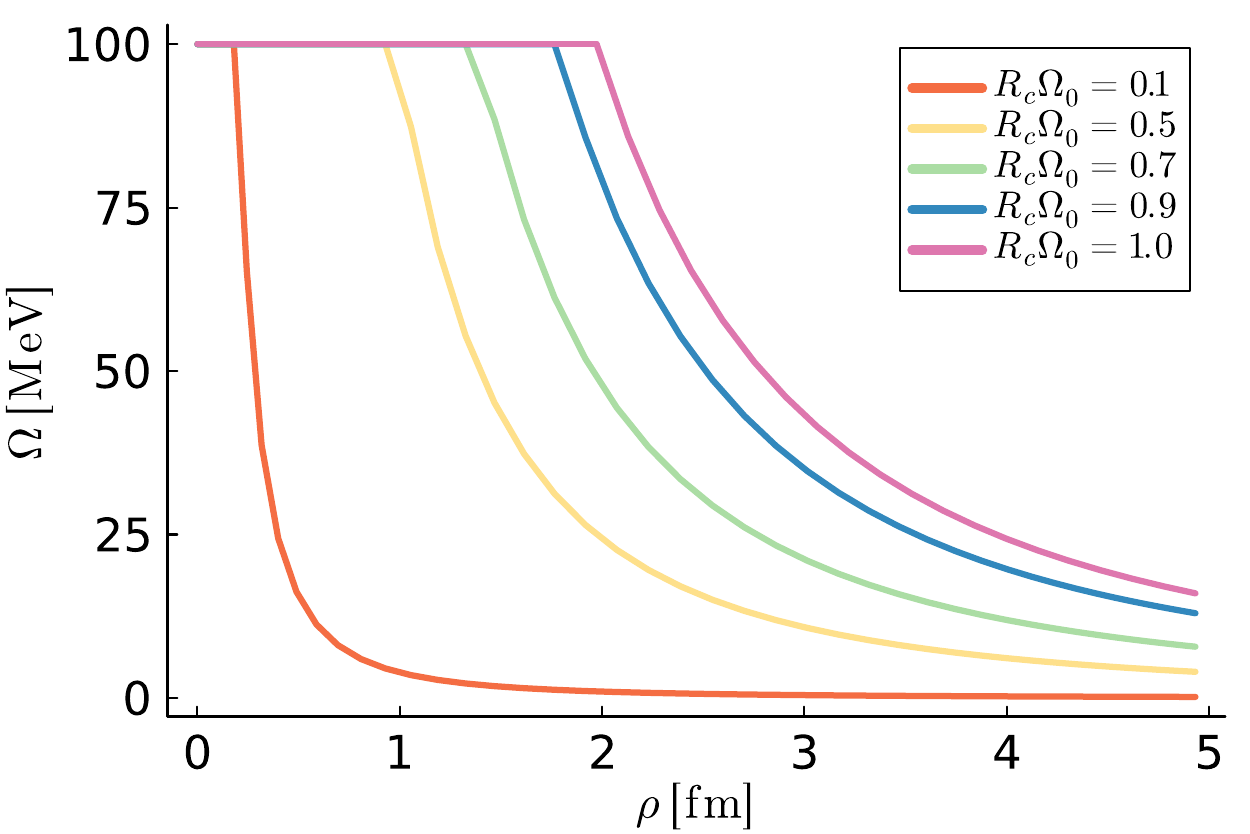} \\
    \includegraphics[width=0.95\linewidth]{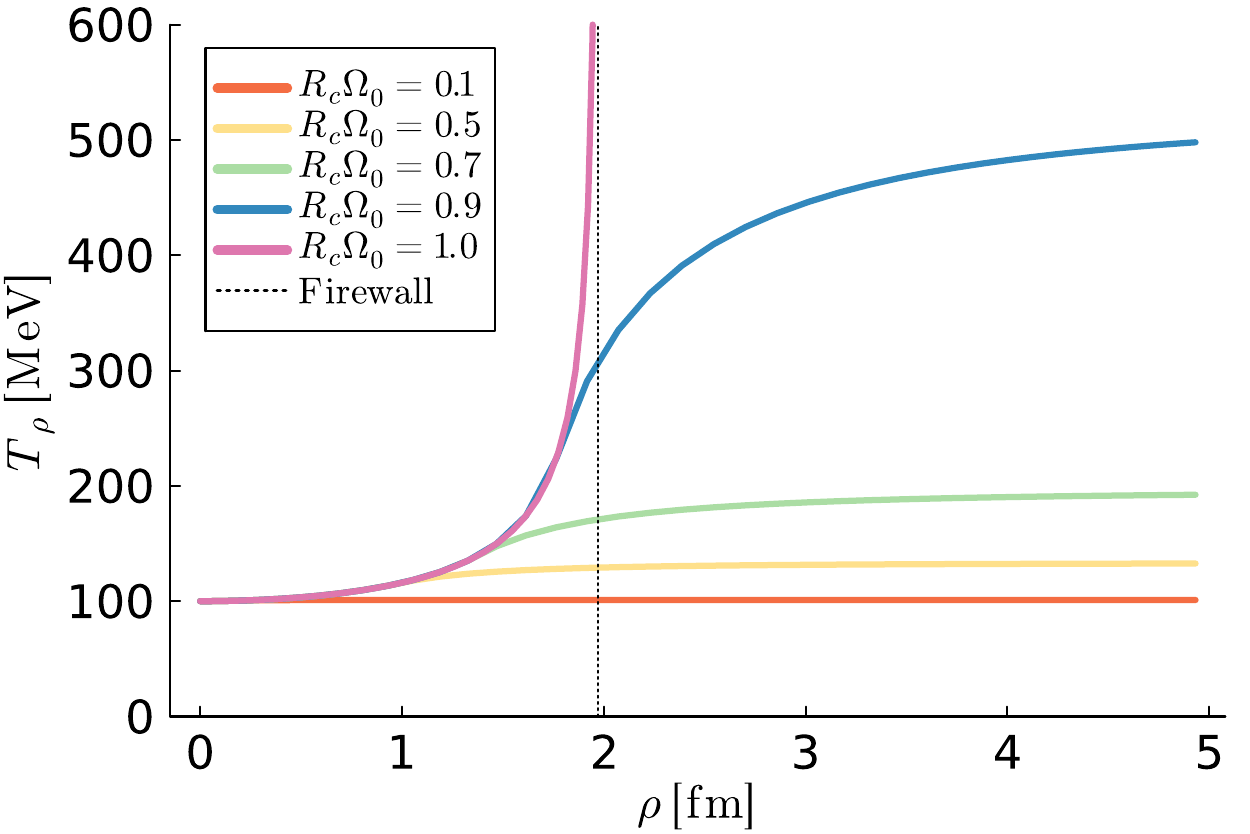}
    \end{tabular}
    \caption{Angular velocity (upper panel) and local temperature $T_\rho$ (lower panel) for the Rankine-like vortices described in Eqs. \eqref{eq:omrankin} and \eqref{eq:trankin}. The limit where $R_c\to 1/\Omega$ corresponds to the firewall scenario.}
    \label{fig:Rankine-OmT}
\end{figure}
The second possibility involves continuously matching the rigidly rotating core to the non-vortical rotational profile described by Eq.~\eqref{eq:omnovort}:\footnote{Note that the derivative of the resulting profile~\eqref{eq:omrankin} is discontinuous at the vortex core radius $R_c$. One can obtain a more physical profile by smoothing the $R_c$ junction. On physical grounds, it is expected that both approaches, differentiable and non-differentiable, lead to very similar conclusions.}
\begin{equation}\label{eq:omrankin}
    \Omega(\rho) = \Omega_0 \Theta(R_c - \rho) + \Omega_0 \dfrac{R_c^2}{\rho^2}\Theta(\rho - R_c)\,.
\end{equation}
This type of configuration is known as the Rankine vortex. Dissipation will eventually smoothen the non-differentiable cusp in this profile, but on a timescale much larger than for the discontinuous alternative of Eq. \eqref{eq:omega_discont}. For this reason, we restrict our analysis to the Rankine vortex configuration~\eqref{eq:omrankin}. 

Using Eq.~\eqref{eq:inh_difT}, it follows that the local temperature is given by
\begin{align}\label{eq:trankin}
    T(\rho) &= \dfrac{T_0}{\sqrt{1-\rho^2 \Omega_0^2}}\Theta(R_c - \rho)\nonumber\\& + \dfrac{T_0}{1-R_c^2\Omega_0^2}\sqrt{1-(R_c\Omega_0)^2\dfrac{R_c^2}{\rho^2}}\Theta(\rho  - R_c)\,,
\end{align}
which is continuous and once-differentiable at the junction point $R_c$. The temperature on the rotation axis is again denoted as $T_0$. It follows from Eq.~\eqref{eq:alpha} that the local chemical potential has a similar functional dependence. With this construction, the firewall scenario is realized as the limit where $R_c\to 1/\Omega$. The system extends to $\rho\to\infty$ provided that $R_c<1/\Omega$. In Fig.~\ref{fig:Rankine-OmT}, we show illustrative examples of the Rankine vortex angular velocity (upper panel) together with the local (effective) temperature (lower panel).

\begin{figure}
    \centering
    \begin{tabular}{c}
        \includegraphics[width=0.95\linewidth]{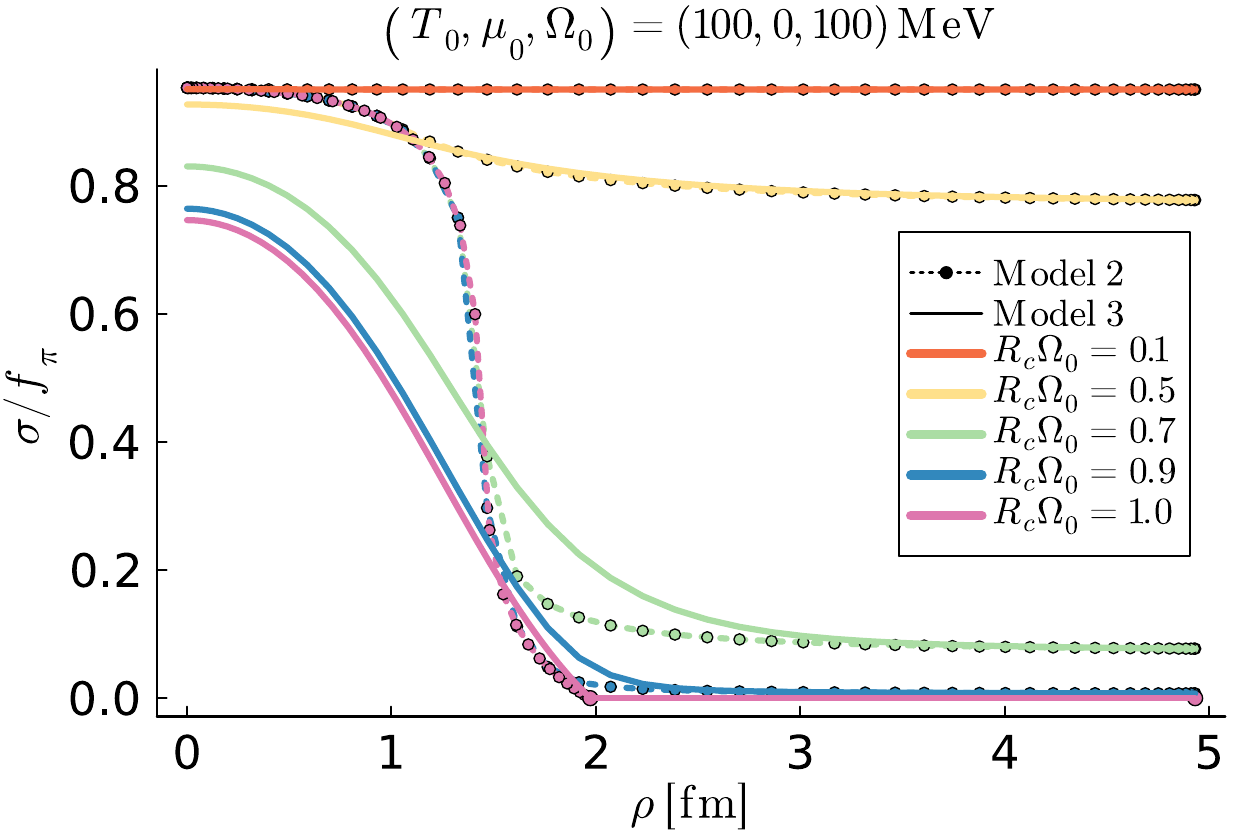} \\
        \includegraphics[width=0.95\linewidth]{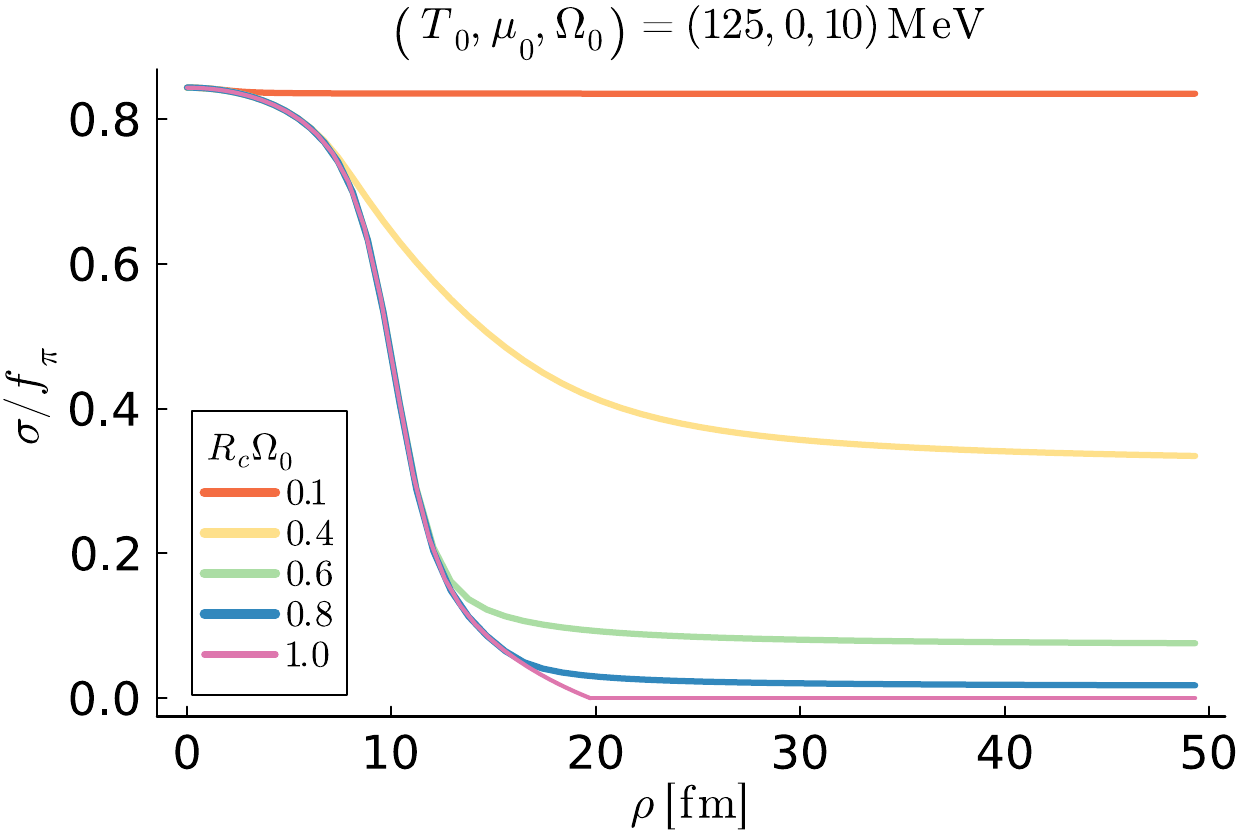}
    \end{tabular}
    \caption{Inhomogeneous condensate $\sigma(\rho)$ in a Rankine-like vortex configuration for different values of the vortex core radius $R_c$, obtained as a solution of the field equation~\eqref{eq:model_diff} with the boundary conditions given in Eq.~\eqref{eq:bdyrankin}.}
    \label{fig:Rankine-sigma}
\end{figure}

The derivation of the differential gap equation \eqref{eq:model_diff} is unaffected by the inhomogeneous rotation, and we can obtain the condensate $\sigma$ by providing appropriate boundary conditions. Close to the rotation axis, the angular velocity is constant, and the asymptotic solution \eqref{eq_sigma_diff_axis_TE} remains unchanged, where again the regularity of the solution requires that the integration constant $\mathcal{C}_0 = 0$. As opposed to the rigidly-rotating case, now the system has infinite extent and the temperature reaches a plateau as $R_c < \rho \to \infty$ (see Fig. \ref{fig:Rankine-OmT}). The value of $\sigma$ at $\rho\to\infty$ is expected to be independent of its behavior at the center $\rho = 0$, which motivates to fix $\sigma(\rho\to\infty)\equiv\sigma_\infty$ to the value that it would take in the absence of rotation for the temperatures and chemical potentials at infinity $T_\infty = T_0/(1-R_c^2\Omega_0^2)$ and  $\mu_\infty = \mu_0/(1-R_c^2\Omega_0^2)$. Therefore, the boundary value problem of model 3 consists of solving the differential equation \eqref{eq:model_diff} supplemented with the boundary conditions 
\begin{equation}\label{eq:bdyrankin}
    \left. \dfrac{\partial\sigma}{\partial\rho}\right|_{\rho= 0 } \hspace{-5pt} = 0\,,\quad \sigma_\infty = \sigma(T=T_\infty,\mu=\mu_\infty,\Omega=0)\,.
\end{equation}
In the limit where $R_c=1/\Omega_0$, the firewall boundary conditions \eqref{eq:variation_bdry} are naturally recovered, since in that case $\sigma_\infty = 0=\sigma(\rho=1/\Omega_0)$.

We now solve the gap equation for models 2 (Sec.~\ref{sec:model2}) and 3 (Sec.~\ref{sec:model3}) for the Rankine vortex configuration. For simplicity in the discussion, we focus on the solutions at vanishing chemical potential $\mu_0=0$. In Fig. \ref{fig:Rankine-sigma} (top panel), we show the condensate $\sigma$ as a function of the radial distance for an illustrative example with $T_0 = \Omega_0 = 100$ MeV. Since $T_0 < T_c \simeq 145$ MeV, the system is in the chirally-broken phase close to the rotation axis. The solid lines correspond to the solutions of the differential equation (model 3), while the dotted lines with circles represent the solutions in the weakly-inhomogeneous approach (model 2). Firstly, we observe that models 2 and 3 always agree sufficiently far from the rotation axis, where the solution becomes homogeneous and the gradients become negligible. Close to the rotation axis, the agreement depends on both the angular velocity $\Omega_0$ (as discussed in Sec. \ref{sec:model3}) and on the vortex core radius $R_c$: the two models are in reasonable agreement if the vortex extends to $R_c\Omega_0<0.5\,$, deviating from each other when $R_c \Omega_0$ is further increased. It is worth noting that the Rankine vortex solution asymptotes to the firewall solution (shown by the $R_c \Omega_0 = 1$ curve) as the vortex radius approaches the light cylinder, $R_c\to 1/\Omega_0$. In the bottom panel of Fig. \ref{fig:Rankine-sigma}, we show again $\sigma(\rho)$ for the Rankine vortex, for phenomenologically relevant values of the temperature, $T_0=125$ MeV, and the angular velocity $\Omega_0 = 10$ MeV. Note that, even for this small value of angular rotation, the system can be considered to be in an inhomogeneous phase, as it happens for $R_c\Omega>0.4$. This is a consequence of the increasing local temperature \eqref{eq:trankin} combined with the boundary condition \eqref{eq:bdyrankin}: if $T_0<T_c$ and $T_\infty>T_c$ the system will be in an inhomogeneous phase, chirally broken close to the rotation axis and chirally restored for $\rho\gg R_c$. 

\begin{figure}
    \centering
    \begin{tabular}{c}
        \includegraphics[width=0.95\linewidth]{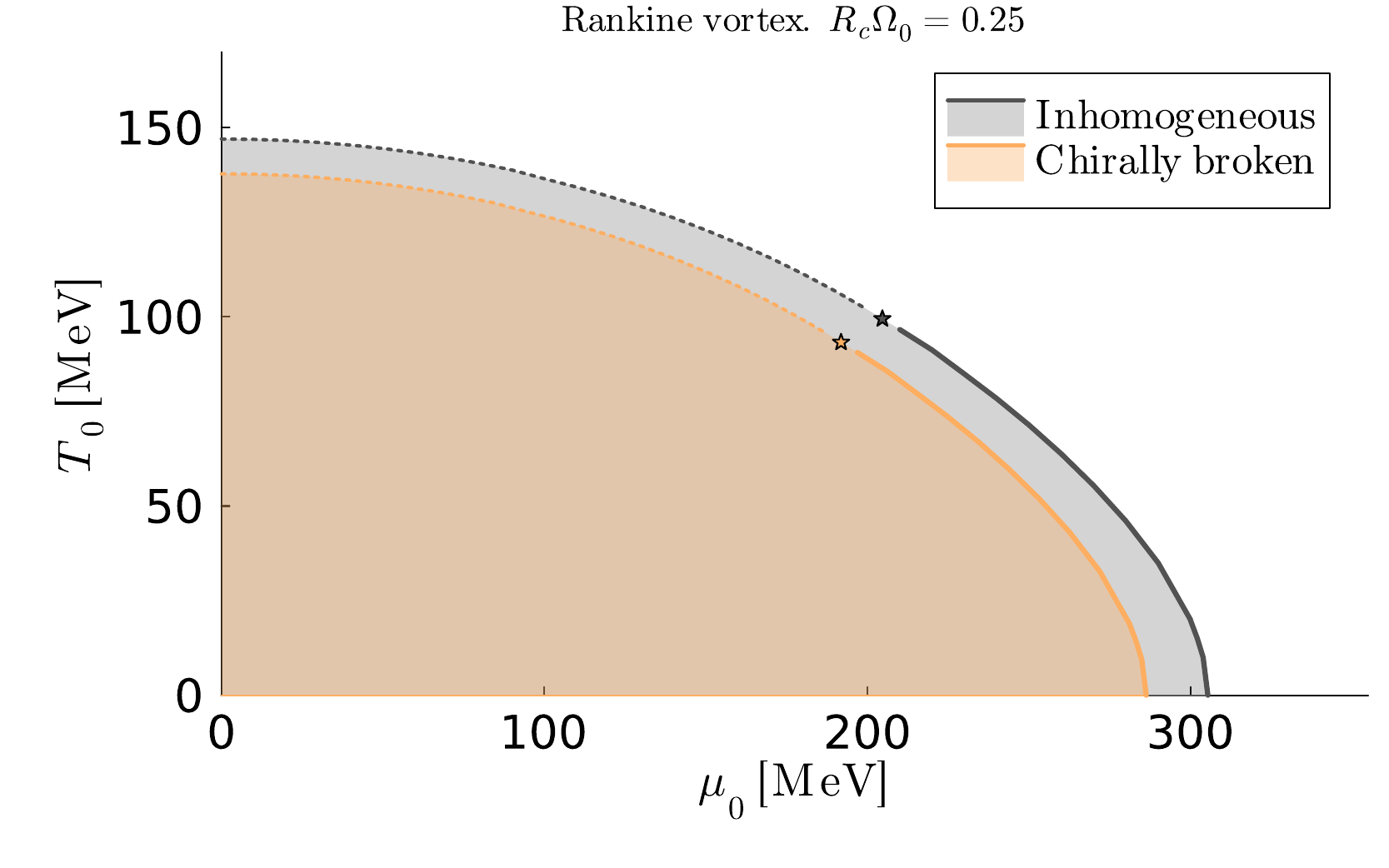} 
    \end{tabular}
    \caption{Expected modification of the phase diagram in the presence of a Rankine vortex with core radius $R_c\Omega_0 = 0.25$ and angular velocity $\Omega_0 = 10$ MeV. The inner (orange) region corresponds to a chirally broken phase, while in the outer (white) region, the system is in a chirally restored phase. The shadowed (gray) band indicates an inhomogeneous phase, where the system is in the chirally broken phase close to the rotation axis and in the chirally restored phase as we depart from the rotating core.
    In this case, models 2 and 3 give indistinguishable results.
    }
    \label{fig:Rankine-phase}
\end{figure}

The above discussion suggests that, for a sufficiently long-lived vortex, the phase diagram should be modified as shown in Fig.~\ref{fig:Rankine-phase}, where the system can reside in an inhomogeneous phase provided that the on-axis temperature $T_0$ and chemical potential $\mu_0$ lie slightly below the transition line. The example depicted in Fig.~\ref{fig:Rankine-phase} applies to phenomenologically relevant values of angular velocity $\Omega_0 = 10$ MeV, where the parameter $R_c\Omega_0 = 0.25$ implies that the vortex radius is about $R_c\simeq5$ fm. We point out that the quantum corrections to the fermion condensate, neglected in this work, would be irrelevant in this scenario, where both the angular velocity and the vortex core radius are sufficiently small.

\section{Discussion and Conclusions}\label{sec:conc}

A rigidly-rotating object must be spatially bounded in all directions normal to the axis of rotation, restricting velocities in the system from exceeding the speed of light and, therefore, from violating causality. In our work, we investigate a rigidly rotating plasma of interacting fermions in a spatially unbounded space, thus formally violating the causality condition. A series of approximations allowed us to study rotating systems even in the absence of boundaries and we argue that the results are qualitatively valid also for the bounded case.

We explain the similarity between the bounded and unbounded approaches by the appearance of what we term ``firewall boundary conditions'', which effectively establish an infinite-temperature state at the spatial surface of radius $R = 1/\Omega$, known as the light cylinder, beyond which the speed of co-rotating particles exceeds the speed of light. 
In the highly-symmetric state that we considered, azimuthal symmetry and vertical homogeneity ensure that the two regions separated by the light cylinder do not exchange any physical quantities like energy or charge, since all radial currents vanish. Nevertheless, the existence of infrared, superhorizon modes that extend beyond the light cylinder, cannot be excluded without the transverse momentum quantization induced by boundary conditions. In the case of bosonic fields, these superhorizon modes render the rigidly-rotating quantum state irregular everywhere \cite{Vilenkin:1980zv,Duffy:2002ss}. For fermions, one may argue that such infrared modes induce a discrepancy between the static (Minkowski) and rotating vacua \cite{Iyer:1982ah}, without affecting the regularity of the state inside the light cylinder. 

In the present paper, within the linear sigma model coupled to quarks (LSM$_q$), we considered the meson fields, $\sigma$ and $\vec{\pi}$, as classical fields. By ignoring their quantum fluctuations, the bosonic part remains unaffected by the problematic superhorizon modes. In the fermionic sector, we computed the finite-temperature observables (grand potential and fermion condensate) within the grand canonical ensemble at finite angular momentum. Here, we approximated the density operator by its local equilibrium version, thereby ignoring all problematic quantum corrections discussed above. In particular, within this local thermal equilibrium (LTE) approximation, the system is locally in a (static) equilibrium, being connected to the ``laboratory'' frame (as seen by a static observer) through a simple Lorentz boost. The grand canonical ensemble thus constructed is characterized by three parameters: the temperature $T_0$ and chemical potential $\mu_0$ on the rotation axis, as well as the rotation angular velocity~$\Omega$. 

Close to the rotation axis, the quantum corrections that were discarded in the LTE approach can be rightfully expected to be negligible for reasonable values of the rotation parameter, being proportional to $(\hbar \Omega / k_B T)^2$, where we restored the Boltzmann constant $k_B$. Close to the light cylinder, however, the quantum corrections become dominant \cite{Ambrus:2017opa}. While it is reasonable to expect that these corrections modify the details of the state, we argue that the main (global and local) features remain qualitatively the same. 

We formulated three approaches to the rotating quark-meson system, with increasing degree of accuracy. In model~1,  we assumed a uniform global condensate, $\bar{\sigma}$, which minimizes the total grand potential $\Phi_{\rm m.f.}$, computed over a cylinder of radius $R \le \Omega^{-1}$. In this approximation, the rotation parameter and the system size enter only through the combination $\Omega R$. When $\Omega R$ is small, the thermodynamics of the system resembles that of the static, non-rotating system. On the contrary, when $\Omega R \to 1$, the system thermodynamics become dominated by the region close to the light cylinder, and $\bar{\sigma} \to 0$. The thermodynamic phase of the system at some value of $\Omega R$ is governed by the average temperature $\overline{T}$ and chemical potential $\bar{\mu}$, computed by suitably averaging the local temperature and chemical potential given by the Tolman-Ehrenfest law within the cylinder of radius $\Omega R$ (see Fig.~\ref{fig_Averaged-crits}).

In model 2, we considered a slowly-varying $\sigma$ condensate. Neglecting the kinetic terms of the meson Lagrangian, we computed the local value $\sigma(\rho)$ of the condensate by demanding the local minimization of the grand potential $\phi_{\rm m.f.}(\rho)$. As in the case of model 1, the rotation parameter and distance to the rotation axis enter only through the combination $\rho \Omega$. The local thermodynamics of the system can be inferred directly by the Tolman-Ehrenfest law. Consequently, the radial profile of the sigma meson coincides with that obtained by taking a diagonal trajectory in the $T$-$\mu$ phase diagram of the static system. Thus, the chiral symmetry is always restored as $\rho \to \Omega^{-1}$. 

Model 2 allows the system to reside in a mixed phase. For small enough $T_0$ and $\mu_0$, the region close to the rotation axis will be in a chirally-broken phase. Chiral symmetry is restored either by a crossover or by a first-order transition with respect to the radial distance from the rotation axis. This particular phase ordering is consistent with that found in Ref.~\cite{Chernodub:2020qah} for the $(2+1)$ compact electrodynamics model, also shown to be consistent with the Tolman-Ehrenfest law.

Finally, in model 3, we obtained the $\sigma$ condensate as a solution of the Klein-Gordon equation, which becomes a second-order differential equation with respect to the radial distance $\rho$. Regularity on the rotation axis fixes one integration constant. The second integration constant is fixed by demanding that the total grand potential of the system is minimal. 
In the particular case when the system extends up to the light cylinder, we uncovered that the ``firewall'' paradigm selects as the thermodynamically-favored profile for $\sigma$ the one for which $\sigma(\rho) \to \sigma_{\rm LC} = 0$ as $\rho \to \Omega^{-1}$ and the light cylinder is approached.

The radial derivatives appearing in model 3 allow the angular velocity $\Omega$ to enter as a separate parameter. Similarly to model 2, $\sigma \to 0$ close to the light cylinder. On the rotation axis, model 3 agrees with model 2 only in the limit $\Omega \to 0$. On the other hand, obtaining $\sigma(\rho)$ as a solution of the radial differential equation precludes the development of sharp jumps. Thus, for any finite value of $\Omega$, systems that are chirally-broken on the rotation axis become chirally-restored via a crossover transition with respect to $\rho$. 

Within model 3, we also characterized the transition from the case when the system is in a mixed (broken$+$restored) phase to the case when the chiral symmetry is fully restored. This transition can be discussed at the level of the condensate on the rotation axis, $\sigma_0$. At small values of $\Omega$, the $T$-$\mu$ phase diagram is identical to that of the static system and of model 2. As $\Omega$ is increased, the phase transition line in the $T$-$\mu$ plane shrinks, moving towards smaller $T$ and $\mu$. When $\Omega$ exceeds a critical value of $\Omega_c \simeq 175$ MeV, the system is chirally restored at any chemical potential and temperature. This unexpected result is supported by the $\Omega$-$T$ phase diagram, constructed at $\mu_0 = 0$; as well as by the $\Omega$-$\mu$ diagram, computed at $T_0 =0$.

We employed Models 2 and 3 to the more realistic case when the angular velocity has an inhomogeneous radial profile, $\Omega_\rho \equiv \Omega(\rho)$. We considered the Rankine vortex configuration, which comprises a vortical core rotating with constant angular velocity $\Omega_0$, of finite radius $R_c$. Outside this core, the flow is vorticity-free and the angular velocity decays as $\Omega_\rho = \mathcal{C}_\Omega / \rho^2$. This configuration ensures  the continuity of the angular velocity, reducing the effects of dissipative damping.

We now comment on the expected effects of enclosing the system in a physical boundary. As shown in Ref.~\cite{Ambrus:2015lfr} for the commonly-used spectral and MIT boundary conditions, the bulk of the system remains unaffected, so long as $R T \gtrsim 1$. On the other hand, the firewall effect is suppressed, even as $\Omega R \to 1$, as the effect of the boundary leads to a suppression of the fermion condensate near the boundary, thereby reversing the effect of rotation. Models 1 and 2 can be expected to give results similar to those obtained here. For model 3, one must find the integration constant $\sigma_0$ by explicitly minimizing the total grand potential, which will now be finite, even in the $\Omega R \to 1$ limit.

Quantum corrections due to the full density operator of the rotating state introduce $\Omega$ as an energy scale and facilitate chiral restoration by increasing the fermion condensate. Moreover, quantum corrections become dominant as $\Omega R \to 1$. For models 1 and 2, it is clear that chiral symmetry restoration will occur at smaller system sizes. In the case of model 3, on one hand, we expect an enhancement of chiral restoration, such that the phase-transition lines of constant $\Omega$ travel towards the lower-left corner of the phase diagram. On the other hand, we can expect that the critical angular velocity $\Omega_c$ beyond which the system is in the chirally-restored phase also decreases. Therefore, it is so far unclear whether the purple region in Fig.~\ref{fig:Tmuphase}, bounded by the phase transition line for $\Omega = \Omega_c$, will increase of decrease due to the addition of the quantum correction terms.

Finally, we comment on more general implications of rotation on the phase diagram of the strongly-interacting matter. All three approaches considered in this paper predict the same qualitative effect: rotation favors chiral symmetry restoration, lowering the critical temperature of the chiral transition. However, first-principle numerical simulations of a purely gluonic plasma, which largely defines the non-perturbative properties of quark-gluon plasma, show a reversed order of phases for relatively slow rotation~\cite{Braguta:2021jgn, Braguta:2023iyx, Braguta:2024zpi}. The apparent inconsistency with the Tolman-Ehrenfest law is ascribed to the evaporation of the magnetic component of the gluonic condensate~\cite{Braguta:2023tqz}, which cannot be captured by the standard formulation of the linear sigma model coupled to quarks.\footnote{At higher rotation, the order seems to be reversed again~\cite{Chernodub:2022veq}, becoming  consistent with the Tolman-Ehrenfest effect. However, the latter observation, made in numerical simulations of the lattice Yang-Mills theory, may be spoiled by artifacts related to a too-fast Euclidean rotation~\cite{Ambrus:2023bid}.} A possible extension of the present work is to consider the Polyakov loop-enhanced LSM$_q$ model, taken with models 1-3. 

\acknowledgments 
We thank P.~Aasha for fruitful discussions. This work was funded by the EU’s NextGenerationEU instrument through the National Recovery and Resilience Plan of Romania - Pillar III-C9-I8, managed by the Ministry of Research, Innovation and Digitization, within the project entitled ``Facets of Rotating Quark-Gluon Plasma'' (FORQ), contract no. 760079/23.05.2023 code CF 103/15.11.2022.

\bibliography{plasma}

\end{document}